\documentclass[aps,rmp,twocolumn]{revtex4}

\usepackage{amsmath}
\usepackage{amssymb}
\usepackage{graphicx}
\usepackage{psfrag}
\usepackage{mathrsfs}
\usepackage[usenames]{color}

\newcommand{\ecoli}{\emph{E.coli\/} }
\newcommand{\tr}{\mathsf{t}}

\newcommand{\PRE}{{\it Phys. Rev. E.} }
\newcommand{\PRL}{{\it Phys. Rev. Lett.} }

\newcommand{\PNAS}{{\it Proc. Natl. Acad. Sci. U.S.A.\/} }
\newcommand{\NAT}{{\it Nature\/} }

\begin{document}

\title{Single DNA conformations and biological function}

\author{Ralf Metzler}
\email{metz@nordita.dk}
\affiliation{NORDITA---Nordic Institute for Theoretical Physics,
Blegdamsvej 17, 2100 Copenhagen \O, Denmark}
\affiliation{Department of Physics, University of Ottawa, 150 Louis
Pasteur, Ottawa, Ontario, K1N 6N5, Canada (New address)}
\author{Tobias Ambj{\"o}rnsson}
\affiliation{NORDITA---Nordic Institute for Theoretical Physics,
Blegdamsvej 17, 2100 Copenhagen \O, Denmark}
\affiliation{Department of Chemistry, Massachusetts Institute of Technology,
77 Massachusetts Avenue, Cambridge, Massachusetts 02139 (New address)}
\author{Andreas Hanke}
\affiliation{Department of Physics and Astronomy, University of Texas at
Brownsville, 80 Fort Brown, Brownsville, TX 78520}
\affiliation{Institute of Biomedical Sciences and Technology, University of
Texas at Dallas, Richardson, TX 75083}
\author{Yongli Zhang}
\affiliation{Department of Physiology and Biophysics, Albert Einstein
College
of Medicine, Bronx, NY 10461}
\author{Stephen Levene}
\affiliation{Institute of Biomedical Sciences and Technology and
Department of Molecular and Cell Biology, University of Texas at Dallas,
Richardson, TX 75083}

\begin{abstract}
Abstract. From a nanoscience perspective, cellular processes and their
reduced in vitro imitations provide extraordinary examples for highly
robust few or single molecule reaction pathways. A prime example are
biochemical reactions involving DNA molecules, and the coupling of these
reactions to the physical conformations of DNA. In this review,
we summarise recent results on the following phenomena: We investigate the
biophysical properties of DNA-looping and the
equilibrium configurations of DNA-knots, whose relevance to biological
processes are increasingly appreciated. We discuss how random DNA-looping
may be related to the efficiency of the target search process of proteins for
their specific binding site on the DNA molecule. And we dwell on the
spontaneous formation of intermittent DNA nanobubbles and their importance
for biological processes, such as transcription initiation. The physical
properties of DNA may indeed turn out to be particularly suitable for the
use of DNA in nanosensing applications.\\[0.2cm]
\end{abstract}

\maketitle

\vspace*{8cm}

\newpage
\clearpage

\tableofcontents

\clearpage

\section{Introduction}

Deoxyribonucleic acid (DNA) is the molecule of life as we know
it.\footnote{Our DNA world during
biotic and prebiotic evolution was supposedly preceded by an RNA world and,
quite likely, by sugarless nucleic acids.} It contains all information of
an entire organism.\footnote{A small fraction of genetic information is stored
on DNA that is kept at other regions of the cell and not replicated on cell
division, such as mitochondrial or ribosomal DNA.}
This information is copied during cell division with an extremely
high fidelity by the replication mechanism. Despite the rather high
chemical and physical stability of DNA, due to constant action of enzymes
and other binding proteins (mismatches, rupture) as well as potential
environmentally induced damage (radiation, chemicals), this low error rate,
i.e., the suppression of the liability to mutations, is only possible with the
constant action of repair mechanisms \cite{alberts,snustad,kornberg,kornberg1}.
Although DNA's structural and mechanical properties are rather well established
for isolated DNA molecules (starting with Rosalind Franklin's X-ray diffraction
images \cite{franklin}), the characterisation
of DNA in its cellular environment, and even in vitro during interaction with
binding proteins, is subject of ongoing investigations.


Recent advances in experimental techniques such as fluorescence methods,
atomic force microscopy, or optical tweezers have leveraged the potential
to both probe and manipulate the equilibrium and out of equilibrium
behaviour of \emph{single\/} DNA molecules, making it possible to explore DNA's
physical and mechanical properties as well as its interaction with other
biopolymers, such as the DNA-protein interplay during gene regulation or repair
processes. An important ingredient is the coupling to thermal activation due
to the highly
Brownian environment. Although mostly performed in vitro, these experiments
provide access to increasingly refined information on the nature of DNA and
its environment-controlled behaviour.

In addition to chromosomal packaging inside the nucleus of eukaryotic cells
and the concentration of DNA in the membraneless nucleoid region of
prokaryotes, the global
structure of the DNA molecule can be affected by topological entanglements.
Thus, by error or design a DNA molecule can attain a knotted or concatenated
state, reducing or inhibiting biologically relevant functions, for instance,
replication or transcription. Such entangled states
can be actively reduced by enzymes of the topoisomerase family. Their precise
action, in particular, how they determine the presence of an entangled state,
is not fully known. Current studies therefore aim at shedding light on
possible mechanisms, in particular, in view of the importance of
topoisomerase action (or better, its inhibition) in tumour proliferation.
Other applications may be directed towards the treatment of viral deceases
by modifying the packaging of viral DNA to create knots in the virus capsid
and prevent ejection of the DNA into a host, and thereby infection.
DNA knots are also being recognised as a potential complication in the use
of nanochannels for DNA separation and sequencing. In such confined geometries
DNA knots are created with appreciable probability, affecting the reliability
of these techniques. Similarly to DNA knots, DNA looping is intimately
connected to the function of DNA. Current results on DNA looping and DNA
knot behaviour are summarised in the first parts of this review.

The Watson-Crick double-helix represents the thermodynamically stable state
of DNA at moderate salt concentrations and below the melting temperature.
This stability is effected by Watson-Crick hydrogen bonding and the
stronger base stacking of neighbouring base-pairs (bps). However, even at room
temperature DNA locally opens up intermittent flexible single-stranded domains,
so-called DNA-bubbles. Their size typically ranges from a few broken
bps, increasing to some 200 broken bps closer to the
melting temperature. The thermal melting of DNA has traditionally been used
to obtain the sequence-dependent stability parameters of DNA. More recently,
the role of intermittent bubble domains has been investigated with respect
to the liability of DNA-denaturation induced by proteins that selectively
bind to single-stranded DNA. It has been speculated that due to the liability
to denaturation of the TATA motif bubble formation may add in
transcription initiation. The dynamics of single bubbles can be monitored
by fluorescence methods, opening a window to both study the breathing of
DNA experimentally, but also to obtain high precision DNA stability data.
Finally, bubble dynamics has been suggested as a useful tool in optical
nanosensing. DNA breathing is the topic of the second part of this work.

Essentially all the biological functions of DNA rely on site-specific
DNA-binding proteins locating their targets (cognate sites) on the DNA
molecule, and therefore require searching through megabases of non-target
DNA in a highly efficient manner. For instance, gene regulation is performed
by specific regulatory proteins. On binding to a promoter area on the DNA,
they recruit or inhibit binding of RNA polymerase and subsequent
transcription of the associated gene. The search for the cognate site is
in fact facilitated by the DNA molecule: in addition to three-dimensional
search it enables the proteins to also move one-dimensionally along the
DNA while being non-specifically bound. Moreover, at points where the DNA
loops back on itself, this polymeric conformation provides shortcuts for
the proteins in the chemical coordinate along the DNA, approximately giving
rise to search-efficient L{\'e}vy flights. Target search is currently a very
active field of research, and single molecule methods have been shown to
provide essential new information. Moreover, the architecture of more complex
promoters relying on the simultaneous presence of several regulatory
proteins is being investigated to create in silico circuits for highly
sensitive chemical probes in small volumes. Such nanosensing applications
are expected to be of great importance in microarrays or other nano- and
microapplications. The third part of this review deals with diffusional
aspects of gene regulation.

At the same time DNA's role in classical polymer physics is increasingly
appreciated. With the possibility to reproduce DNA with extremely low
error rate by the PCR\footnote{Polymerase Chain Reaction: thermal
denaturation of a DNA molecule into two single strands and subsequent
cooling in a solution of single nucleotides and invariable primers,
produces two new complete double-stranded DNA molecules. Cycling of this
process produces large, monodisperse quantities of DNA.}, monodisperse
samples can be prepared. While shorter single-stranded DNA can be
used as a model for flexible polymers, the double strand exhibits a
semiflexible behaviour with a persistence length, that can be easily
probed experimentally. Moreover, DNA
is orders of magnitude longer than conventional polymers. Combined with
the potential of single molecule probing, DNA is advancing as a model
polymer.

After an introduction to the properties of DNA we address these functional
properties of DNA from the perspective of biological relevance, physical
behaviour and nanotechnological potential. Most emphasis will be put on
the single molecular aspects of DNA. We note that this is not intended to
be an exhaustive review on the physical properties of DNA. Rather, we
present some important features and their consequences from a personal
perspective.

\section{Physical properties and biological function of DNA}

Biomolecules, that occur naturally in biological systems, can be grouped
into unspecific oligo- and macromolecules and biopolymers in the stricter sense
\cite{alberts}. Unspecific biomolecules are
produced by biological organisms in a large range of molecular weight and
structure, such as polysaccharides (cellulose, chitin, starch, etc.),
higher fatty acids, actin filaments or microtubules. Also
the natural `india-rubber' from the {\em Hevea Brasiliensis\/} tree,
historically important for both industrial purposes and the development of
polymer physics \cite{treloar} belongs to this group.

\begin{figure}
\includegraphics[width=6.4cm]{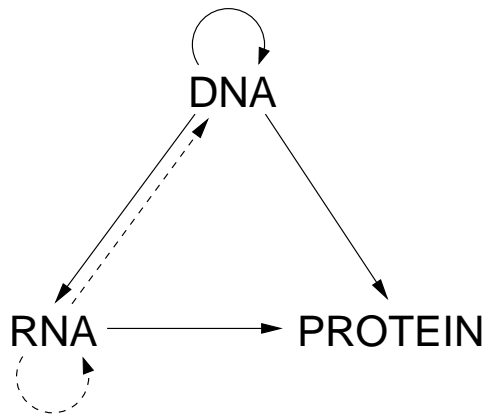}
\caption{Central dogma of molecular biology after F. Crick:
Potentially, information flow is completely symmetric
between the three levels of cellular biopolymers (DNA, RNA, proteins).
However, the recognised pathways are only those represented here,
where solid lines represent probable transfers, and dotted lines for
(in principle) possible transfers \protect\cite{crick}.
\label{crick}}
\end{figure}

Biopolymers in the stricter sense we are going to assume here comprise
the polynucleotides DNA and RNA consisting of the four-letter nucleotide
alphabet with A-T and G-C (A-U and G-C for RNA) bps, and the
polypeptidic proteins consisting of 20 different amino acids, each coded for
by 3 bases (codons) in the RNA \cite{alberts,snustad,kornberg,kornberg1}.
We will come back to proteins later when reviewing binding protein-DNA
interactions. Biopolymers are copied and/or created according
to the information flow sketched in figure \ref{crick}, the so-called central
dogma of molecular biology, a term originally coined by Frances Crick
\cite{crick1}. Accordingly, starting from the genetic code stored in the DNA
(in some cases in RNA) DNA is copied by DNA polymerase (replication), and
the proteins as the actually task-performing biopolymers are created via
messenger RNA (created by DNA transcription through RNA polymerase) and
further by translation in ribosomes to proteins.\footnote{Alternatively, the
genetic code can be transcribed into transfer and ribosomal RNA that is not
translated into proteins.}

\begin{figure}
\includegraphics[height=7.68cm]{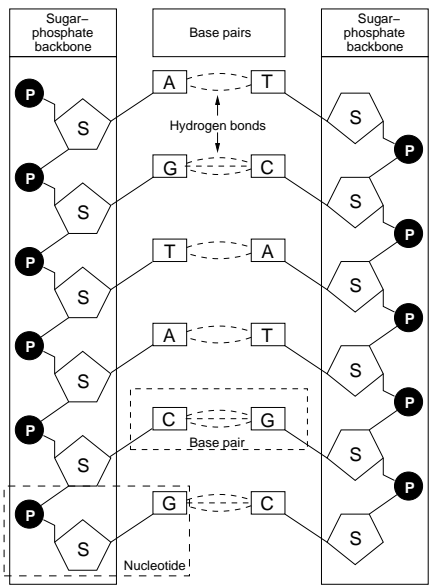}
\caption{Ladder structure of the DNA formed by its four building-blocks
A, G, C, and T, giving rise to the typical double-helical structure
of DNA. A-T bps establish 2 H-bonds, G-C bps 3 H-bonds.
\label{dh}}
\end{figure}

DNA is made up of the four bases
\cite{alberts,snustad,kornberg,kornberg1,bloomfield,rnaworld}:
A(denine), G(uanine), C(ytosine), and
T(hymine) that form the DNA ladder structure shown in figure \ref{dh}.
These building-blocks A, G, C, T bp according to the key-lock principle
as A-T and G-C, where the AT bond is weaker than the GC bond in terms of
stability. Apart from the Watson-Crick base-pairing energy, the stability
of dsDNA is effected by the stacking interactions, the specific matching
of subsequent bps along the double-strand, i.e., bp-bp
interactions. In standard literature, the stacking interactions are listed
for pairs of bps (e.g., for AT-GC, AT-AT, AT-TA, etc.), see
below.\footnote{Longer ranging bp-bp interactions are most likely small in
comparison.}

\begin{figure*}
\includegraphics[width=18cm]{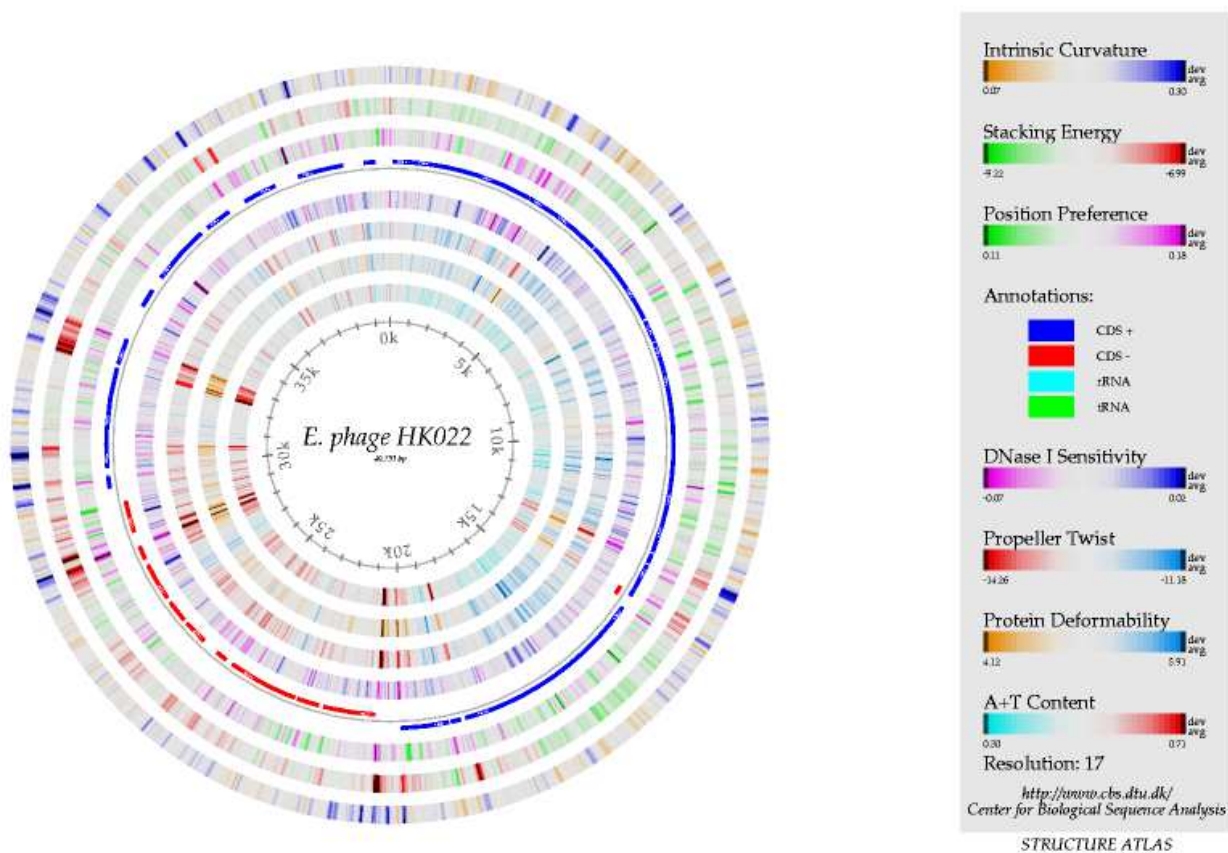}
\caption{Structure atlas of the \ecoli genome. Figure courtesy David Ussery,
Technical University of Denmark. The structure atlas is available under
the URL {\tt www.cbs.dtu.dk/services/GenomeAtlas/}.
\label{structure}}
\end{figure*}

Based on this AGCT alphabet, the
primary structure of DNA can be specified. DNA's six local structural elements
twist, tilt, roll, shift, slide, and rise are effected by the stacking
interactions between vicinal bps. In figure \ref{structure}, we show a map
with the structure elements of the entire \ecoli genome, demonstrating the
degree of structural information currently available. These structural
elements define the local geometrical structure of DNA within a typical
correlation (persistence) length\footnote{The persistence length of a polymer
chain defines the characteristic length scale above which the polymer is
susceptible to bending induced by thermal fluctuations, i.e., it is the
length scale above which the tangent-tangent correlation decays along the
chain, see the Appendix.} of about 150 bps corresponding to 50 nm
(the bp-bp distance measures 3.4 {\AA}, reflecting the rather complex chemical
structure of a nucleotide in comparison to the monomer size of man-made
polymers such as polyethylene) \cite{frank,frank1,marko,marko1}.
On a larger scale, much longer than the persistence length,
DNA becomes flexible. On this level, tertiary structural elements come
into play. One example is DNA looping, that is the formation of polymeric
lasso loops induced by chemical bonds between binding proteins attached to
the DNA at specific bps which are remote along the DNA backbone
\cite{alberts,snustad,ptashne1,revet,bell,bell1,hame_looping}.
An extreme limit of tertiary structure is the
packaging of DNA onto histones and further wrapping into the chromosomes
of eukaryotic cells \cite{schiessel,kreth,alberts}. At the same time, dsDNA may
locally open into floppy ssDNA bubbles, with a persistence length of a few
bases.\footnote{In fact, it has been questioned whether there is a meaningful
value of the persistence length of ssDNA at all, due to its significant
apparent sequence dependence \cite{goddard}.} These fluctuation-induced bubbles
increase their statistical weight at higher temperatures, until the dsDNA fully
denatures (melts). We will come back to DNA denaturation
bubbles below. Depending on the
external conditions, DNA occurs in several configurations. Under physiological
conditions,
one is concerned with B-DNA, but there are other states such as A, B', Z, ps,
triplex DNA, quadruplex DNA, cruciform, and H, reviewed, for instance, in
\cite{frank,frank1}. DNA occurs naturally in a large
range of length scales. In viruses, DNA is of the order of a few $\mu$m
long. In bacteria, it already reaches lengths of several mm, and in mammalian
cells it can reach the order of a few m, roughly 2 m in a human cell and 35 m
in a cell of the South American lungfish, albeit split up into the
individual chromosomes \cite{kornberg}. DNA in bacteria in
vivo, or extracted from bacteria and higher cells for our purposes can
therefore be viewed a fully flexible polymer with a persistence length of
roughly 50 nm, being governed by generic effects independent of the detailed
sequence. On short scales DNA becomes semiflexible and governed by the
worm-like chain model (Kratky-Porod model) \cite{grosberg}; on even shorter
scales, local structural elements become important (in particular, for
recognition by binding proteins \cite{alberts}), and eventually molecular
resolution is reached.

Stacking interactions govern the local structure of dsDNA. Globally,
an additional constraint arises due to the circular nature of the DNA,
since it has to satisfy the conservation law \cite{calu,white1,fuller}
\begin{equation} \label{ltw}
{\rm Lk}={\rm Tw}+{\rm Wr},
\end{equation}
where Lk stands for the linking number, Tw for the twist, and Wr for the
writhe of the double helix. The linking number $Lk$ is an integer and
formally
given by one-half the number of signed crossings of one DNA strand with the
other in any regular projection of the molecule. $Lk$ is a topological
property, and no deformation of a closed DNA, without breaking and rejoining
the DNA strands, will alter it. $Tw$ is equal to the number
of times that the two strands of DNA wind about the central
axis of the molecule, and $Wr$ is a number whose absolute value equals
approximately the number of times that the DNA axis winds about
itself.\footnote{For details about the calculation of Tw and Wr for
representative models of DNA, see \cite{white}.}
Whereas Tw is a property of the double-helical structure of DNA,
Wr is a property of the DNA axis alone. Tw and Wr do not need to
be integers and are not conserved, but coupled through
Lk by equation (\ref{ltw}).
A nicked circular DNA, i.e., when the twist can fully relax, carries
${\rm Lk}_0=N/h$ links, where $N$ is the number of bp and $h$ ($h\simeq 10.5$
in B-DNA) the number of bps per turn.

\begin{figure}
\includegraphics[width=7.6cm]{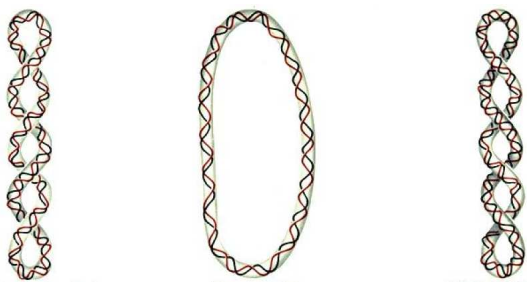}
\caption{Right-handed (negative), normal, and left-handed (positive)
superhelix. The DNA of virtually all terrestrial organisms is
negatively supercoiled.}
\label{wind1}
\end{figure}

\begin{figure}
\includegraphics[height=6.0cm]{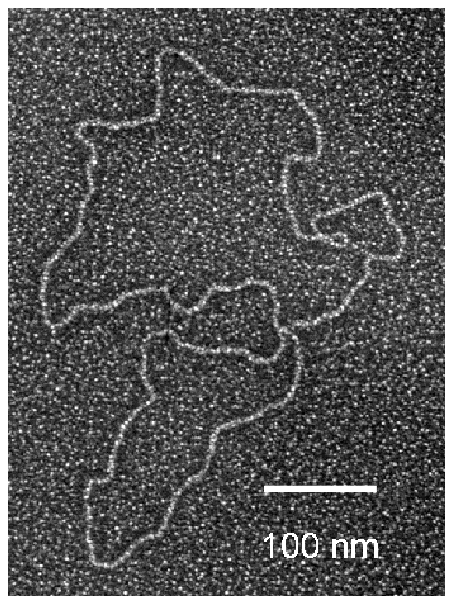}
\includegraphics[height=6.0cm]{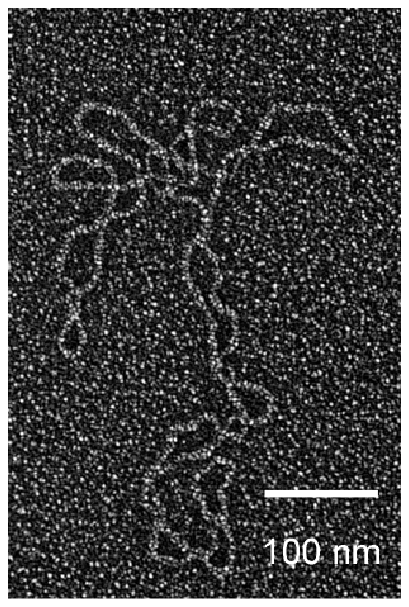}
\caption{Electron micrographs of nicked (left) and supercoiled (right)
6996-bp plasmid DNAs. The supercoiled example is from a population of DNA
molecules with an average superhelix density, $\bar {\sigma }$= -0.027,
close to the value expected \textit{in vivo}.}
\label{wind2}
\end{figure}

\begin{figure}
\includegraphics[width=7.2cm]{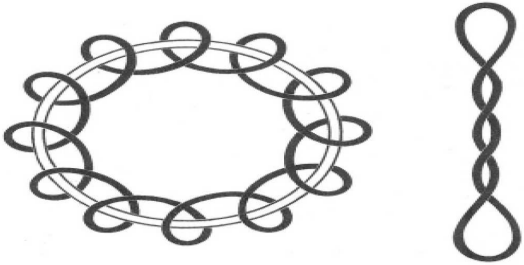}
\caption{Toroidal (left) and plectonemic (right) forms of supercoiled DNA.}
\label{wind3}
\end{figure}

The degree of \emph{supercoiling\/} of DNA can be expressed in terms of
the linking number difference, $\Delta Lk=Lk-Lk_0$.
The DNA of virtually all terrestrial organisms is underwound or
{\em negatively supercoiled}, i.e., $\Delta Lk<0$
(figures \ref{wind1} and \ref{wind2}).\footnote{An exception are
thermophilic organisms living near
undersea geothermal vents that have positively supercoiled DNA in order to
stabilise the double helix at extreme temperatures.}
Often, the superhelical density $\sigma=\Delta Lk/Lk_0$ is used; most
supercoiled DNA molecules isolated from either prokaryotes or eukaryotes
have $\sigma$ values between $-0.05$ and $-0.07$ \cite{bauer}.
Negative supercoiling is regulated in prokaryotes by DNA gyrase; eukaryotes
lack gyrase but maintain negative supercoiling through winding of DNA around
nucleosomes and interactions with DNA-unwinding proteins.
There are two forms of intracellular supercoiling, the {\it plectonemic}
form, characteristic of plasmid DNA and accessible, nucleosome-free
regions of chromatin, and the {\it toroidal} or {\it solenoidal} form,
where supercoiling is
attained by DNA wrapped around histone octamers or prokaryotic non-histone
DNA-binding proteins (figure \ref{wind3}).
The former is the active form of supercoiled DNA and
is freely accessible to proteins involved in transcription, replication,
recombination and DNA repair. The latter is the stored form of supercoiled
DNA and is largely responsible for the extraordinary degree of compaction
required to condense typical genomes into the cell's nucleus.\footnote{The
nucleus of a human cell has a radius of circa 5 $\mu$m and stores the 2 m
of the human genome \cite{sun}.}
Negative supercoiling facilitates the local unwinding of DNA by providing
a ubiquitous source of free energy that augments the unwinding free energy
accompanying the interactions of many proteins with their cognate DNA
sequences. The local unwinding of DNA, in turn, is an integral part of many
biological processes such as gene regulation and DNA replication
(see section \ref{generegulation}). Therefore, understanding the interplay of
supercoiling and local helical structure is essential to the
understanding of biological mechanisms
\cite{marko1,benham,goetze,levene,levene1}.

Ribonucleic acid (RNA)
consists of the same building blocks as DNA, with the exception
that T(hymine) is replaced by U(racile) \cite{rnaworld}. RNA typically occurs
in single-stranded form. Therefore, its secondary structure is richer, being
characterised by sequences of hairpins: Smaller regions in which
chemically remote sequences
of bases match, pair and form hairpins which are stiff and energy-dominated,
similar to dsDNA. The remaining regions form entropy-dominated floppy loops,
analogous to the ssDNA bubbles. Additional
tertiary structure in RNA comes about by the formation of so-called
pseudoknots, chemical bonds established between bases sitting on chemically
distant segments of the secondary structure. In RNA-modelling
the incorporation of pseudoknots is a non-trivial problem, which currently
receives considerable interest; see, for instance,
references \cite{rnaworld,orland,orland1,baiesips}.

\section{DNA-looping}

The formation of DNA loops mediated by proteins bound at distant sites 
along a single molecule is an essential mechanistic aspect of many biological 
processes including gene regulation, DNA replication, and recombination 
(for reviews, see \cite{loop1,loop2}).
In {\em E. coli}, DNA looping represses gene expression 
at the {\em ara}, {\em gal}, {\em lac}, and {\em deo} 
operons \cite{loop3,loop4,loop5,loop6}
and activates transcription from 
the {\em gln}ALG operon \cite{loop7}.  
The size of DNA loops formed in these systems varies 
between approximately 100 and 600 bps.  In eukaryotes, a variety of
transcription factors bind to enhancers that are hundreds to several 
thousand bps away from their promoters and interact with RNA 
polymerases directly or through mediators in order to achieve combinatorial 
gene regulation \cite{loop8}.
DNA looping is required to juxtapose two recombination 
sites in intramolecular site-specific recombination 
\cite{loop9,loop10,loop11} and is also
employed by a number of restriction endonucleases such 
as {\em Sfi}I and {\em Ngo}MIV, 
which recognise and cut two copies of well-separated cognate sites 
simultaneously \cite{loop12,loop13,loop14}.
Here we describe a recent statistical-mechanical theory of 
loop formation that 
connects global mechanical and geometric properties of both DNA and 
protein and demonstrates the importance of protein flexibility in 
loop-mediated protein-DNA interactions \cite{loop26,loop51}.

\subsection{Biological significance of DNA looping}
\label{loopbio}

The biological importance of DNA loop formation 
is underscored by the abundance of architectural proteins in the cell such 
as HU, IHF, and HMG, which facilitate looping by bending the intervening
DNA between protein-recognition sites \cite{loop15}.
Moreover, DNA looping has been 
shown to be subject to regulation through the binding of effector 
molecules that alter protein conformation or protein-DNA 
interactions \cite{loop16}.

Two characteristics of DNA looping have been demonstrated by {\em in vitro}
and {\em in vivo} experiments. One is cooperative binding of a protein 
to its two cognate sites, which can be demonstrated by footprinting 
methods \cite{loop17}.  
DNA looping can increase the occupancies of both binding sites; in particular,
it can significantly enhance protein association to the lower-affinity site
because of the tethering effect of DNA looping.  This is a general mechanism 
by which many transcription factors recruit RNA polymerases in gene
regulation. Another hallmark is the helical dependence of loop formation
\cite{loop1,loop3}, which arises because of DNA's limited torsional 
flexibility and the requirement for correct 
torsional alignment of the two protein-binding sites.
Although many methods have been developed to directly observe DNA looping 
{\em in vitro}, such as scanning-probe \cite{loop7}
and electron microscopy \cite{bloomfield}, and single-molecule techniques
\cite{loop19}, assays based on helical dependence have 
been the only way to identify DNA looping {\em in vivo}. In these experiments, 
the DNA length between two protein binding sites is varied and the yield 
of DNA loop formation is monitored, for example by the repression or 
activation of a reporter gene \cite{loop20}.
Using this helical-twist assay, DNA 
looping in the {\em ara} operon was first discovered \cite{loop3}.

Our knowledge about the roles of DNA bending, twist, and their respective 
energetics in DNA looping has come largely from analyses of DNA 
cyclisation \cite{loop1,loop21,loop22}.
Circularisation efficiencies of DNA fragments, which are 
quantitatively described by $J$-{\em factors}, 
oscillate with DNA length and therefore torsional 
phase \cite{loop23,loop24}. The $J$-factor is defined as the
ratio of the partition function of a circularised polymer chain
to that of an open chain. Since there is a dimension reduction due to
circularisation constraints (two polymer ends have to meet), the ratio
has a unit of concentration, or $1/ L^3$ with $L$ representing length;
see \cite{loop26} for details.
In the present context, the $J$-factor is equal to the free DNA-end
concentration whose bimolecular ligation efficiency
equals that of the two ends of a cyclising DNA molecule \cite{loop25}.
For short DNA fragments $J$-factors are 
limited by the significant bending and twisting energies required to form 
closed circles, whereas for long DNA, the chain entropy loss during
circularisation exceeds the elastic-energy decrease and reduces the 
$J$-factor.  Because of this competition between bending and twisting 
energetics and entropy, there is an optimal DNA length for 
cyclisation \cite{loop26}.
Analogous behaviour has been expected for DNA looping, especially with 
respect to the helical dependence discussed above.

Quantitative analyses of DNA looping and cyclisation are challenging 
problems in statistical mechanics and have been largely limited to 
Monte Carlo or Brownian dynamics simulations 
\cite{loop27,loop28,loop29,loop30,loop31}. Analytical 
solutions are available only for some ideal and special cases.
An important contribution in this area is the theory of Shimada and 
Yamakawa \cite{loop32}, which is based on a 
homogeneous and continuous elastic 
rod model of DNA.  This theory has been applied extensively to DNA 
cyclisation \cite{loop23,loop33} and also DNA looping 
\cite{loop21,loop22,loop34}. The Shimada-Yamakawa
theory makes use of a perturbation approach, in which small configurational 
fluctuations of a DNA chain around the most probable configuration are 
accounted for in the evaluation of the partition function.

The elastic-equilibrium conformation is obvious for the homogeneous DNA 
circle studied by Shimada and Yamakawa \cite{loop32}.
However, the search for the 
elastic-energy minimum of homogeneous DNA molecules with complex geometry, 
such as in DNA looping, supercoiling, and the case of inhomogeneous DNA 
sequences containing curvature and nonuniform DNA flexibility, is not 
trivial \cite{loop4,loop35,loop36}. 
Recently, a statistical-mechanical theory for sequence-dependent 
DNA circles has been developed \cite{loop26} and applied to the
problem of DNA cyclisation \cite{loop26} and DNA looping \cite{loop51}. 
In this model, the DNA configuration is described by 
parameters defined at dinucleotide steps, i.e., tilt, roll, and twist, 
which allows straightforward incorporation of intrinsic or protein-induced 
DNA curvature at the bp level. Following Shimada and Yamakawa's
method, the theory first determines the mechanical equilibrium configuration
in small DNA circles (i.e., less than $\sim 1000$ bp) under certain
constraints; fluctuations around the equilibrium configuration are then
taken into account using an harmonic approximation.
The new method is much more computationally 
efficient than Monte Carlo simulation, has comparable accuracy, 
and has been applied successfully to analyse experimental results from 
DNA cyclisation \cite{loop26}. 

The basis of the extension of the model
to DNA looping \cite{loop51} is 
to treat the protein subunits as connected rigid bodies and to allow for 
a limited number of degrees of freedom between the subunits.  Motions of 
the subunits are assumed to be governed by harmonic potentials and an 
associated set of force constants, neglecting the anharmonic terms often 
required for proteins undergoing large conformational fluctuations among 
their modular domains.  Indeed, the use of a harmonic approximation is 
supported by the success of continuum elastic models that are based only 
on shape and mass-distribution information in descriptions of protein 
motion \cite{loop37}.
Similar to the description used for individual DNA bps 
in the model, protein geometry and dynamics are described by three 
rigid-body
rotation angles (tilt, roll, and twist).  Therefore, DNA looping can be 
viewed as a generalisation of DNA cyclisation in which the protein component 
is characterised by a particular set of local geometric constraints and 
elastic constants. This treatment not only unifies the theoretical 
descriptions of DNA cyclisation and looping, but also allows consideration 
of flexibilities at protein-DNA and protein-protein interfaces and 
application of the concepts of linking number and writhe. In previous 
work, proteins were considered rigid and their effects on DNA configuration 
were represented by a set of constraints applied to DNA ends 
\cite{loop1,loop38,loop39}.
With the present approach, programs developed for analysing DNA cyclisation 
can be used to analyse DNA looping with only minor modifications.

The new method \cite{loop26,loop51} is most applicable to the problem 
of short DNA loops, in which 
the free energy of a wormlike chain is dominated by bending and torsional 
elasticity \cite{loop26,loop51}.
Possible modes of 
DNA self contact and contacts between protein and DNA at positions other 
than the binding sites are not considered.
For large loops 
contributions to the free energy from chain entropy and DNA-DNA contacts 
can become highly significant. Several alternative treatments of DNA 
looping have appeared recently. One of these addresses the excluded-volume 
contribution to DNA looping within large open-circular molecules 
\cite{hame_looping},
whereas two others consider the effect on looping of traction at the ends 
of a DNA chain \cite{loop41,loop42}.
None of these treatments includes helical phasing
effects on DNA looping.  In contrast, a method based on the Kirchhoff 
elastic-rod model, which includes the helical-phase dependence, has been 
presented \cite{loop39,loop43}.
However, this approach does not include thermal 
fluctuations {\em per se} and therefore is not directly applicable to 
calculations of the $J$-factor.  The comprehensive treatment of small DNA
loops described in \cite{loop26,loop51}
is thus far unique to the extent that it accounts for 
sequence- and protein-dependent conformational and flexibility parameters, 
thermal fluctuations, and helical phasing effects.

\subsection{DNA loop model}

\begin{figure}
\begin{center}
\includegraphics[width=7.6cm]{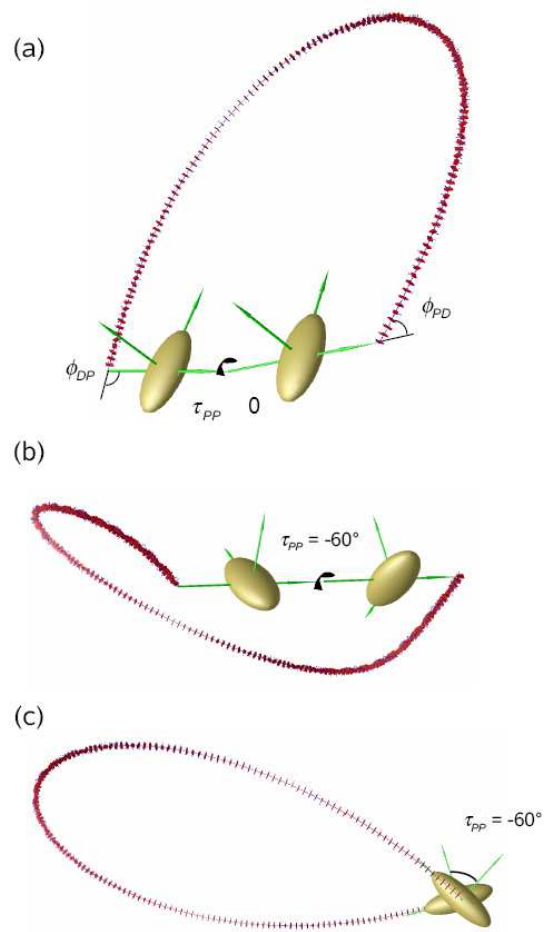}
\end{center}
\caption{Rigid-body models for studies of protein-mediated DNA looping.
(a) A prototype 137-bp DNA loop generated by interactions with a pair of 
rigid, DNA-binding protein subunits is shown.  DNA bps are 
represented by rectangular slabs (red) with axes (blue) that indicate 
the orientation of the local Cartesian coordinate frame whose origin 
lies at the centre of each bp.  Two sets of coordinate axes 
(green) represent the local coordinate frames embedded in the protein 
subunits (gold ellipsoids) that mediate DNA looping.  The coupling of 
protein and DNA geometry is characterised by tilt, roll, and twist 
values for the DNA-protein, protein-protein, and protein-DNA interfaces.  
Three of these variables are shown here: the DNA-protein roll angle, 
$\phi_{DP}$; the protein-protein twist angle, $\tau_{PP}$; 
and the protein-DNA roll angle, $\phi_{PD}$.  
(b) Prototype 179-bp loop with protein-protein twist angle, $\tau_{PP}$,
equal to $-60$ degrees.  The view is from the base of the loop toward 
the DNA apex.  (c) Loop conformation shown in (b) viewed from the side, 
perpendicular to the loop dyad axis.
\label{loopfig1}}
\end{figure}

The protein subunits that mediate loop 
formation are modelled as two identical and connected rigid 
bodies, as shown in figure \ref{loopfig1} \cite{loop51}.
There are three additional sets of rigid-body 
rotation angles that are defined in addition to those for dinucleotide 
steps: two sets for the interfaces between protein and the last (DP) and 
first bps (PD) of the DNA and one set for the interface between the 
two protein domains (PP), where the symbols in parentheses are used to 
indicate the corresponding angles through subscripts. The local 
Cartesian-coordinate frames for protein subunits are defined such that 
their origins coincide with vertices of a circular chain and their 
$z$-axes point toward the next vertex in succession.  Thus protein 
dimensions can be modelled in terms of a non-canonical value for the helix 
rise corresponding to particular segments within a circular polymer chain.

Angles are expressed in degrees, and length in units of the DNA helical 
rise, $\ell_{bp} = 3.4$ {\AA}. All 
calculations used canonical mechanical parameters for duplex DNA: helical 
twist $\tau_{0} = 34.45^{\circ}$, a sequence-independent twist-angle 
standard deviation, or twisting flexibility, $\sigma_{\tau} = 4.388^{\circ}$,
and standard deviations, or bending flexibilities, for all tilt and roll 
angles,
$\sigma_{\theta}$ and $\sigma_{\phi}$, respectively, of $4.678^{\circ}$
(equivalent to a persistence length of 150 bp). Except for specific cases 
where intrinsic DNA bending is considered, the average values of tilt and 
roll are taken to be zero.

\subsection{Simplified protein geometries and flexibility parameters}
\label{loopsection}

For DNA loops with either zero or nonzero end-to-end distances, constraints 
are directly applied to the DNA ends, as in the case of DNA cyclisation.  
We modelled DNA loops formed during site synapsis using protein-dependent 
parameters
${\rm{roll }} = \phi_{DP}  = \phi_{PD} = 90^{\circ}$ and
${\rm{twist}} = \tau_{DP}  = \tau_{PD} = 34.45^{\circ}$.
The angle was considered an adjustable parameter that we denote the 
{\em axial angle} and, unless specified, all other protein-related angular 
parameters were set equal to $0^{\circ}$. In these cases the DNA ends 
(the centres of two protein-binding sites on DNA) are separated by twice 
the protein-arm length $\ell_p$ and displaced from one another along 
the $+x$ direction, or toward the major groove of DNA. Projected along 
the $x$-axis, the axial angle is the included angle between the tangents 
to the DNA at the two protein binding sites and is altered by varying the 
twist between protein subunits (figure \ref{loopfig1} b, c). 
An axial angle equal to $0^{\circ}$
corresponds to antiparallel axes at the ends as shown in 
figure \ref{loopfig1}a.  
The case of a rigid protein assembly is modelled by setting the standard 
deviations of the DP, PP, and PD sets of rigid-body rotation angles to 
$1 \cdot 10^{-8}$ deg.

\subsection{DNA loops having zero end-to-end distance and antiparallel 
helical axes}

DNA loops containing $N$ bps in which the two ends meet in an 
antiparallel orientation can be empirically described by the following 
formula:
\begin{eqnarray} \label{1}
& & {\rm{Tilt:}}  \quad \, \, \, \, 
\theta_i = - A_i \, \cos(180 + \delta) \\ \nonumber
& & {\rm{Roll:}}  \quad \, \, \, \phi_i = A_i \, 
\sin (180 + \delta ) \\ \nonumber
& & {\rm{Twist:}} \quad \tau_i  = \tau^0 \nonumber
\end{eqnarray}
where $\tau^{0}$ is the intrinsic DNA twist and $\delta$ an arbitrary angle 
related to the unconstrained torsional degree of freedom of DNA. The 
coefficients $A_i$ are given by
\begin{equation} \label{2}
A_i = \frac{1}{N} \, f\left( {\frac{i}{{N - 1}}} \right) \, ,
\quad i = 0, \ldots, N - 1
\end{equation}
with
\begin{equation} \label{3}
f(x) = \left\{ \begin{array}{l}
g(x) \, \, , \qquad \quad \, \, \, 0 \le x \le 0.5 \\
g(0.5 - x) \, \, , \quad 0.5 < x \le 1 \\
\end{array} \right.
\end{equation}
where
\begin{equation} \label{4}
g(x) = \sum\limits_{i = 1}^5 {a_{i\,} x^i } ,\quad 0 \le x \le 0.5 \, \, .
\end{equation}
\begin{figure}
\begin{center}
\includegraphics[width=7.6cm]{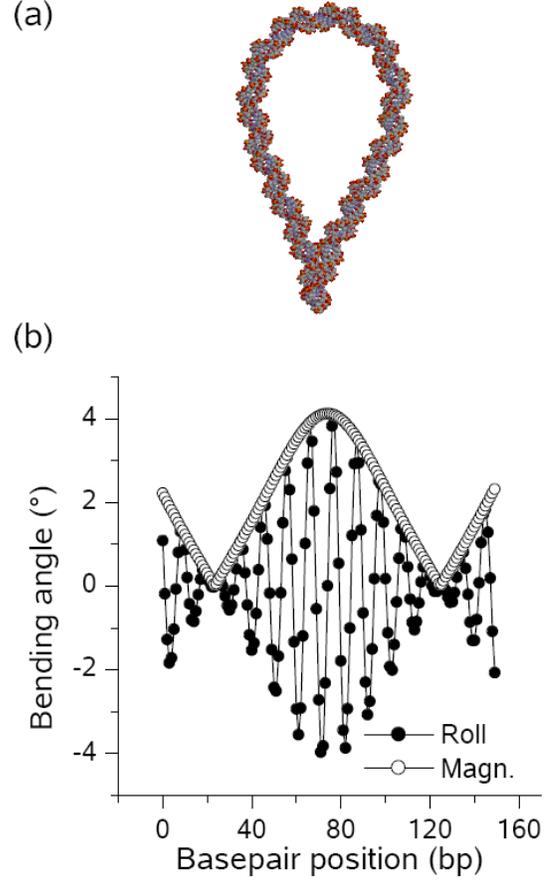}
\end{center}
\caption{Conformation of an antiparallel, 150-bp DNA loop with zero 
end-to-end distance.  (a) Computed space-filling model of the loop 
generated with 3DNA \cite{loop49}.  
The ends of the DNA juxtapose exactly with 
antiparallel helical axes and exact torsional phasing. (b) Equilibrium 
roll and magnitude of the loop shown in (a).  The bending magnitude of 
each dinucleotide step is defined as $\sqrt{\theta_i^2 + \phi_i^2}$
where $\theta_i$ and $\phi_i$ are the tilt and roll of $i$-th 
dinucleotide step, respectively.
\label{loopfig2}}
\end{figure}
The coefficients in equation (\ref{4}) were obtained by fitting the space 
curve corresponding to the DNA helical axis that gives the minimum elastic 
energy conformation of DNA loops of different sizes and are as follows:
$a_0 = -335.0142$, $a_1 = 2318.881$, $a_2 = -1299.164$, 
$a_3 = -4483.366$, $a_4 = 38169.74$, $a_5 = -54753.5$. 
The error for end-to-end distances computed using 
equation (\ref{1}) is less
than $2\%$ of DNA length from 50 bp to 100 bp, and less than $0.5\%$ 
from 100 bp to 500 bp. The torsional phase angle between two ends is
$\xi  =  - \left( {N - 2} \right)\tau  - 2\delta$. The entire loop lies 
in a plane, and the angle between the normal vector of the plane and
the $x$-axis of the external coordinate can be shown to be 
$\psi  = 180 + \tau  - \delta$. The expressions for $\xi$ and $\psi$ 
suggest that $\delta$ is related to DNA bending isotropy. Loop 
configurations with different $\delta$ values are related to each other 
by globally twisting DNA molecules. Since the orientation of the first 
bp is fixed, this global twist is equivalent to rotation of the 
loop plane, which corresponds to the rotational symmetry met in DNA 
cyclisation of homogeneous DNA with bending isotropy \cite{loop26}. 
Therefore, $J$-factors for configurations with different $\delta$ values 
are identical.

If DNA looping needs to be torsionally in-phase, only two degenerate 
loop configurations are available, breaking the rotational symmetry. 
These loop geometries can be expressed
by equation (\ref{1}) with two different $\delta$ values:
$\delta_1 = - (N - 2) \tau / 2$ and
$\delta_2  = 180 - (N - 2) \tau / 2$,
which satisfy the torsional phase requirement 
$\xi  = 360 \cdot n,\;n = 0, \pm 1, \pm 2, \ldots$
In contrast to DNA cyclisation, no twist change is involved in forming these 
ideal DNA loops for any DNA length and thus the helical dependence vanishes 
in this case.  From the expression given above for $\psi$ it is clear that 
the helical axes of the two loops are coincident and their directions are 
reversed. Figure \ref{loopfig2}
shows the bending profile of the loop configuration 
corresponding to $\delta_1$ for a 150 bp DNA. Surprisingly, the maximal 
$J$-factor occurs at approximately the same DNA length, or 460 bp (data 
not shown), as in DNA cyclisation \cite{loop26}.
This can be partly explained by 
the fact that the total bending magnitude of the loop is $290$ degrees, 
close to a full circle, instead of $180$ degrees.

\subsection{DNA looping with finite end-to-end distance, antiparallel 
helical axes, and in-phase torsional constraint}

\begin{figure}
\begin{center}
\includegraphics[width=6.8cm]{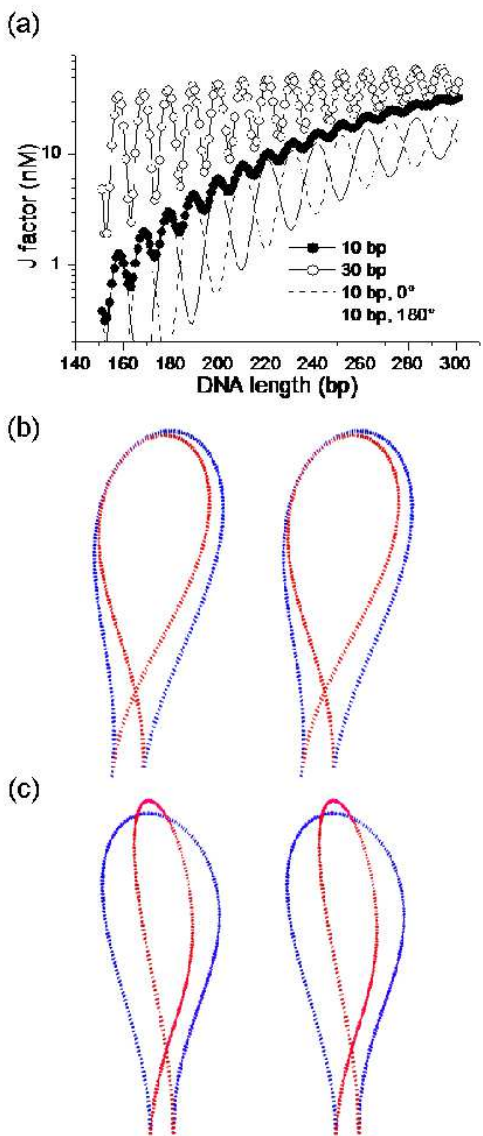}
\end{center}
\caption{The DNA-length-dependent $J$-factor and loop configuration as a 
function of end-to-end separation (the $J$-factor is defined
in section \ref{loopbio}).
(a) The helical dependence of DNA looping 
is shown for values of the end-to-end separation equal to 10 bp and 30 bp. 
The two configurations for the 10-bp separation are obtained from 
corresponding configurations with zero end-to-end separation by using an 
iterative algorithm. Therefore the two configurations are designated by the 
initial configurations with phase angles 
$\delta = -(N-2) \tau / 2 + 0$ ($0^{\circ}$, dashed line) and
$\delta = -(N-2) \tau / 2 + 180$ ($180^{\circ}$, solid line)
as described in the text. (b) and (c) show
stereo models of the two equilibrium 
configurations for 210-bp (b) and 215-bp (c) antiparallel DNA loops with 
end-to-end separation equal to 10 bp. The 210- and 215-bp DNA correspond 
to an adjacent peak and valley of the curve in (a), respectively. 
Conformations shown in blue correspond to $\delta = 0$; 
those shown in red are for $\delta = 180^{\circ}$. Note that for $N$-bp DNA, 
the chain contour length is equal to $(N-1)\ell_{bp}$.
\label{loopfig3}}
\end{figure}

Separation of the DNA ends breaks the rotational symmetry, restoring the 
dependence on helical twist. Figure \ref{loopfig3}a 
shows the $J$-factor as a function 
of DNA length for end-to-end distances of $10$ bp and $30$ bp. The helical 
dependence increases with end-to-end separation.  Starting from the two 
loop configurations (corresponding to $\delta_1$ and $\delta_2$) with zero
end-to-end distance and in-phase torsional alignment as initial
configurations, two mechanical equilibrium configurations are obtained by 
using the iterative algorithm described in \cite{loop26}.
The $J$-factor 
in figure \ref{loopfig3}a is the sum of separate $J$-factors 
calculated for the two 
configurations. Note that in all cases involving configurations that differ 
in linking number, equilibration between the two forms requires breakage of
at least one of the protein-DNA interfaces. The contributions from each of 
these configurations are shown in detail for the case where the ends are 
separated by $10$ bp. Interestingly, the length dependence of $J$ computed 
from the individual configurations are out of phase and have a periodicity 
of 2 helical turns, which results from the half-twist dependence of the 
phase angles $\delta_1$ and $\delta_2$.  However, their sum displays a 
periodicity of one helical turn. Figures \ref{loopfig3} b and c
show two such configurations for DNA molecules that are torsionally 
in-phase ($N = 210$ bp) or out-of-phase ($N = 215$ bp).

In the case of cyclisation, the helical-phase dependence of the $J$-factor 
persists at DNA lengths well beyond that corresponding to the maximum 
value of $J$, which lies near $500$ bp. This is clearly not the case for 
DNA looping.  In figure \ref{loopfig3}a, 
the periodic dependence of $J$ on DNA length 
for $10$-bp end-to-end separation decays nearly to zero well before the 
maximum $J$ value is reached.  Although the periodicity of $J$ is not 
attenuated quite as strongly for $30$-bp end separation, there is less 
than four-fold variation in the value of $J$ near $300$ bp, as opposed 
to the more than ten-fold variation in cyclisation $J$-factors expected
in this length range.  The differences between looping and cyclisation 
are largely due to substantial differences in the relative contributions 
of DNA writhe in the two processes, as discussed below.

\subsection{DNA looping in synapsis}
\label{loopsyn}

Intramolecular reactions of most site-specific recombination systems 
\cite{loop9,loop10,loop11}
and a number of DNA restriction endonucleases such as {\em Sfi}I and 
{\em Ngo}MIV \cite{loop12},
proceed through protein-mediated intermediate structures 
in which a pair of DNA sites are brought together in space and the
intervening DNA is looped out.  The intermediate nucleoprotein complex 
involved in site pairing and strand cleavage (and also exchange, in the 
case of recombinases) is termed the {\em synaptic complex}. In these systems, 
two characteristic geometric parameters are of interest: the average 
through-space distance between the sites and the average crossing angle 
between the two ends of the loop, which we denote the axial angle
(see section \ref{loopsection}).
The latter quantity can be described in terms of the twist angle between 
the protein domains, $\tau_{PP}$ (figure \ref{loopfig1}b), 
and we use these terms interchangeably.

\begin{figure}
\begin{center}
\includegraphics[width=7.6cm]{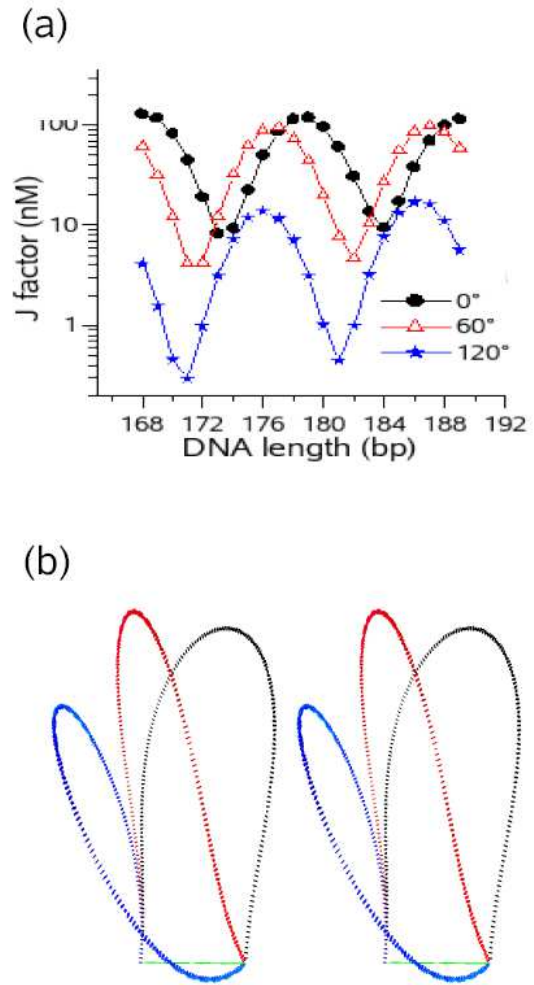}
\end{center}
\caption{Dependence of the $J$-factor on axial angle 
(the $J$-factor is defined in section \ref{loopbio},
and the axial angle is defined as 
the average crossing angle between the two 
ends of the loop, see section \ref{loopsection}).
(a) DNA-length dependence of $J$ for axial angles of 
$0^{\circ}$, $60^{\circ}$, and $120^{\circ}$ 
with the end-to-end separation set equal to 40 bp.
Note that the positions 
of the extrema shift to the left with increasing values of the axial angle.  
(b) Stereo models of minimum elastic-energy conformations of 179-bp loops 
colour coded in accord with the corresponding axial-angle values in (a).
\label{loopfig4}}
\end{figure}

Figure \ref{loopfig4} shows the helical dependence of looping 
(figure \ref{loopfig4}a) and the 
elastic-minimum configuration of DNA loops 
(figure \ref{loopfig4}b) for different 
values of the axial angle. The most prominent feature of these results 
is that the phase of the helical dependence is shifted as a function of the
axial angle, characterised by a relative global shift of the curve along 
the $x$-axis. This implies that DNA looping does not always occur most 
efficiently when two sites are separated by an integral number of helical 
turns, as has been suggested for some simple DNA looping systems studied 
previously. The axial angle also globally modulates $J$-factors, which 
is apparent from the vertical shift in the $J$ versus length curve and 
effects on the amplitude of the helical dependence. 
The torsion-angle-independent value of $J$, averaged over a full helical 
turn, decreases with increasing axial angle, whereas the amplitude of 
the helical dependence increases.  The above observations can be 
qualitatively explained by analogous results from DNA cyclisation.  
As in cyclisation, DNA forms loops most efficiently when the number of 
helical turns in the loop is close to an integer value. It is therefore 
appropriate to consider this issue in terms of the linking number for the
looped conformation, $Lk$, which involves contributions from the geometries 
of both the protein and DNA.

\begin{figure}
\begin{center}
\includegraphics[width=7.6cm]{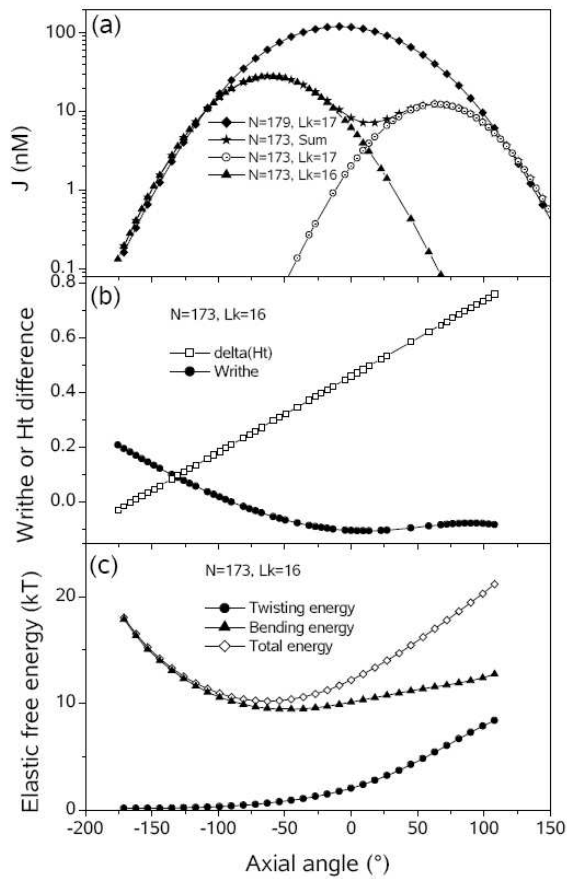}
\end{center}
\caption{$J$-factor, loop-geometry parameters, and elastic-free energies as 
functions of axial angle; compare figure \ref{loopfig4}.
(a) $J$-factor values for loop topoisomers 
corresponding to 179-bp and 173-bp loops in figure \ref{loopfig4}.
The principal contribution to $J$ for $N = 179$ bp comes from a single 
loop topoisomer with $Lk = 17$.  For $N = 173$ bp, the overall $J$-factor 
is the sum of contributions from two loop topoisomers with $Lk$ values of 
16 and 17, generating a bimodal dependence of $J$ on axial angle as 
described in the text. (b) Excess helical twist, $\Delta H_t$, 
and writhe of the loop formed by the $Lk = 16$ topoisomer for $N = 173$ bp
 as a function of axial angle.  Excess twist is computed from the expression 
$H_{t, loop} - 16$, where $H_{t, loop}$ is the loop helical turn value 
described in the text, and depends linearly on the axial angle.  
The writhing number of the loop was calculated using the method of 
Vologodskii \cite{loop28,loop50}. (c) Elastic-free energies of the 
$Lk = 16$ loop topoisomer for $N = 173$ bp calculated according to 
equation 38 of Zhang and Crothers \cite{loop26}. The individual 
contributions of bending and twisting energies are shown along 
with their sum.
\label{loopfig5}}
\end{figure}

We define the {\em loop helical turn} $H_{t,loop}$ as the sum of the 
DNA twist and the twist introduced by the protein subunits, divided by 
$360$.  Therefore, changing the twist angle, the axial angle will 
shift the phase of the helical dependence relative to that of
the DNA alone.  For a loop with $N = 179$ bp and $\tau_{PP} = 0$, the 
total twist is simply equal to that for the DNA loop.  Because this loop 
has $17.0$ helical turns, only one loop topoisomer contributes to the 
$J$-factor.  The value of $J$ is a local maximum at $\tau_{PP} = 0$ and, 
as shown in figure \ref{loopfig5}a,
decreases monotonically for both $\tau_{PP} > 0$
and $\tau_{PP} < 0$.  Contributions to $J$ from other topoisomers of 
the $179$-bp loop are less than $5$ percent over the range 
$-135^{\circ} < \tau_{PP} < +120^{\circ}$. The twist for the planar 
equilibrium conformation of a $173$-bp loop is $16.5$ helical turns;
thus there are two alternative loops that can be efficiently formed 
(figure \ref{loopfig5}a): 
either a loop with $H_{t,loop} = 17.0$ and $\tau_{PP} > 0$, 
or a loop with $H_{t,loop} = 16.0$ and $\tau_{PP} < 0$. The $J$ value 
at $\tau_{PP} = 0$ is a local minimum and there is a bimodal dependence 
on axial angle for loops in which the DNA twist is half-integral. 
We investigated the phase shift of the $J$-factor and found that 
this quantity is a non-linear function of the axial angle.  From 
figure \ref{loopfig4}a, 
the calculated phase shifts for $60^{\circ}$ and $120^{\circ}$ 
axial angles relative to $0^{\circ}$ are approximately $52^{\circ}$ 
and $103^{\circ}$, respectively.  Moreover, the local maxima for the total
$J$ curve for $N = 173$ shown in figure \ref{loopfig5}a
are located at $-58.5^{\circ}$ 
and $63^{\circ}$, positions that are not in agreement with predicted 
angle values based solely on $H_{t,loop}$ ($-166^{\circ}$ and 
$194^{\circ}$, respectively).

These deviations can be explained by the fact that writhe makes an 
important contribution to the overall $Lk$ for the loop.  This aspect 
of DNA looping is dramatically different from that in the cyclisation 
of small DNA molecules.  The conformations of small DNA circles are
close to planar and the writhe contribution is small relative to DNA 
twist \cite{loop26,loop30,loop44,loop45}.
In the case of protein-mediated looping, nonzero 
values of the axial angle impose an intrinsically nonplanar conformation 
on the DNA.  The relative contributions of loop writhe and twist for 
the $Lk = 16$ topoisomer of a $173$-bp loop are shown as a function of 
axial angle in figure \ref{loopfig5}b.

In figure \ref{loopfig5}c, 
we plot the axial-angle-dependent values of the bending and 
twisting free energies for the $Lk = 16$ topoisomer and their sum, which 
is the total elastic-free energy of the loop. The minimum value of the 
total elastic energy occurs at $\tau_{PP} = -58.5^{\circ}$, coincident 
with the position of the $J$-factor maximum for this topoisomer 
(figure \ref{loopfig5}a).
This mechanical state can be achieved with very little twist deformation 
of the loop, but at the expense of significant bending energy.  Further 
reduction of the axial angle requires even less twisting energy; however, 
the bending energy increases monotonically. In contrast, for 
$\tau_{PP} > -58.5^{\circ}$, somewhat less bending energy is required,
but the twisting energy begins to increase significantly with increasing 
axial angle. Since the sense of the bending deformation for 
$\tau_{PP} > 0$ opposes the needed reduction in loop linking number, 
the elastic energy cannot be decreased by increasing the axial angle.
The only way that the loop geometry can compensate for this is through 
twist deformation. This asymmetry arises because we are considering the 
contribution of only one loop topoisomer to the elastic free energy.

\subsection{Conclusion}

The statistical-mechanical theory for DNA looping discussed above
\cite{loop26,loop51} 
suggests that the helical dependence of DNA looping is 
affected by many factors and leads to the conclusion that whereas a 
positive helical-twist assay can often confirm DNA looping, a negative 
result cannot exclude DNA looping.  Since it is difficult to explore 
the architecture of DNA loops with current experimental techniques, 
this theory will be useful for more reliably analysing DNA looping
with limited experimental data.  The model has advantages over 
previous approaches based exclusively on DNA mechanics, particularly 
when protein flexibility is taken into account. In these cases, entropy 
effects become important and are responsible for the observed decay of 
looping efficiency with DNA length.

\section{DNA knots and their consequences: entropy and targeted knot
removal}

\begin{figure}
\includegraphics[height=5.4cm]{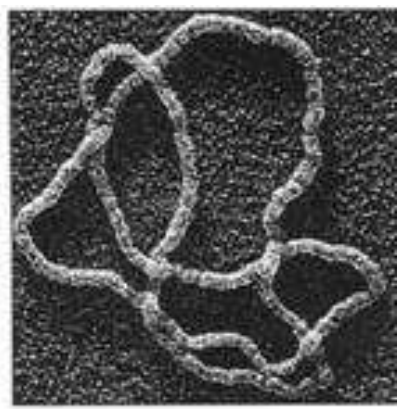}
\caption{Electron microscope image of a DNA trefoil knot, from
\cite{wassermann2}. $\copyright$ Science, with permission.
\label{dna_trefoil}}
\end{figure}

Bacterial DNA occurs largely in circular form. Notably, instead of a simply
connected ring shape (the unknot), the DNA often exhibits permanently
entangled states, such as catenated and knotted DNA. An example for a DNA
trefoil knot is shown in figure \ref{dna_trefoil}. Such configurations have
potentially devastating effects on the cell development. Conversely, however,
knots might have designed purposes in gene regulation, separating different
regions of the genome, or, alternatively, locking chemically
remote parts of the genome proximate in geometrical space. In eukaryotic
cells additional topological effects occur in the likely entanglement of
individual chromosomes. Here, we concentrate on the prokaryotic case.

\subsection{Physiological background of knots}

The discovery how one can use molecular biological tools to create knotted
DNA resolved a long-standing argument against the
Watson-Crick double helix picture of DNA \cite{frank}, namely that the
replication of DNA could not work as the opening up of the double helix
would produce a superstructure such that the two daughter strands could
not be separated. In fact, the topology of both ssDNA and dsDNA is continuously
changed in vivo, and this can readily be mimicked in vitro, although the
activity of enzymes in vivo is much more restricted than in vitro
\cite{deibler,dna_topo}: Different concentrations of enzymes versus knotted
DNA molecules accessible in vitro,
that is, makes it possible to probe topology-altering effects by enzymes
which in vivo do not contribute to such effects.

\begin{figure}
\includegraphics[width=8.6cm]{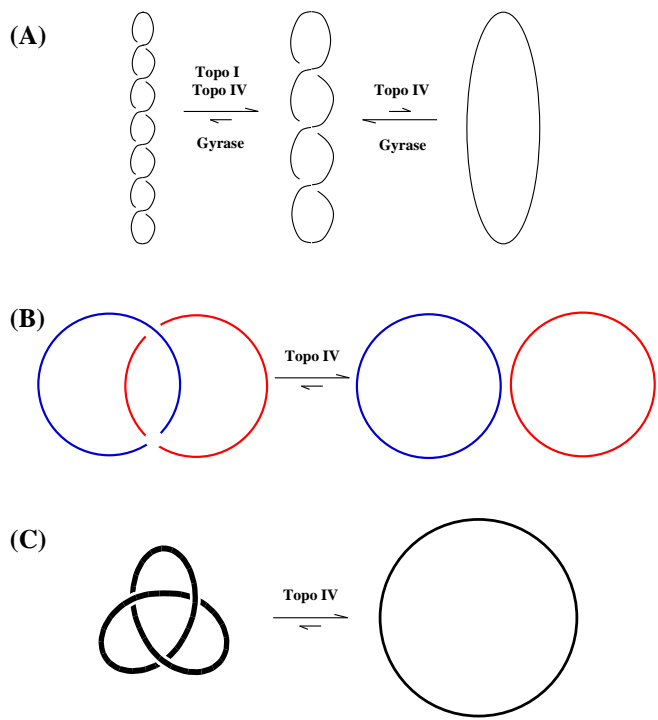}
\caption{Enzymes changing the topology of dsDNA by cutting and pasting of one
or both strands (example for \ecoli): (A) Torsional stress resulting from
the Lk deficit causes the DNA double helix to writhe about itself (negative
supercoiling). In \ecoli, gyrase introduces negative supercoils into DNA
and is countered by topoisomerase I (topo I) and topo IV, which relax negative
supercoils. (B) Topo IV unlinks catenanes generated by replication or
recombination in vivo. (C) Topo IV unknots DNA in vivo. After \cite{deibler}.
\label{topogyro}}
\end{figure}

Although it would be likely
with a probability of roughly $\frac{1}{2}$ that the linear DNA injected by
bacteriophage $\lambda$ into its host \ecoli would create a knot before
cyclisation, it turned out to be difficult to detect \cite{frank}. First studies
therefore concentrated on the fact that under physiological conditions knots
are introduced by enzymes, DNA replication and recombination, DNA repair, and
topoisomerisation, using these enzymes to prove both knotting and unknotting
\cite{liu,mizuuchi,pollock,spengler,wassermann,wassermann1,wassermann2}.
DNA-knotting is also prone to occur behind a stalled replication fork
\cite{viguera,sogo}. Some of the typical topology-altering reactions
undergoing in \ecoli are summarised in figure \ref{topogyro}.
Knots can efficiently be created from nicked\footnote{One of the two strands
is cut.} dsDNA under action of
topoisomerase I at non-physiological concentrations \cite{topoknot}.
Another possibility is by active packaging of a DNA mutant into phage
capsids \cite{arsuaga}, and then denaturing the capsid proteins. Both
methods produce a distribution of different knot types. They can be
separated by electrophoresis \cite{electro}.

The existence of DNA-knots has far-reaching effects on physiological processes,
and knottedness of DNA has therefore to be eliminated in order to
maintain proper functioning of the cell. Among other possible effects, it
is immediately clear that the presence of a knot in a circular DNA impedes
replication of the DNA, i.e., the full separation of the two daughter
strands \cite{alberts,frank}. Moreover, even transcription is impaired
\cite{portugal}. The presence of knots inhibits the assembly of chromatin
\cite{rodriguez}, knotted chromosomes cannot be separated during mitosis
\cite{alberts}, and knots in a chromosome may serve as topological barriers
between different sections of chromosomes, such that the genomic structural
organisation is altered, and certain sections of the chromosomal DNA may no
longer interact \cite{staczek}. Conversely, it is conceivable that knots,
analogously to protein induced DNA looping, lock remote segments of the genome
close together in geometric space. Finally, knots may lead to double-strand
breaks, as they weaken biopolymers considerably due to creation
of localised sharp bends \cite{arai,pieranski,saitta,stasiak1}
as well as macroscopic lines and ropes \cite{mcnally}.\footnote{The weakness
of strings at the site of the knot can be experienced easily by pulling apart
a linear nylon string in comparison to a knotted one \protect\cite{pieranski}.}

\begin{figure}
\includegraphics[width=6.8cm]{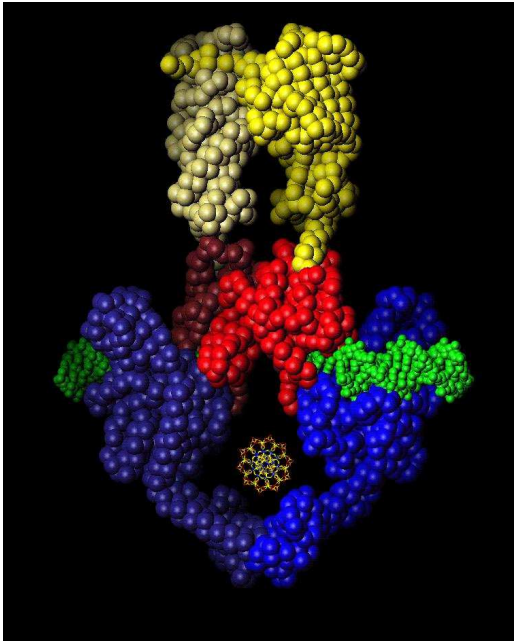}
\caption{Topoisomerase II. This enzyme can
actively change the topology of DNA by cutting the double-strand and
passing another segment of double-stranded DNA through the gap before
resealing it. The image depicts a short stretch of DNA (horizontally
at the bulge of the enzyme, as well as another segment in the lower
clamp (perpendicular to the image) after passage through the gap from
the upper clamp. This mechanism makes sure that no additional strand
passage through the open gap can take place \cite{berger,wang_to}.
Figure courtesy James M Berger, UC Berkeley.
\label{topoII}}
\end{figure}

Above we said that knots can be introduced, inter alia, by the different
enzymes of the topoisomerase family. To remove a knot
from a dsDNA, it is necessary to cut both strands, and then pass one
segment through the created gap, before resealing the two open ends. In
vivo, this is usually achieved by topoisomerases II and IV. A reconstruction
of topo II is shown in figure \ref{topoII}, indicating the upper clamp
holding a segment of the DNA, while the bulge-clamp introduces the cut
through which the upper segment is passed. In the figure, the segment
visible in the pocket of the lower clamp has already been passed through
the gap. After resealing, topo II detaches. This process requires energy,
provided by ATP. Notably, topo II is extremely efficient, for circular dsDNA
of length $\simeq 10$ kbp it was found that topo reduced the knotted state in
between 50 and 100-fold, in comparison to a `dumb' enzyme, which would simply
pass segments through at
random \cite{rybenkov}. We note that the step-wise action of topoisomerase
II was recorded in a single molecule setup using magnetic tweezers
\cite{stasiak,strick}. Topoisomerases are surveyed in the review of
\cite{wang}.

\subsection{Classification of knots}

Knottedness can only be defined on a closed (circular) chain. This is
intuitively clear as in an open linear chain a knot can always be tied, or
an existing knot released. Mathematically, this means that knot invariants
are only well-defined for a closed space-curve. However, a linear chain
whose ends are permanently attached to one, or two walls, or whose ends are
extended towards infinity, can be considered as (un)knotted in the proper
mathematical sense, i.e., their knottedness cannot change. In a loser
sense, we will also speak of knots on an open piece of DNA, appealing to
intuition.

The classification of knots, or graphs in general, in terms of invariants
can essentially be traced back to Euler, recalling his graph theoretical
elaboration in connection with the Bridges of K{\"o}nigsberg problem
\cite{euler}, determining a closed path by crossing each K{\"o}nigsberg bridge
exactly once. However, the first investigations of topological problems in
modern science is most probably due to Kepler, who studied surface tiling to
great detail (therefore the notion of Kepler tiling in mathematical literature)
\cite{kepler}. Further initial steps
were due to Leibniz, Vandermonde and Gauss, in whose collection of papers
drawings of various knots were found\footnote{Probably copies from an English
original.} whose linking (`Umschlingungen'=windings) number is indeed a knot
invariant \cite{adams,kauffman,reidemeister}. Gauss' student, Listing, in fact
introduced the term `topology', and his work on knots may be viewed as the
real starting point of knot theory \cite{listing}, although his complexions
number was proved by Tait not to be an invariant.

Inspired by Helmholtz' theory of an ideal fluid and building on Listing's
early contributions to knot theory, Scotsmen and chums Maxwell, Tait and
Thomson (Lord Kelvin) started to discuss the possible implications of
knottedness in physics and chemistry, ultimately distilled into Thomson's
theory of vortex atoms \cite{thomson,thomson1}. Out of this endeavour emerged
Tait's interest in knots, and he devoted most of his career on the
classification of knots. Numerous charts and still unresolved conjectures
on knots document his pioneering work \cite{tait,tait1,tait2,tait3}. The
studies were carried on by Kirkman and Little
\cite{kirkman,kirkman1,little,little1}.
A more detailed historical account of knot theory may be found in the review
article by van de Griend \cite{degriend}, and on the St. Andrews history of
mathematics webpages\footnote{The MacTutor History of Mathematics
archive, URL: {\tt http://turnbull.mcs.st-and.ac.uk/~history/}}.

Planar projections of knots were rendered unique by Listing's introduction
of the handedness of a crossing, i.e., the orientational information assigned
to a point where in the projection two lines intersect. With this information,
projections are the standard representation for knot studies. On their basis,
the minimum
number of crossings (`essential crossings') can be immediately read off as
one of the simplest knot invariants. To arrive at the minimum number, one
makes use of the Reidemeister moves, three fundamental permitted moves of the
\begin{figure}
\includegraphics[width=8cm]{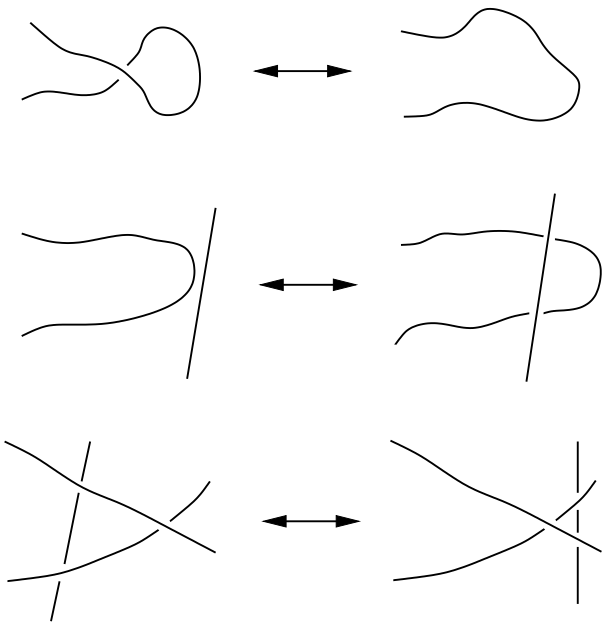}
\caption{The three Reidemeister moves. All topology-preserving moves of a
knot projection can be decomposed into these three fundamental moves.
\label{reidemeister}}
\end{figure}
lines in a knot projection, as shown in figure \ref{reidemeister}. More complex
knot invariants include polynomials of the Alexander, Kauffman and HOMFLY types
\cite{adams,kauffman,reidemeister}.\footnote{These polynomials all start to
be degenerate for higher order knots, i.e., above a certain knot complexity
several knots may correspond to one given polynomial \cite{adams,kauffman}.
In the case of the simpler knots attained in most DNA configurations and in
knot simulations, the Alexander polynomials are unique, in contrast to the
Gauss or Edwards invariant, compare, e.g., reference \protect\cite{volo}.}
Here, we will only employ the number of essential crossings as
classification of knots, in particular, we do not concern ourselves with the
question of degeneracy for a given knot invariant. However, the bookkeeping
of knot types is vital in knot simulations.

\subsection{Long chains are almost always entangled.}

During the polymerisation and final cyclisation of a polymer grown
in a solvent under freely floating conditions, a knot is created
with probability 1. This Frisch-Wassermann-Delbr{\"u}ck conjecture
\cite{frisch,delbrueck} could be mathematically proved for a self-avoiding
chain \cite{sumners,pippenger}, compare also \cite{vanderzande}. This is
consistent with numerical findings
that the probability of unknot formation decreases dramatically with
chain length \cite{frank3,volo}. Indeed, recent simulations results indicate
that the probability of finding the unknot in such a cyclisised polymer
decays exponentially with chain length \cite{koniaris,michels,michels1,janse}:
\begin{equation}
P_\emptyset(N)\propto \exp\left(-\frac{N}{N_c}\right).
\end{equation}
However, there exist theoretical arguments and simulations results indicating
that the characteristic number of monomers $N_c$ occurring in this relation
may become
surprisingly large \cite{frank2,grosberg4,klenin,shimamura1}. The probability
to find a given knot type ${\cal K}$ on random circular polymer formation has
been fitted with the functional form \cite{katritch,dobay,shimamura1}
\begin{equation}
P_{\cal K}(N)=a\big(N-N_0\big)^b\exp\left(-\frac{N^c}{d}\right),
\end{equation}
where $a$, $b$, and $d$ are free parameters depending on ${\cal K}$, and $c
\approx 0.18$. $N_0$ is the minimal number of segments required to form a
knot ${\cal K}$, without the closing segment \cite{dobay}. The tendency
towards knotting during polymer cyclisation creates problems in industrial
and laboratory processes.

\subsection{Entropic localisation in the figure-eight slip-link structure.}

\begin{figure}
\includegraphics[height=4cm]{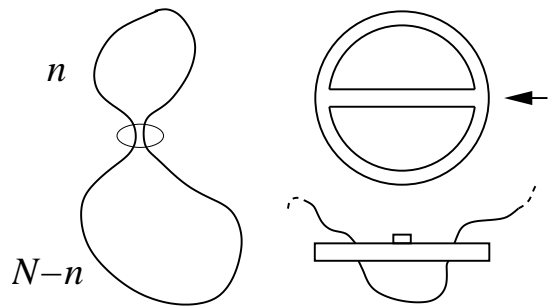}
\caption{Figure-eight structure, in which a slip-link separates two loops
of size $n$ and $N-n$, such that they can freely exchange length among
each other, but none of the loops can
completely retract from the slip-link. On the right, a schematic drawing
of a slip-link, which may be thought of as a small belt buckle.
\label{fig8}}
\end{figure}

To obtain a feeling for how and when entropy leads to the localisation of
a permanently entangled structure, we consider the simplest polymer object
with non-trivial (non-unknot) geometry, the 
figure-eight structure (F8) displayed in figure \ref{fig8}. In this compound,
a pair contact is enforced by a slip-link, separating off two loops in the
circular polymer, such that none of the loops can fully retract, and both
loops can freely exchange length among each other. We denote the loop sizes
by $n$ and $N-n$, where $N$ is the (conserved) total length of the polymer
chain. For such an object, we can actually perform a closed statistical
mechanical analysis based on results from scaling theory of polymers,
and compare the result with Monte Carlo simulations of the F8.

The statistical quantities that are of particular interest are the gyration
radius, $R_g$, and the number of degrees of freedom, $\omega$
\cite{degennes}. $R_g$, as defined in equation (\ref{appgyr}),
measures the root mean squared distance of the monomers along the chain to the
gyration centre,
and is therefore a good measure of its extension. It can, for instance, be
measured by light scattering experiment. The degrees of freedom $\omega$ count all possible
different configurations of the chain. For a circular polymer (i.e., a polymer
with $\mathbf{r} (0)=\mathbf{r}(N)$), the gyration radius becomes
\begin{equation}
\label{gyrra}
R_g\simeq AN^{\nu}
\end{equation}
with exponent $\nu=1/2$ for a Gaussian chain, and $\nu=0.588$ in the 3D
excluded volume case ($\nu=3/5$ in the Flory model, and $\nu=3/4$
in 2D). Whereas in 2D this scaling contains truly a ring polymer, in 3D the
exponent $\nu$ emerges from averaging over all possible topologies, and
necessarily includes knots of all types \cite{duplantier1,deutsch,degennes}.
For a circular chain, the number of degrees of freedom contains the number of all possible
ways to place an $N$-step walk on the lattice with connectivity $\mu$
(e.g., $\mu=2d$ on a cubic lattice in $d$ dimensions),
$\mu^N$, and the entropy loss for requiring a closed loop, $N^{-d\nu}$,
involving the same Flory exponent $\nu$. For the Gaussian case, we recognise
in this entropy loss factor the returning probability of the random walk. In
the excluded volume case, $N^{-d\nu}$ is an analogous measure
\cite{degennes,grosberg,hughes}. Thus, for a circular chain embedded in
$d$-dimensional space, the number of degrees of freedom is
\begin{equation}
\label{dof}
\omega\simeq\mu^NN^{-d\nu}.
\end{equation}
Let us evaluate these measures for the F8 from figure \ref{fig8}.

As a first approximation, consider the F8 as a Gaussian (phantom) random walk,
demonstrating that, like in the charged knot case \cite{paul}, entropic effects
give rise to long-range interactions. The two loops correspond to returning
random walks, i.e., the number of degrees of freedom for the F8 in the phantom chain case
becomes \cite{degennes,flory1,grosberg}\footnote{Here and in the following we
consider two configurations of a polymer chain different if they cannot be
matched by translation. In addition, the origin of a given structure is fixed
by a vertex point (see below), i.e., a point where several legs of the polymer
chain are joint. In the F8-structure, this vertex naturally coincides with the
slip-link. For a simply connected ring polymer, such a vertex is a two-vertex
anywhere along the chain. \label{counting}}
\begin{equation}
\label{fig8id}
\omega_{\rm F8,PC}\simeq\mu^Nn^{-d/2}(N-n)^{-d/2},
\end{equation}
where $d$ is the embedding dimension. We note that normalisation of this
expression produces the probability density function for finding
the F8 with a given loop size $\ell=na$ ($L=Na$),
\begin{equation}
\label{f8_pdf}
p_{\rm F8,PC}(\ell)\simeq\mathscr{N}\ell^{-d/2}(L-\ell)^{-d/2},
\end{equation}
where $\mathscr{N}$ denotes a normalisation factor.
The conversion from expressing the chain size in
terms of the number of monomers to its actual length is of advantage in what
follows, as it allows to more easily keep track of dimensions. Here, we use
the length unit $a$, which may be interpreted as the monomer size (lattice
constant), or as the size of a Kuhn statistical segment.

\begin{figure}
\includegraphics[width=6.8cm]{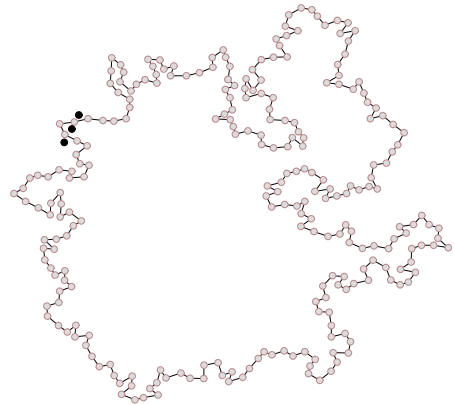}
\caption{Bead-and-tether chain used in Monte Carlo simulation, showing a
typical equilibrium configuration for a self-avoiding chain: the localisation
of the smaller loop is distinct. Note that in this 2D simulation the slip-link
is represented by the three tethered black beads.
\label{sl}}
\end{figure}

To classify different grades of localisation, we follow the convention from
references \cite{slili2d,paraknot}.
The average loop size $\langle \ell\rangle$ determined through
$\langle \ell \rangle = \int_{a}^{L-a} \ell p(\ell) d\ell$
is trivially $\langle\ell\rangle=L/2$ by symmetry of
the structure. Here, we introduce a short-distance cutoff set by the lattice
constant $a$. However, as the probability density function
is strongly peaked at $\ell=0$ and $\ell=L$,
the two poles caused by the returning probabilities, and therefore a {\em
typical\/} shape consists of one small ({\em tight\/}) and one large ({\em
loose\/}) loop, compare figure \ref{sl}. This can be quantified in terms of
the average size of the smaller loop,
\begin{equation}
\label{sls_f8}
\langle \ell \rangle_<\equiv 2\int_a^{L/2} \ell p(\ell) d\ell \, .
\end{equation}
In $d=2$, we obtain
\begin{equation}
\langle \ell \rangle_<\sim \frac{L}{|\log(a/L)|},
\end{equation}
such that with the logarithmic correction the smaller loop is only marginally
smaller than the big one. In contrast, one observes {\em weak localisation}
\begin{equation}
\label{fig8d3}
\langle \ell \rangle_<\sim a^{1/2}L^{1/2}
\end{equation}
in $d=3$, in the sense that the relative size $\langle \ell \rangle_</L$
tends to zero for large chains. By comparison, for $d>4$ one encounters
$\langle \ell \rangle_<\sim a$, corresponding to {\em strong localisation},
as the size of the smaller loop does not depend on $L$ but is set by the
short-distance cutoff $a$. Above four dimensions, excluded volume effects
become negligible, and therefore both Gaussian and self-avoiding chains are
strongly localised in $d\ge 4$.\footnote{Consideration of higher than the
physical 3 space dimensions is often useful in polymer physics.}

To include self-avoiding interactions, we make use of results for general
polymer networks obtained by Duplantier \cite{duplantier,duplantier1}, which
are summarised in the appendix at the end of this review. In terms of such
networks, our F8-structure corresponds to the following parameters: the
number ${\cal N}=2$ of polymer segments with lengths $s_1=\ell=na$ and $s_2=
L-\ell=(N-n)a$, forming ${\cal L}=2$ physical loops, connected by $n_4=1$
vertex of order four.
By virtue of equation (\ref{network}), the number of configurations
of the F8 with fixed $\ell$ follows the scaling form
\begin{equation}
\label{t_scaling}
\omega_{\rm F8}(\ell)\simeq \mu^L (L-\ell)^{\gamma_{\rm F8}-1}
{\cal X}_{\rm F8}\left(\frac{\ell}{L-\ell}\right),
\end{equation}
with the configuration exponent $\gamma_{\rm F8}=1-2d\nu+\sigma_4$. In the
limit $\ell\ll L$, the contribution of the large loop
in equation (\ref{t_scaling})
should not be affected by a small appendix, and therefore should exhibit the
regular Flory scaling $\sim (L-\ell)^{-d\nu}$ \cite{kafri,kafri1,duplantier3}.
This fixes the scaling behaviour of the scaling function ${\cal X}_{\rm F8}
(x)\sim x^{\gamma_{\rm F8}-1+d\nu}$ in this limit ($x \to 0$ in dimensionless
variable $x$), such that
\begin{equation}
\label{sl_limit}
\omega_{\rm F8}(\ell)\simeq\mu^L(L-\ell)^{-d\nu}\ell^{-c},
\quad \ell\ll L,
\end{equation}
where $c=-(\gamma_{\rm F8}-1+d\nu)=d\nu-\sigma_4$. Using $\sigma_4=-19/16$
and $\nu = 3/4$ in $d=2$ \cite{duplantier,duplantier1}, we obtain
\begin{equation}
\label{expon}
c=43/16=2.6875, \quad d=2.
\end{equation}
In $d = 3$, $\sigma_4\approx -0.48$ \cite{schaefer,kafri,kafri1} and
$\nu\approx 0.588$, so that
\begin{equation}
c\approx 2.24.
\end{equation}
In both cases the result $c>2$ enforces that the loop of length
$\ell$ is strongly localised in the sense defined above.
This result is self-consistent with the a priori assumption $\ell\ll L$.
Note that for self-avoiding chains, in $d=2$ the
localisation is even {\em stronger\/} than in $d=3$, in contrast
to the corresponding trend for ideal chains.

\begin{figure}
\includegraphics[width=6.8cm]{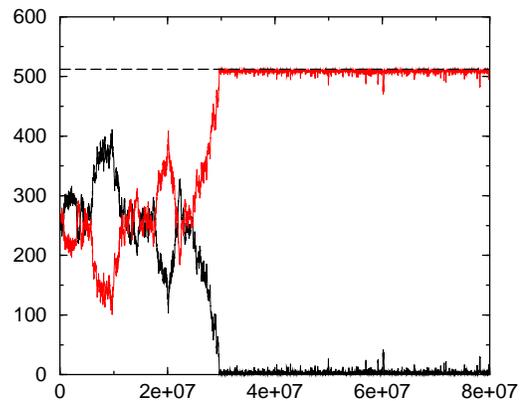}
\caption{Monte Carlo simulation of an F8-structure in 2D: loop
sizes $\ell$ and $L-\ell$ as a function of Monte Carlo steps for a chain
with 512 monomers. The symmetry breaking after the symmetric initial
condition is distinct.
\label{fig8_sim}}
\end{figure}

\begin{figure}
\includegraphics[width=6.8cm]{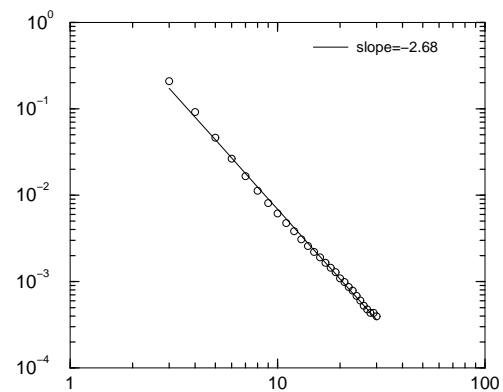}
\caption{Power-law fit to the probability density function of the smaller
loop. The fit
produces a slope of -2.68, in excellent agreement with the calculated value.
\label{sl1}}
\end{figure}

We performed Monte Carlo (MC) simulations of the 2D figure-eight structure,
in which the slip-link was represented by three tethered
beads enforcing a sliding pair contact such that the loops cannot fully
retract (see figure \ref{fig8_sim}). We used a 2D hard core bead-and-tether
chain with 512 monomers, starting off from a symmetric initial condition
with $\ell=L/2$.
Self-crossings were prevented by keeping a maximum bead-to-bead
distance of 1.38 times the bead diameter, and a maximum step length of 0.15
times the bead diameter.
As shown in figure \ref{sl1}, the size distribution
for the small loop can be fitted to a power law with exponent $c=2.68$
in good agreement with equation (\ref{expon}).

An experimental study of entropic tightening of a macroscopic F8-structure
was reported in reference \cite{bennaim}. There, a granular chain consisting
of hollow steel spheres connected by steel rods was once twisted and then
put on a vibrating table. From digital imaging, the distribution of loop
sizes could be determined and compared to a power-law with index 43/16
as calculated for the 2D excluded volume chain. The agreement was found to
be consistent \cite{bennaim}.

\subsection{Simulations of entropic knots in 2D and 3D.}

Much of our knowledge about the interaction of knots with thermal fluctuations
is based on simulations of knotted chains. Before going further into the
theoretical modelling of knotted chains, we report some of the results based
on simulations studies of both Gaussian and self-avoiding
walks.\footnote{Although per se a Gaussian chain cannot have a fixed topology
due to its phantom character, such simulations introduce a fixed topology
by rejecting moves that result in a different knot type.}
Such simulations
either start with a given knot configuration and then perform moves of
specific segments, each time making sure that the topology is preserved;
or, each new configuration emanates from a new
random walk, whose correct topology may be checked by calculating the
corresponding knot invariant, usually the Alexander polynomial, and
created configurations that do not match the desired topology are discarded.
We note that it is of lesser significance that knot invariants such as the
Alexander polynomials in fact are no longer unique for more complex knots,
because for typical chain lengths with the highest probability simpler knots
are created, for which
the invariants are unique. For more details we refer to the works
quoted below.

In fact, the fixed topology turns out to have a highly non-trivial effect
on chains without self-excluded volume. As conjectured in \cite{descloizeaux},
a Gaussian circular chain, whose permitted set of configurations is restricted
to a fixed topology, will exhibit self-avoiding behaviour. This was proved in a
numerical analysis in \cite{deutsch}. The required number of monomers to
reach this self-avoiding exponent was estimated to be of the order of 500.
Keeping this non-trivial scaling of a Gaussian chain at fixed topology in mind,
knot simulations on the basis of phantom Gaussian chains were performed in
\cite{dobay1}, always making sure that the configurations taken into the
statistics fulfil the desired knot topology.

The dependence of the gyration radius $R_g$ on the knot type was investigated
for simpler knots in 3D in reference \cite{janse}. On the basis of the
expansion
\begin{equation}
R_g^2\simeq A_{\cal K}\Big(1+B_{\cal K}N^{-\Delta}+C_{\cal K}N^{-1}+o(1/N)\Big)
N^{2\nu_{\cal K}},
\end{equation}
including a confluent correction \cite{janse,leguillou,zinn} in comparison to
the standard expression (\ref{gyrra}), it was found that the Flory exponent
$\nu_{\cal K}$ is independent of the knot type ${\cal K}$ and has the
3D value $0.588$. This was interpreted via a {\em localisation\/} of knots
such that the influence of tight knots on $R_g$ is vanishingly small. In fact,
$\Delta$ is
of the order of $0.5$ according to the investigations in references
\cite{leguillou,zinn,janse1,li}.
Based on longer chains in comparison to reference
\cite{janse}, the study of \cite{orlandini} thus corroborates the
independence of $\nu_{\cal K}\approx 0.588$ of the knot type ${\cal K}$.
In recent AFM experiments analysing single DNA knots, the Flory scaling
$R_g\simeq N^{\nu}$ was confirmed for both simple and complex knots
\cite{valle}.

For the number of degrees of freedom $\omega_{\cal K}$, it was found for the
form\footnote{Note that
we changed the exponent by 1 in comparison to the original work, making the
counting of non-translatable configurations consistent with the counting
convention specified in footnote \protect\ref{counting}.}
\begin{equation}
\label{orland}
\omega_{\cal K}\simeq A_{\cal K}N^{\alpha_{\cal K}-2}\mu_{\cal K}^N
\left(1+\frac{B_{\cal K}}{N^{\Delta_{\cal K}}}+\ldots\right)
\end{equation}
with confluent corrections, that while for the unknot with $\alpha_{\emptyset}
\approx 0.27$ expression (\ref{orland}) is consistent with the standard
result (\ref{dof}) ($[0.27-2]/3\approx -0.58\approx \nu$), for prime knots
$\alpha_{\cal K}=\alpha_{\emptyset}+1$, and for composite knots with $N_f$
prime components,
\begin{equation}
\alpha_{\cal K}=\alpha_{\emptyset}+N_f.
\end{equation}
This finding is in agreement
with the view that each prime component of a knot ${\cal K}$
is tightly localised and statistically able to move around one central loop,
each prime component counting an additional factor $N$ of degrees of freedom. The fact that
for a chain of finite thickness the size of the big central loop is in fact
diminished by the size of the tight knot is a confluent effect, such that the
confluent exponent $\Delta$ should be related to the size distribution of
the knot region.
Not surprisingly, the connectivity factor $\mu_{\cal K}\approx 4.68$ was found
to be independent of ${\cal K}$, assuming the standard value for a cubic
lattice \cite{guttmann}. Also the amplitude $A_{\cal K}$ and the exponent
$\Delta_{\cal K}$ of the confluent correction turned out to be ${\cal
K}$-independent. We note that a similar analysis in (pseudo) 2D\footnote{The
simulated polymer chain moves in 2D, however, crossings are permitted at
which one chain passes underneath another. In that, the simulated polymers
are in fact equivalent to knot projections with a certain orientations of
individual crossings.} also strongly points towards tight localisation of
the knot \cite{guitter}.

In contrast to the above results, 3D simulations undertaken in \cite{quake}
(also compare \cite{quake1}) show the dependence
\begin{equation}
\label{quake}
R_g\simeq N^{3/5}C^{-4/15}
\end{equation}
of the gyration radius on the knot type, characterised by the number $C$ of
essential crossings. $R_g$, that is, decreases as a power-law with $C$,
where the exponent $-4/15=1/3-\nu$ \cite{quake}. The functional form
(\ref{quake})) was derived from a Flory-type argument for a polymer construct
of $C$ interlocked loops of equal length $N/C$ by arguing that each loop
occupies a volume $\simeq (N/C)^{3\nu}$, and the volume of the knot is given
by $V\simeq C(N/C)^{3\nu}$ (i.e., assuming that due to self-avoiding repulsion
the volume of individual loops adds up to the total volume). Equation
(\ref{quake}) then follows immediately. This model of equal loop sizes is
equivalent to a completely delocalised knot. It may therefore be speculated,
albeit rather long chain sizes of up to 400 were used, whether the numerical algorithm
employed for the simulations in \cite{quake} causes finite-size effects that,
in turn, prevent a knot localisation. We note that the Flory-type
scaling assumed to derive expression (\ref{quake}) is consistent with a
modelling brought forward in reference \cite{grosberg2}, in which the knot is
quantified by the aspect ratio in a configuration corresponding to a maximally
inflated tube with the given topology (i.e., a state corresponding to
complete delocalisation). In reference \cite{quake}, the temporal relaxation
behaviour of a given knot was also studied. While regular
Rouse behaviour was found for the case of the unknot, the knotted chains
displayed somewhat surprising long time contributions to the relaxation
time spectrum \cite{treloar,ward,ferry,erman}, a phenomenon
already pointed out by de Gennes within an
activation argument to create free volume in a tight knot in order to move
along the chain \cite{degennesma}. Note that relatively lose knots in shorter
chains do not appear to exhibit such extremely long relaxation time behaviour
\cite{lai1}.

Simulation of a 3D knot with varying excluded volume showed, if only the
excluded volume becomes large enough, the gyration radius of the knot is
independent of the knot type \cite{shimamura}. The picture of tight knots
is further corroborated in the study by Katritch {\it et al.} using a Gaussian
chain model with fixed topology to demonstrate that the size distribution
of the knot is distinctly peaked at rather small sizes \cite{katritch}.

Apart from determining the statistical quantities $R_g$ and $\omega_{\cal K}$
from simulations, there also exist indirect methods for quantifying the size
of the knot region in a knotted polymer. One such method is to confine an
open chain containing a knot between two walls, and measuring the finite
size corrections of the force-extension curve due to the knot size. This is
based on the idea that the gyration radius for a system depending on more
than one length scale (i.e., apart from the chain length $N$) shows above
mentioned confluent corrections, such that \cite{farago}
\begin{equation}
R_g=AN^{\nu}\Phi\left(\frac{N_0}{N},\frac{N_1}{N},\ldots\right)\simeq
AN^{\nu}\Big(1-BN^{-\Delta}\Big)
\end{equation}
when only the largest correction is considered, and in 3D $\Delta\approx 0.5$
is supposed to be universal \cite{leguillou,zinn,janse1,li}.
If this leading correction is due to the argument $N_0/N$ in the scaling
function $\Phi$, the length scale $N_0$ depends on $N$ through the scaling
$N_0\sim N^t$ with $\Delta=1-t$. From Monte Carlo simulations of a bead and
tether chain model, it could then
be inferred that the size of the knot scales like \cite{farago}
\begin{equation}
N_k\sim N^t, \quad t=0.4\pm0.1
\end{equation}
This, in turn, enters the force-extension curve $f'=G(R')$ with the
dimensionless force $f'=fAN^{\nu}/(k_BT)$ and distance $R'=R/(AN^{\nu})$ of
the walls, in the form with confluent correction
\begin{equation}
f'\simeq G(R')\Big(1+g(R')N^{-\Delta}\Big).
\end{equation}
From the simulation, $t=0.4$ corresponds to the best data collapsing, assuming
the validity of
the scaling arguments. An argument in favour of this approach is the
consistency of the exponent $t=0.4$ with the inferred $\Delta=0.6$, which is
close to the known value. Note that the force-extension of a chain with a
slip-link was discussed in reference \cite{pull} and shown that a loop
separated off by a slip-link is confined within a Pincus-de Gennes blob.
We also note that results
corresponding to delocalisation in force-size relations were reported in
\cite{sheng,sheng1}. An entropic scale was conceived in \cite{roya}: Separating
two chains with fixed topology but allowing them to exchange length (e.g.,
through a small hole in a wall) would enable one to infer the localisation
behaviour of a knot by comparing the equilibrium balance of this knot with
a slip-link construct of known degrees of freedom until the
average length on both sides coincides. The preliminary results in \cite{roya}
are shadowed by finite-size effects of the accessible system size,
as limited by computation power. The analysis in reference \cite{marcone} of a
self-avoiding polygon model uses the
method of closure of a short fragment of the knot and subsequent determination
of its Alexander polynomial to obtain the scaling exponent $t=0.75$; in a
second variant, the authors find a consistent result by a variant of the
knot scale method. Another recent study uses a more realistic model for a
polymer chain, namely, a simplified model of polyethylene; with up to 1000
monomers in the simulation, the exponent $t\approx0.65$ is found (and
delocalisation is obtained in the dense phase) \cite{virnau}.

Thus, there exist simulations results pointing in both directions, knot
localisation and delocalisation. As the latter may be explained by finite
size effects, it seems likely that (at least simple) knots in 2D and 3D
localise in the sense that the knot region occupies a portion of the chain
that is significantly smaller in comparison to the entire chain. In particular,
this would imply that the average size of the knot region $\langle\ell\rangle$
scales with the chain length $Na$ with an exponent less than one, such that
\begin{equation}
\lim_{N\to\infty}\frac{\langle\ell\rangle}{Na}=0.
\end{equation}
Below, we show from analytical grounds that such a localisation is a natural
consequence of interactions of a chain of fixed topology with fluctuations.
We note, however, that conclusive results for knot localisation may in fact
come from experiments: Manipulation of single chains such as DNA can be
performed for rather long chains, making it possible to reach beyond the
finite-size corrections inherent in, e.g., the force-extension simulations
mentioned above. The aforementioned AFM studies on single DNA knots indeed
reveal knot localisation of flattened knots \cite{valle}; due to experimental
limitations, presently only one DNA length was investigated, such that the
scaling exponent $t$ currently cannot be obtained.

Before proceeding to these analytical approaches, we note that there have
also been performed simulations of knotted chains under non-dilute conditions
\cite{stella,orlanddense}.
In (pseudo) 2D, these have found delocalisation of the knot, i.e., $\lim_{
N\to\infty}\langle\ell\rangle/N=const$. We come back to these
simulations below in connection with the modelling of dense and $\Theta$-knots.

\subsection{Flattened knots in dilute and dense phases.}

Analytically, knots are a hard problem to tackle. Statistical mechanical
treatments of permanently entangled polymers are so difficult to treat
since topological restrictions cannot be formulated as a Hamiltonian problem
but appear as hard constraints partitioning the phase space
\cite{degennes,grosberg,vilgis,kholodenko}.\footnote{For comparison,
self-avoidance in 3D is usually treated as a perturbation, i.e., as a ``soft
constraint'', in analytical studies \protect\cite{degennes}.} A segment of a 3D
knot, in other words, can move without feeling the constraints due to the
non-trivial topology of a knotted state, until it actually collides with
another segment. The accessible phase space of degrees of freedom is therefore characterised
by inequalities.\footnote{Although a similar statement is true for polymer
networks in 3D, the field theoretical results for their critical exponents are
in fact obtained as {\em averages\/} over {\em all\/} topologies. For
instance, the exponent $\nu$ entering the gyration radius of a a 3D polymer
ring counts all knotted states \protect\cite{duplantier1}.}

Consequently, only a relatively small range of problems have
been treated analytically, starting with the seminal papers by Edwards
\cite{edwards1,edwards2}, in which he considers the classification of
topological constraints in polymer physics. De Gennes addressed the problem
of tight knot motion along a polymer chain using scaling arguments for the
activation of free length inside the knot region, producing a double-exponential
expression for the corresponding time scale \cite{degennesma}, which might
explain the extreme long-time contributions in the relaxation time
spectrum of permanently entangled polymers \cite{doi,treloar,ward,erman,ferry}.
Some analytical results were obtained for a pair, or an `Olympic' gel of
entangled polymer rings, see for instance,
\cite{otto,otto1,vilgis1,ferrari,ferrari1}. In a mean field approach based on
the Kauffman invariant the entropy of knots was investigated in references
\cite{grone,grone1,grone2}. Similarly, some statistical properties of random
knot diagrams were investigated in \cite{nechaev,nechaev1}. However, some
insight
can be gained on the basis of phenomenological models, which we will come back
to below. Here, we continue with an analytical study of flat knots.

\begin{figure}
\begin{center}
\includegraphics{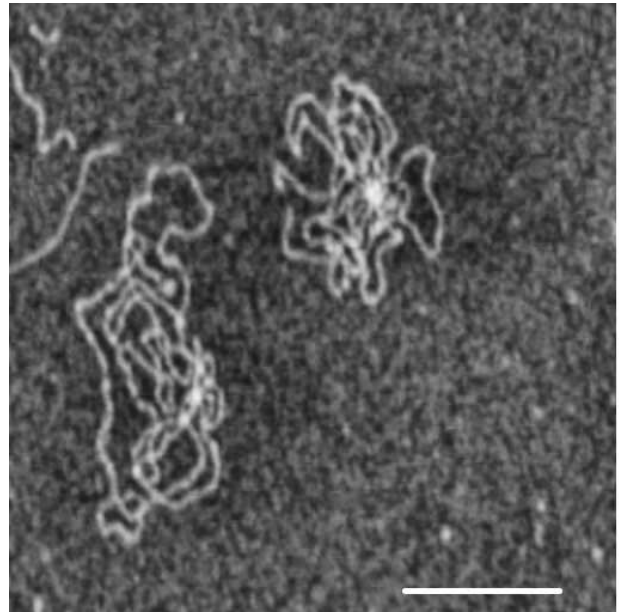}
\end{center}
\caption{AFM tapping-mode images of flattened, complex DNA-knots with
approximately 30-40 essential crossings, see \protect\cite{valle}.
The substrate surface used is AP-mica (freshly cleaved mica
reacted with an amino terminal silane to obtain a positively charged surface).
The DNA knots used are extracted from bacteriophage P4; the DNA is a 11.4
kbp molecule (with a 1.4 kbp deletion resulting in a final length of 10 kbp)
which has two cohesive ends. They are not covalently closed, thus no
supercoiling is present. The knot adsorbed out of the 3D bulk
on to the surface is strongly trapped, i.e., the knot is `projected' onto the
surface without any equilibration. The knot appears rather delocalised.
Courtesy F. Valle and G. Dietler.
\label{valle}}
\end{figure}

One possibility to treat knotted polymer chains analytically is to confine
the degrees of freedom of the knot to motion in 2D, only. The knot, that is, is
preserved, as at the crossings the chain is allowed to form an
over-/underpassing, while the rest of the knot is confined to 2D. Such a
confinement can in fact be experimentally realised in various ways. Thus,
the chain can be confined between two close-by glass slabs, as demonstrated
in \cite{craighead}; it can be pressed flat on a surface by gravitation or
similar forces, for instance in macroscopic systems \cite{bennaim,bennaim1};
the chain can be adhesively bound to a membrane and still reach
configurational equilibrium, as experimentally shown for DNA in references
\cite{valle,maier}. Or it can be adsorbed to a mica surface either by APTES
coating or by providing bivalent Mg ions in solution, as shown in
figure \ref{valle}. From such flat knots as discussed in
the remainder of this section, we will
be able to infer certain generic features also for 3D knots.

A flat knot therefore corresponds to a polymer network in 2D, but the
orientation of the crossings is preserved, such that the network graph
actually coincides with a typical knot projection
\cite{kauffman,reidemeister,adams}, as shown in figure \ref{tref_proj}
on the left. This projection of the trefoil, and similar projections for
all knots, displays the knot with the essential crossings. A flat knot can,
in principle acquire an arbitrary number of crossings by Reidemeister moves;
for instance, the bottom left segment of the flat trefoil can slide under
the vicinal segment, creating a new pair of vertices, and so on. However,
we suppose that such transient additional loops are sufficiently short-lived
so that we can neglect them in our analysis. Then, we can apply results from
scaling analysis of polymer networks of the most general type shown in
figure \ref{network}, see the primer in the appendix. We note that from the
Monte Carlo simulations we performed it may be concluded that such additional
vertices can in fact be neglected.

\subsubsection{Flat knots in dilute phase.}

\begin{figure}
\begin{center}
\includegraphics[width=7.2cm]{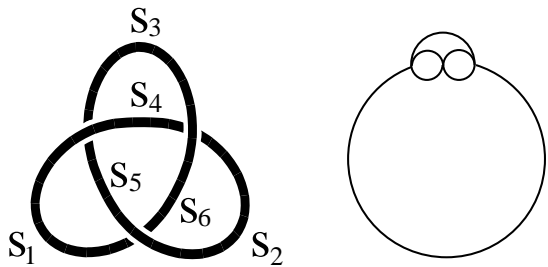}
\end{center}
\caption{Flat trefoil knot with segment labels. On the right, a schematic
representation of a localised flat trefoil with one large segment is
shown.
\label{tref_proj}}
\end{figure}

We had previously found that for the F8-structure the probability density
function for the size of
each loop is peaked at $\ell\to 0$ and $\ell\to L$.
From the scaling analysis for
self-avoiding polymer networks, we concluded strong localisation of one
subloop. For more complicated structures, the joint probability to find
the individual segments with given lengths $s_i$ is expected to peak at
the edges of the higher-dimensional configuration hyperspace. Some analysis
is necessary to find the characteristic shapes. Let us consider here the
simplest non-trivial knot, the (flat) trefoil knot $3_1$ shown in figure
\ref{tref_proj}.
Each of the three crossings is replaced with a vertex with four
outgoing legs, and the resulting network is assumed to separate into
a large loop and a multiply connected region which includes the vertices.
Let $\ell=\sum_{i=1}^5 s_i$ be the total length of all segments contained
in the multiply connected knot region. Accordingly, the length of the large
loop is $s_6=L-\ell$.

\begin{figure*}
\begin{center}
\includegraphics{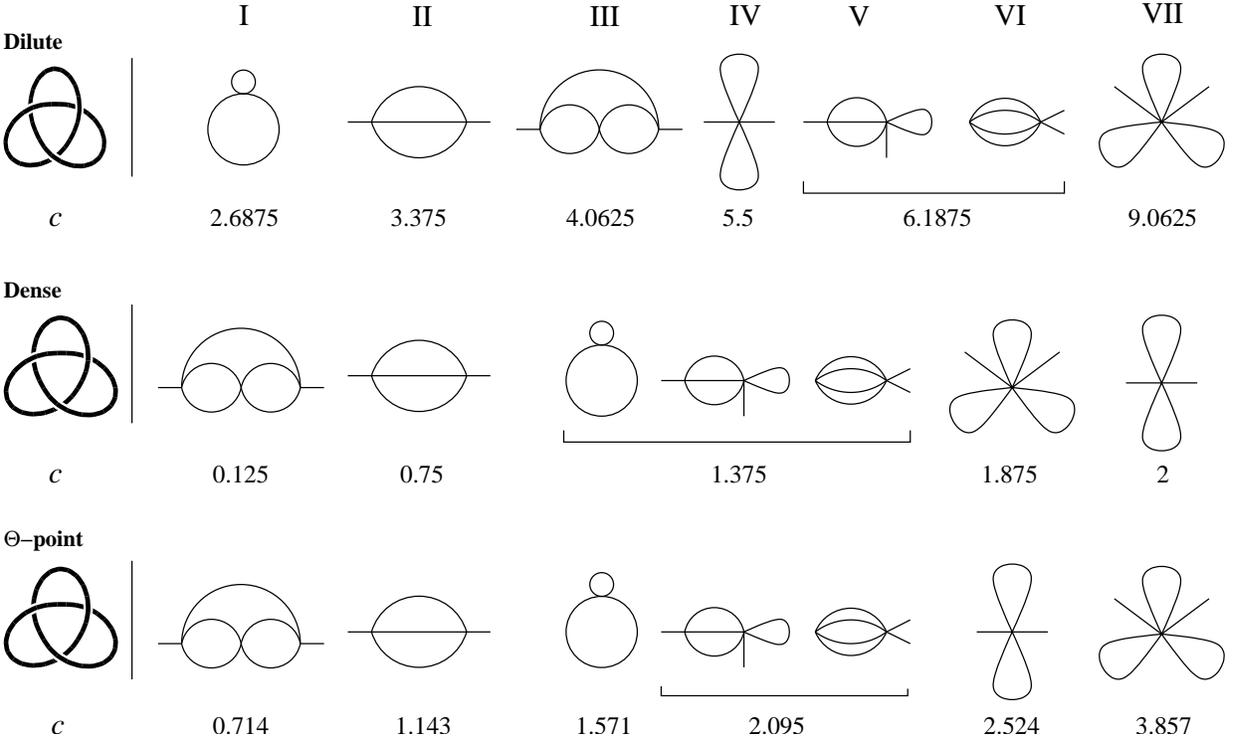}
\end{center}
\caption{Hierarchy of the flat trefoil knot $3_1$. Upper row: dilute phase.
Middle row: $\Theta$-phase. Bottom row: dense phase. To the left of each
row, the trefoil projection is shown. It splits up into the hierarchies of
configurations, with exponents $c$ below each contraction. The small protruding
legs represent the big central loop, compare, for instance, figure
\protect\ref{tref_proj} on the right with position III of the top row. See
text for details.
\label{knot_zoo}}
\end{figure*}

In the limit $\ell\ll L$, the number of configurations of this network can be
derived in a similar way as in the scaling approach followed for the F8. This
procedure determines the concrete behaviour of the scaling form
\begin{equation}
\omega_{\rm III}\simeq\mu^L{\cal W}_{\rm III}\left(L-\ell,\ell,
\frac{s_1}{\ell},\frac{s_2}{\ell},\frac{s_3}{\ell},\frac{s_4}{\ell}\right)
\end{equation}
including the scaling function ${\cal W}$ that depends on altogether six
arguments. The index III is chosen according to figure \ref{knot_zoo},
where the flat trefoil configuration in the dense phase appears at
position III of the scheme (explained below). After
some manipulations, the number of degrees of freedom yields in the form \cite{slili2d}
\begin{equation}
\label{t_limit}
\omega_{\text{III}}(\ell,L)\sim\mu^L (L-\ell)^{-d \nu}\ell^{-c},
\end{equation}
with the scaling exponent
\begin{equation}
c=-(\gamma_{\text{III}}-1+d\nu)-m, \qquad m=4.
\end{equation}
Here, $m=4$ corresponds to the number of independent integrations over the
segments $s_i$ ($i=1,\ldots,4$) of the knot region, as we only retain the
cumulative size $\ell=\sum_{i=1}^5s_i$ of the knot region. Putting numerical
values, we find $c=65/16$, i.e., {\em strong localisation}.

However, some care is necessary in performing these integrations,
since the scaling function ${\cal W}_{\rm III}$ may
exhibit non-integrable singularities if one or more of the arguments
$s_i/\ell$ tend to $0$. The geometries corresponding to these
limits (edges of the configuration hyperspace)
represent {\em contractions\/} of the original
trefoil network ${\cal G}_{\text{III}}$ in the sense that the length of
one or more of the segments $s_i$ is of the order of the
short-distance cutoff $a$. If such a short segment connects
different vertices, they cannot be resolved on larger length
scales, but appear as a single, new vertex. Thus, each
contraction corresponds to a different network ${\cal G}$,
which may contain a vertex with up to eight outgoing legs. For
the flat trefoil knot, there exist six different contractions,
as grouped in figure \ref{knot_zoo} around the original flat
trefoil at position III. As an example, in the top row of figure \ref{knot_zoo}
contraction VI follows from the original trefoil III if the uppermost segment
becomes very small, and similarly the network VII emanates from contraction VI
if one of the four symmetric segments becomes very small.
For each of these networks, one can calculate the corresponding
exponent $c$ in a similar way as above, leading to the general
expression
\begin{equation}
\label{euler}
c=2+\sum_{N\ge 4}n_N\left\{\frac{N}{2}\Big(d\nu-1\Big)+
\Big(|\sigma_N|-d\nu\Big)\right\}.
\end{equation}
The $\sigma_N$ are given in equation (\ref{top_coeff}). In figure
\ref{knot_zoo}, the various contractions are arranged in increasing
exponent $c$.

Our scaling analysis relies on an expansion in
$a/\ell\ll 1$, and the values of $c$ determine a sequence of
contractions according to higher orders in $a/\ell$: The {\em smallest\/}
value of $c$ corresponds to the most likely contraction, while the
others represent corrections to this leading scaling behaviour, and
are thus less and less probable (see figure \ref{knot_zoo}).
To lowest order, the trefoil behaves like a large ring polymer at whose fringe
the point-like knot region is located. At the next level of resolution,
it appears contracted to the figure-eight shape ${\cal G}_\mathrm{I}$.
For more accurate data, the higher order
shapes II to VII may be found with decreasing
probability. Interestingly, the original uncontracted trefoil
configuration ranks third in the hierarchy of shapes.

\begin{figure}
\begin{center}
\includegraphics[width=8cm]{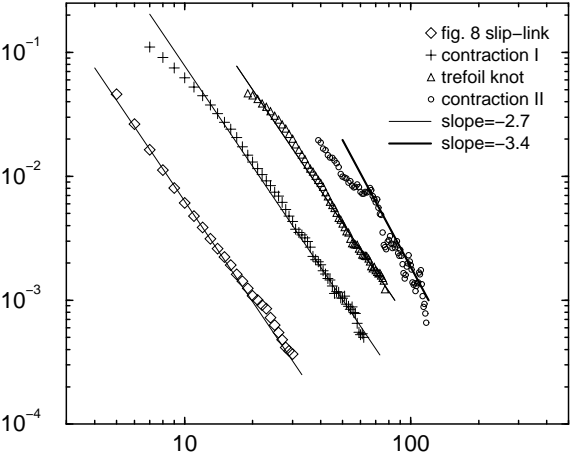}
\end{center}
\caption{Power law tails in probability density functions
for the size $\ell$ of tight segments:
As defined in the figure, we show results for the smaller loop in a
figure-eight structure, the overall size of the trefoil knot,
as well as the two leading contractions of the latter.
\label{figsl1}}
\end{figure}

These predictions were checked by MC simulations with the same conditions as
described above, to prevent intersection. The flat trefoil knot was prepared
from a symmetric, harmonic 3D representation with 512 monomers, which was
collapsed and then kept on a hard wall by the
``gravitational'' field $V=-k_BTh/h^*$ perpendicular to the wall, where $h$ is
the height of a monomer, and $h^*$ was set to 0.3 times the bead diameter.
Configurations corresponding to contraction I
are then selected by requiring that besides a large loop, they contain
only one segment larger than a preset cutoff length (taken to be 5 monomers),
and similarly for contraction II.
The size distributions for such contractions, as well as for all
possible knot shapes are shown in figure \ref{figsl1}.
The tails of the distributions are indeed consistent with the predicted
power laws, although the data (especially for contraction II) is too
noisy for a definitive statement.

\begin{figure}
\begin{center}
\includegraphics[height=4.4cm]{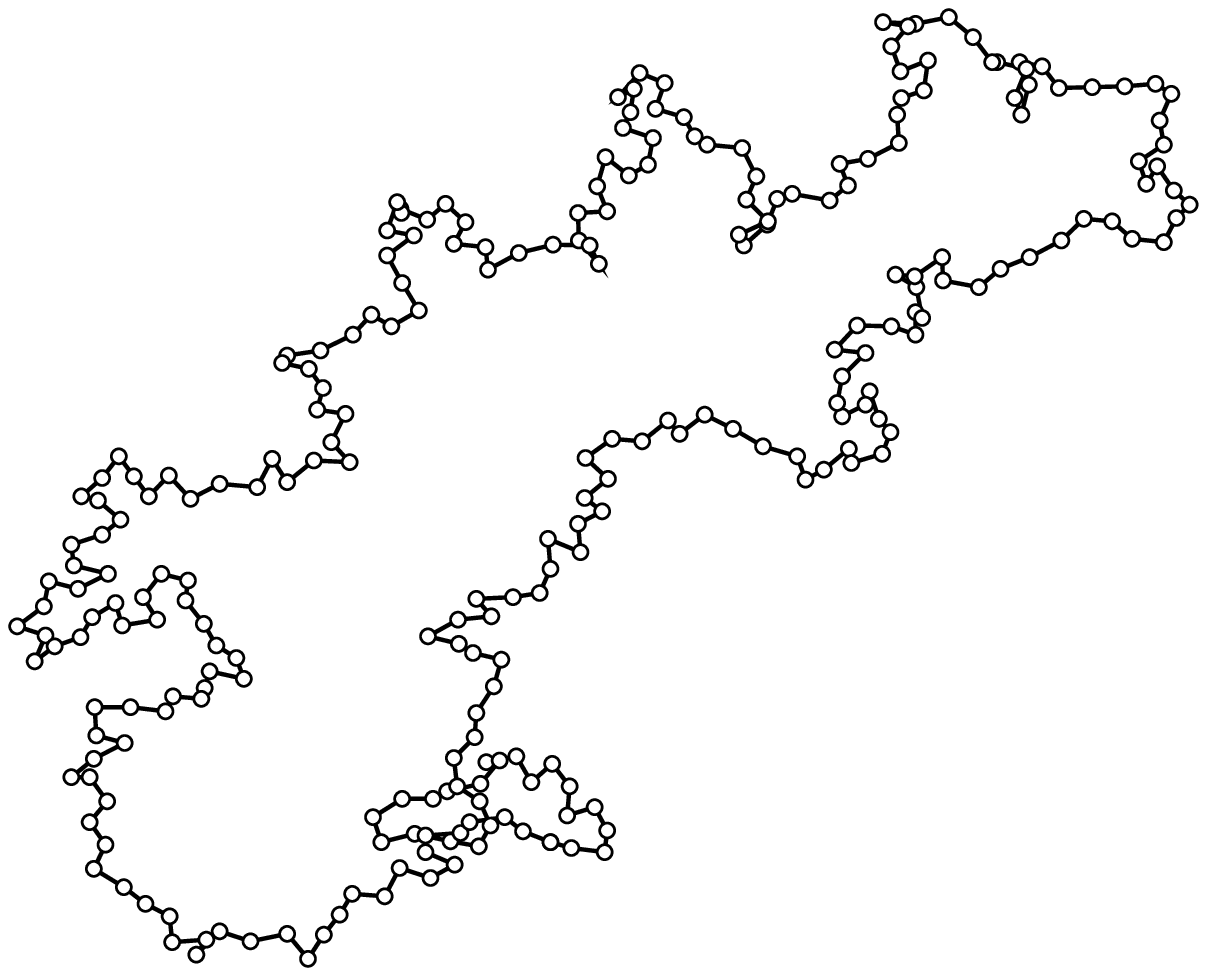}
\includegraphics[height=4.4cm]{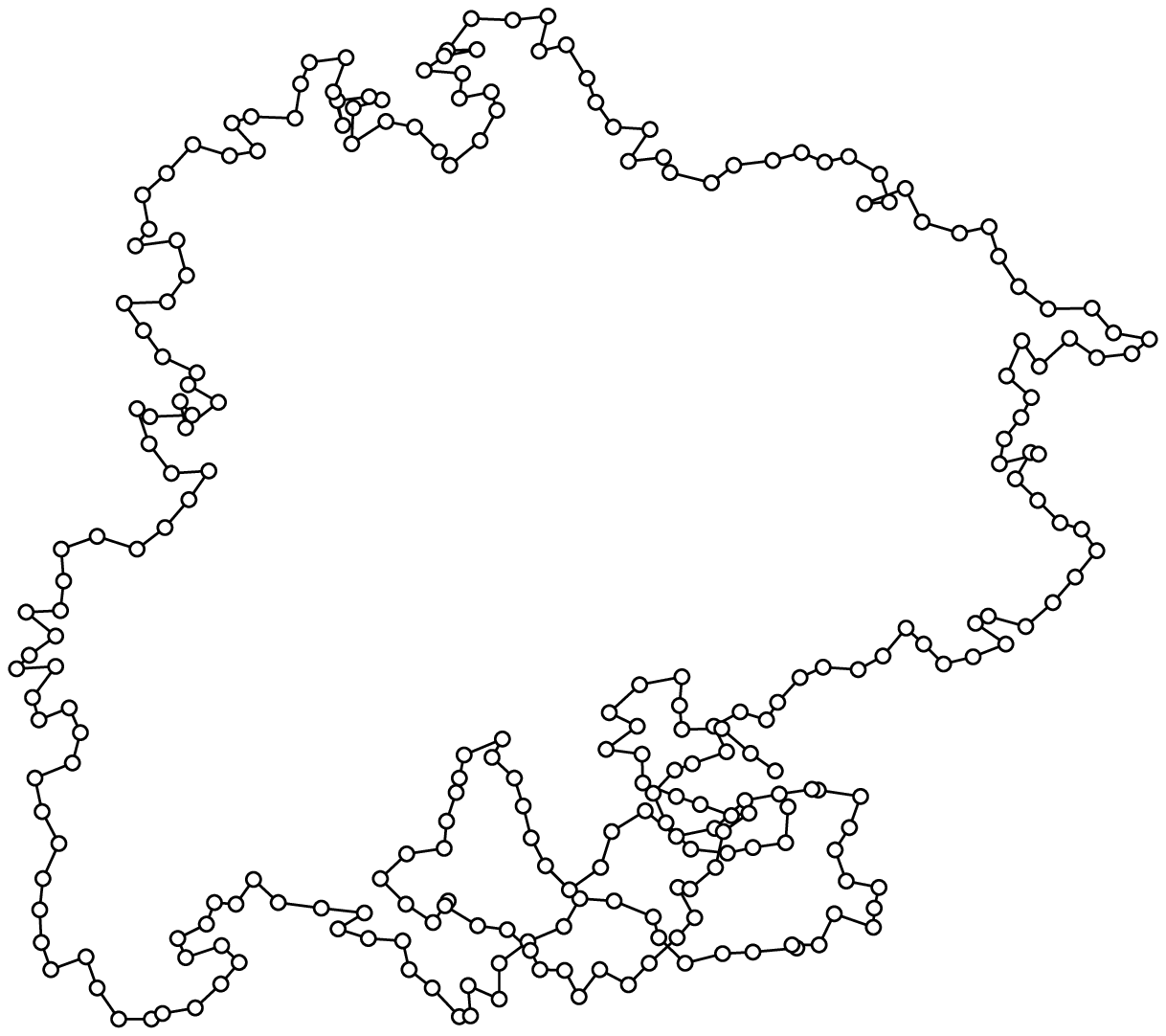}
\includegraphics[height=4.4cm]{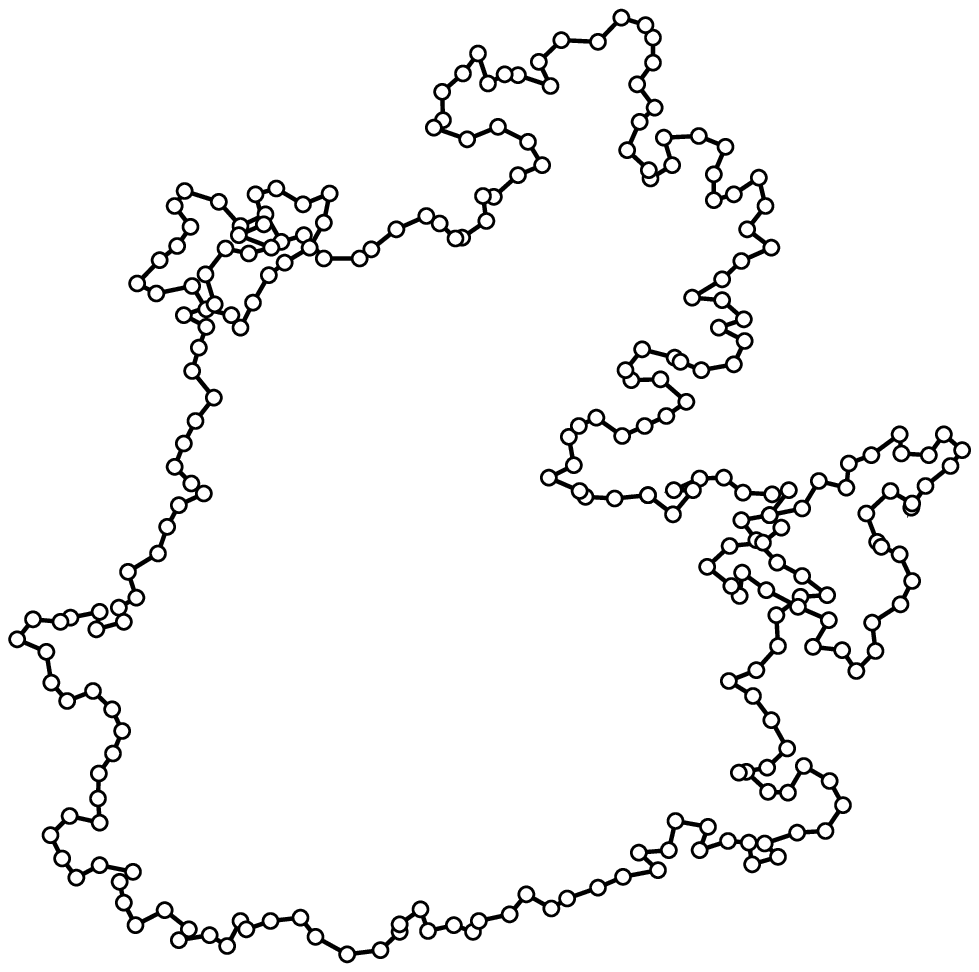}
\end{center}
\caption{Typical configurations of 256-mer chains for the trefoil $3_1$,
the prime knot $8_{19}$, and the composite knot $3_1$\#$3_1$
consisting of two trefoils, in $d=2$. The initial conditions were symmetric
in all cases.
\label{fig_mcknots}}
\end{figure}

\begin{figure}
\includegraphics[width=6.8cm,angle=90]{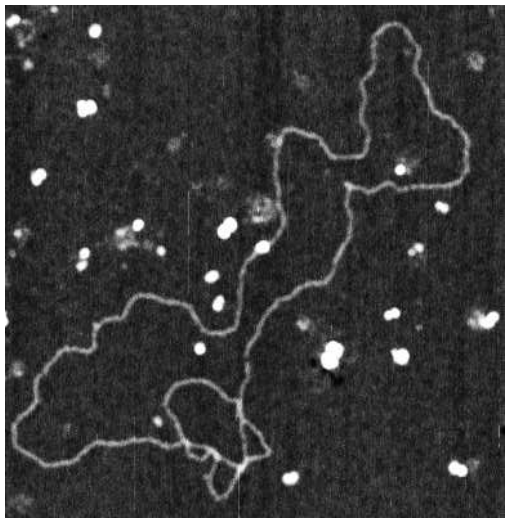}
\caption{Flat knot imaged by AFM, similar to the one shown in figure
\ref{valle}. However, this knot is rather simple (likely a trefoil)
and was allowed to relax while attaching to freshly cleaved mica in
the presence of bivalent Mg counterions. Courtesy E. Ercolini, J. Adamcik,
and G. Dietler. Note the close resemblance to
the trefoil configuration shown in figure \ref{fig_mcknots}.
\label{erika}}
\end{figure}

Our scaling results pertain to all flat prime knots. In particular, the
dominating contribution for {\em any} prime knot corresponds to the
figure-eight contraction ${\cal G}_{\text{I}}$, as equation (\ref{euler}) predicts
a larger value of the scaling exponent $c$ for any network ${\cal G}$ other
than ${\cal G}_{\text{I}}$. Accordingly, figure \ref{fig_mcknots} demonstrates
the tightness of the prime knot $8_{19}$. Composite knots, however, can
maximise the number of configurations by
splitting into their prime factors as indicated
in figure \ref{fig_mcknots} for $3_1$\#$3_1$. Each prime factor is
tight and located at the fringe of one large loop, and accounts for
an additional factor of $L$ for the number of configurations
as compared to a ring of length $L$ without a knot.
Indeed, this gain in entropy leads to the tightness of knots.
Flat knots can experimentally be produced by `projecting' a dilute 3D knot
from the bulk onto a mica surface, on which the knot is adsorbed. Variation
of the ionic strength in the solution determines whether the knot is going to
be strongly trapped on the surface such that, once captured on the surface, it
is completely immobilised (small ionic strength); or whether the adsorption
is weaker such that the knot can (partially) equilibrate while being confined
to 2D, i.e., equilibrate as a flat knot.
Figure \ref{valle} shows a strongly trapped complex knot, whereas figure
\ref{erika} depicts a weakly adsorbed simple knot, compare \cite{valle}.

\subsubsection{Flat knots under $\Theta$ and dense conditions.}

In many situations, polymer chains are not dilute.
Polymer melts, gels, or rubbers exhibit fairly high densities of
chains, and the behaviour of an individual chain in such systems
is significantly different compared to the dilute phase
\cite{degennes,doi,erman}. Similar considerations apply to biomolecules:
in bacteria, the gyration radius of the almost freely floating
ring DNA may sometimes be larger than the cell radius itself.
Moreover, under certain conditions, there is a non-negligible
osmotic pressure due to vicinal layers of protein molecules,
which tends to confine the DNA \cite{walter,vasilevskaya,lerman}.
In protein folding studies, globular proteins in their native
state are often modelled as compact polymers on a lattice
(see \cite{garel} for a recent review).

A polymer is considered dense if, on a lattice, the fraction $f$ of occupied
sites has a finite value $f > 0$. This can be obtained by considering a single
polymer of total length $L$ inside a box of volume $V$ and taking the limit
$L \to \infty$, $V \to \infty$ in such a way that $f = L/V$ remains
finite \cite{duplantier2,duplantier3,jacobsen}.
Alternatively, dense polymers can be obtained in an infinite volume
through the action of an attractive force between monomers.
Then, for temperatures $T$ below the collapse
(Theta) temperature $\Theta$, the polymers collapse to a dense phase,
with a density  $f > 0$, which is a function of $T$
\cite{duplantier3,duplantier4,owczarek,owczarek1}.
For a dense polymer in $d$ dimensions, the exponent
$\nu$, defined by the radius of gyration $R_g \sim L^{\nu}$, becomes
$\nu = 1 / d$.
The limit $f = 1$ is realised in Hamiltonian paths,
where a random walk visits every site of a given
lattice exactly once \cite{duplantier7,kondev}. Dense polymers may be related
to 2D vesicles and lattice animals (branched polymers)
\cite{fisher,seno,dekeyser,cardy}.

\begin{figure}
\begin{center}
\includegraphics[width=4.2cm]{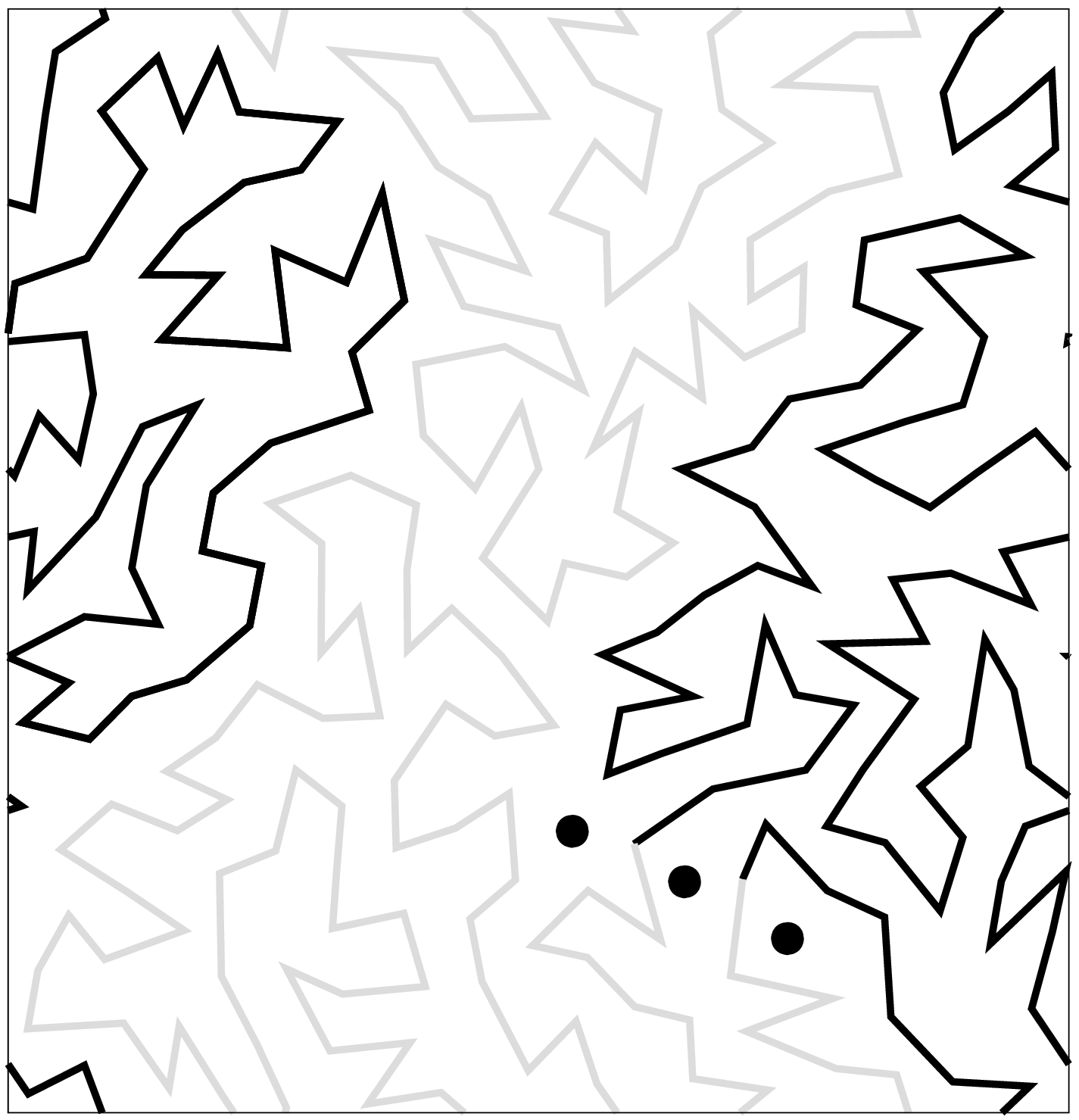}
\includegraphics[width=4.2cm]{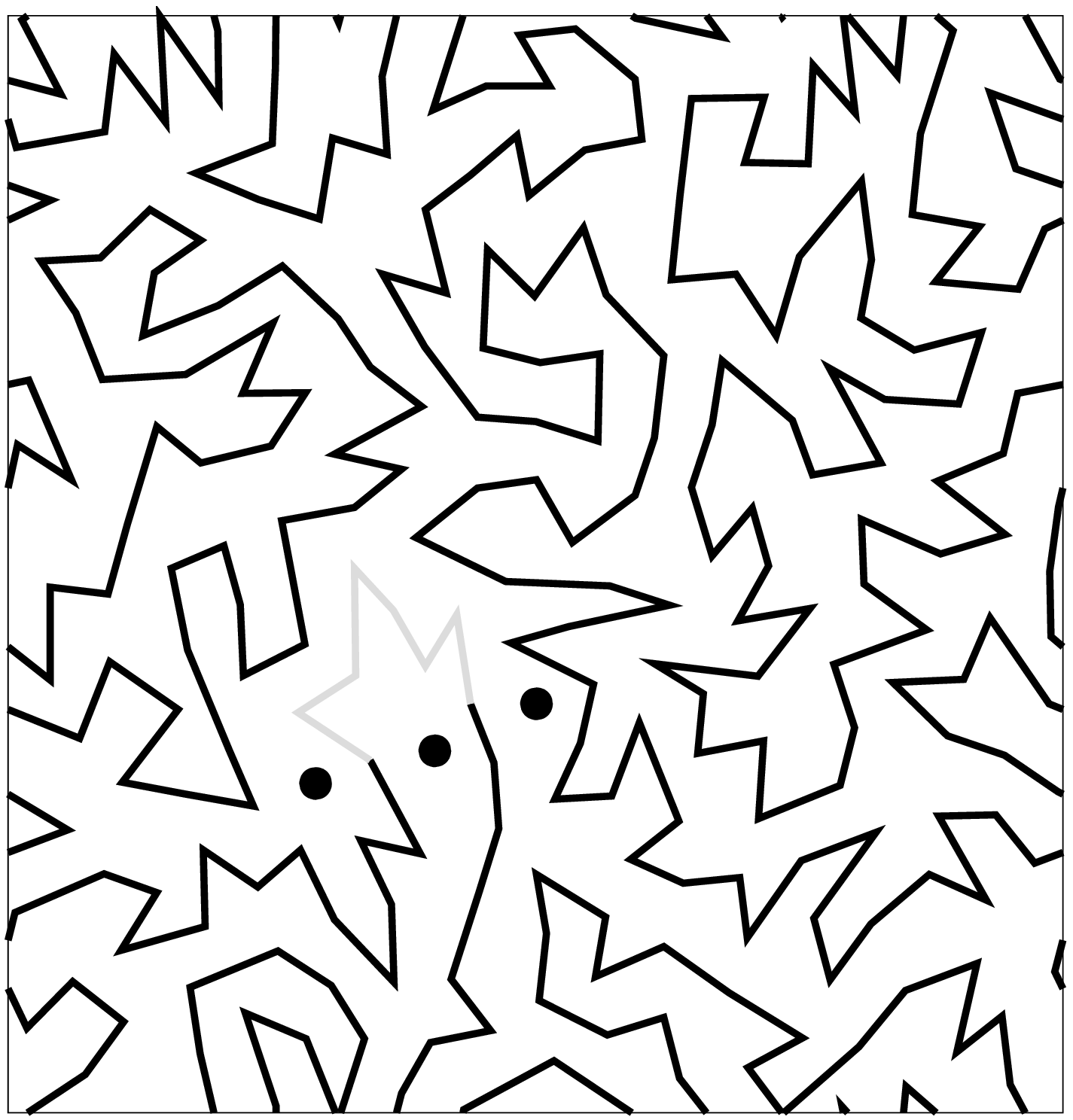}
\end{center}
\caption{Symmetric ($\ell=L/2=128$) initial configuration of a 2D dense
F8 (left) and equilibrium configuration (right) with
periodic boundary conditions. The two different grey values correspond to the
two subloops created by the slip-link. The slip-link itself is
represented by the three (tethered) black dots.\label{fig-8}}
\end{figure}

\begin{figure}
\begin{center}
\includegraphics[height=6cm]{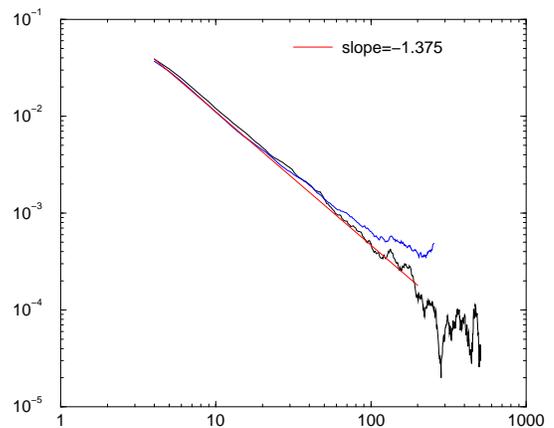}
\end{center}
\caption{The loop size probability distribution  $p(\ell)$ at $\rho=55\%$ area
coverage, for the F8 with 512 (top) and 1024 (bottom) monomers.
The power law with with the predicted exponent $c=1.375$ in equation (\ref{c8})
is indicated by the dotted line.
\label{fig-8_1}}
\end{figure}

As studied in reference \cite{dense},
the value of the exponent $c$ for the 2D dense F8 is (compare to the appendix)
\begin{equation} \label{c8}
c=-\gamma_{\rm F8}=11/8=1.375,
\end{equation}
implying that the smaller loop is {\em weakly localised\/}. This means
that the probability for the size of each loop is peaked
at $\ell=0$ and, by symmetry, at $\ell=L$.
An analogous reasoning for the 2D F8 at the $\Theta$ point gives
\begin{equation} \label{c8_theta}
c=11/7=1.571.
\end{equation}
In both cases the smaller loop is weakly localised in the sense that $\langle
l\rangle_</L\to 0$. Figure \ref{fig-8} shows
the symmetric initial and a typical equilibrium configuration for periodic
boundary conditions obtained from Monte Carlo (MC) simulations, see reference
\cite{dense} for details.
In figure \ref{fig-8}, the lines represent the bonds (tethers) between the
monomers (beads, not shown here). The three black dots mark the locations
of the tethered beads forming the slip-link in 2D.
The initial symmetric configuration soon gives way to a
configuration with $\ell\ll L$ on approaching equilibrium.
Figure \ref{fig-8_1} shows the development of this symmetry breaking as a
function of the number of MC steps.
We note, however,
that the fluctuations of the loop sizes in the ``stationary'' regime
appear to be larger in comparison to the dilute case studied in reference
\cite{paraknot}, compare figure \ref{fig8_sim}.
We checked that for densities (area coverage) above
40\% the scaling behaviour becomes independent of the density.
(The above simulation results correspond to a density of 55\%.)
The size distribution data is well fitted to a power law (for over 1.5 decades
with 1024 monomers),
and the corresponding exponent  with 512 and 1024
monomers in figure \ref{fig-8_1}
is in good agreement with the predicted value (\ref{c8}).

For our MC analysis, we again used a hard core
bead-and-tether chain, in which self-crossings were prevented by keeping a
maximum bead-to-bead distance of 1.38 times the bead diameter, and a maximum
step length of 0.15 times the bead diameter. To create the dense F8 initial
condition, a free F8 is squeezed into a quadratic box with hard
walls. This is achieved by starting off from the free F8, surrounding it by a
box, and turning on a force directed towards one of the edges. Then,
the opposite edge is moved towards the centre of the box, and so on.
During these steps, the slip-link is locked,
i.e., the chain cannot slide through it, and the two loops are of equal length
during the entire preparation. Finally, when the envisaged density is
reached, the hard walls are replaced by periodic boundary conditions,
and the slip-link is unlocked.
After each step, the system is allowed to relax for times larger than the
localisation times occurring at the main stage of the run.

A similar analysis as for the dense/$\Theta$-F8 structure and the dilute flat
trefoil above, reveals the number of degrees of freedom for the flat dense trefoil
in the form \cite{dense}
\begin{equation}
\label{3_limit}
\omega_{3}(\ell,L)\sim\omega_{0}(L)\ell^{-c}
\end{equation}
with $c=-\gamma_3-m$, where $\gamma_3=-33/8$ from equation
(\ref{nexp}) in the appendix (${\cal L} = 4$, $n_4 = 3$)
and $m = 4$ is the number of independent integrations over chain
segments. Thus, $c = 1/8 < 1$ which implies that the dense 2D
trefoil is {\em delocalised}.
As above, we have to consider the various possible contractions of the flat
knot. For dense polymers, the present scaling results show
that both the original trefoil shape ($c = 1/8 < 1$, see above)
and position II ($c = 3/4 < 1$) are in fact {\em delocalised\/} and
represent equally the leading scaling order (cf.~top part of figure
\ref{knot_zoo}). The F8 is only found at the third position and is weakly
localised ($c = 11/8 > 1$).
In an MC simulation of the dense 2D trefoil, we predict that
one mainly observes delocalised shapes corresponding to the original
trefoil and position II in figure \ref{knot_zoo}, and further,
with decreasing probability, the weakly localised F8 and the other
shapes of the hierarchy (top part) in figure \ref{knot_zoo}.

These predictions are consistent with the numerical simulations of
reference \cite{stella}, who observe that the mean
value of the second largest segment of the simulated 2D dense trefoil
configurations grows linearly with $L$, and conjecture the same
behaviour also for the other segments, corresponding to the
delocalisation of the trefoil obtained above.

An analogous reasoning can be applied to the 2D trefoil in the
$\Theta$ phase. We find that
in this case that the leading shape is again the
original (uncontracted) trefoil, with $c = 5/7 < 1$. This
implies that the 2D trefoil is {\em delocalised\/} also at the
$\Theta$ point. All other shapes are at least weakly localised,
and subdominant to the leading scaling order represented by the
original trefoil. The resulting hierarchy of shapes is shown
in figure \ref{knot_zoo} (bottom part).

\subsection{3D knots defy complete analytical treatment.}

As already mentioned, 3D knots correspond to a problem involving hard
constraints that defy a closed analytical treatment. It may be possible,
however, that by a suitable mapping to, for instance, a field theory,
an analytical description may be found. This may in fact be connected
to the study of knots in diagrammatic solutions in
high energy physics \cite{gambini}. There exists a fundamental relation
between knots and gauge theory as knot projections and Feynman graphs
share the same basic ingredients corresponding to a Hopf algebra
\cite{kauffman}. However, up to now no such mapping has been found, and
a theoretical description of 3D knots based on first principles is
presently beyond hope. To obtain some insight into the statistical
mechanical behaviour of knotted chains, one therefore has to resort
to simulations studies or experiments. In addition, a few phenomenological
models for both the equilibrium and dynamical behaviour of knots have been
suggested such as in references
\cite{quake,slili3d,lai1,sheng1,sheng,grosberg1,grosberg2,grosberg4}.

When discussing numerical knot studies, we already mentioned the Flory-type
model brought leading to equation \ref{quake} \cite{quake,quake1}.
One may argue that the
differences in the knot size for the different knot types corresponding to
the same $C$ may be included in the prefactor, that is independent of the
chain length $N$. Obviously, this model of equal loop sizes is equivalent to
a completely delocalised knot.
This statement is in fact equivalent to another Flory-type approach to
knotted polymers reported in \cite{grosberg2}. In this model, the knot is
thought of as an inflatable tube: for a very thin tube diameter, the
tube is equivalent to the original knot conformation; inflating the tube more
and more will increasingly smoothen out the shape until a maximally inflated
state is reached. The knot is then characterised by the aspect ratio
\begin{equation}
p=\frac{L}{D}\,\,, \, \mbox{therefore} \,\,
1\le p\le N,
\end{equation}
between length $L$ and maximum tube diameter $D$. It appears that $p$ is a
(weak) knot invariant, and can be used to characterise the gyration radius of
the knot. It is clear that, by construction, the aspect ratio described a
totally delocalised knot, and indeed it turns out that in good solvent, the
gyration radius shows the dependence $R_g\simeq AN^{3/5}\tau^{1/5}p^{-4/15}$,
where $\tau$ is the (dimensionless) deviation from the $\Theta$ temperature
\cite{grosberg2}. Obviously, the aspect ratio appears to be proportional to
the number of essential crossings in comparison to expression (\ref{quake}).
We note that similar considerations are employed in reference \cite{grosberg1},
including a comparison to the entropy of a tight knot, finding comparable
entropic likelihood. The modelling based on the aspect ratio $p$ is further
refined in \cite{grosberg4}.

Knot localisation is a subtle interplay between the degrees of freedom of one big loop,
and the internal degrees of freedom of the various segments in the knot region. Under
localisation, the number of degrees of freedom
\begin{equation}
\omega\simeq\mu^NN^{1-d\nu}
\end{equation}
includes an additional factor $N$ from the knot region encircling the
big loop. For flat knots, the competition between the single big loop and
the knot region is indeed won by the big loop.
In the case of 3D knots, this balance is presently not resolved for
knots of all complexity. Probably only detailed simulations studies of higher
order knots will make it possible to decide for the various models of 3D knots.
Major contributions are also expected from single molecule experiments, for
instance, from force-extension measurements along the lines of the simulations
study in \cite{farago}, the advantage of experiments being the fact that it
should be possible to go towards rather high chain lengths that are inaccessible
in simulations.
To overcome similar difficulties in the context of the entropic
elasticity for rubber networks, Ball, Doi, Edwards and coworkers
replaced permanent entanglements by slip-links
\cite{edwards2a,edwards2b,edwards2c,edwards2d}.
Gaussian networks containing slip-links have been successful in the prediction
of important physical quantities of rubber networks \cite{erman}, and they
have been used to study a small number of entangled chains \cite{sommer}. In a
similar fashion, one may investigate the statistical behaviour of single polymer
chains in which a fixed topology is created by a number of slip-links. Such
`paraknots' can be studied analytically using the Duplantier scaling results
\cite{paraknot}. As mentioned previously, knowledge of the statistical
behaviour of paraknots can be used to create a knot scale for calibrating
the degrees of freedom of real knots, and therefore also important to understand or design
indirect experiments on knot entropy, such as by force-extension measurements
\cite{pull}. Paraknots may also be useful in the design of entropy-based
functional molecules \cite{hame_cpl,meamb_ctn}.

\section{DNA breathing: local denaturation zones and biological implications}

"A most remarkable physical feature of the DNA helix, and one that is
crucial to its functions in replication and transcription, is the ease
with which its component chains can come apart and rejoin. Many techniques
have been used to measure this melting and reannealing behaviour. Nevertheless,
important questions remain about the kinetics and thermodynamics of denaturation
and renaturation and how these processes are influenced by other molecules in
the test tube and cell'' \cite{kornberg}. This remarkable quotation, despite
30 years old, still summarises the challenge of understanding local and global
denaturation of DNA, in particular, its dynamics. In this section, we report
recent findings on the spontaneous formation of intermittent denaturation
zones within an intact DNA double helix. Such denaturation \emph{bubbles\/}
fluctuate in size by (random) motion of the zipper forks relative to each
other. The opening and subsequent closing of DNA bubbles is often called
\emph{DNA breathing}.

\begin{figure}
\includegraphics[width=8.6cm]{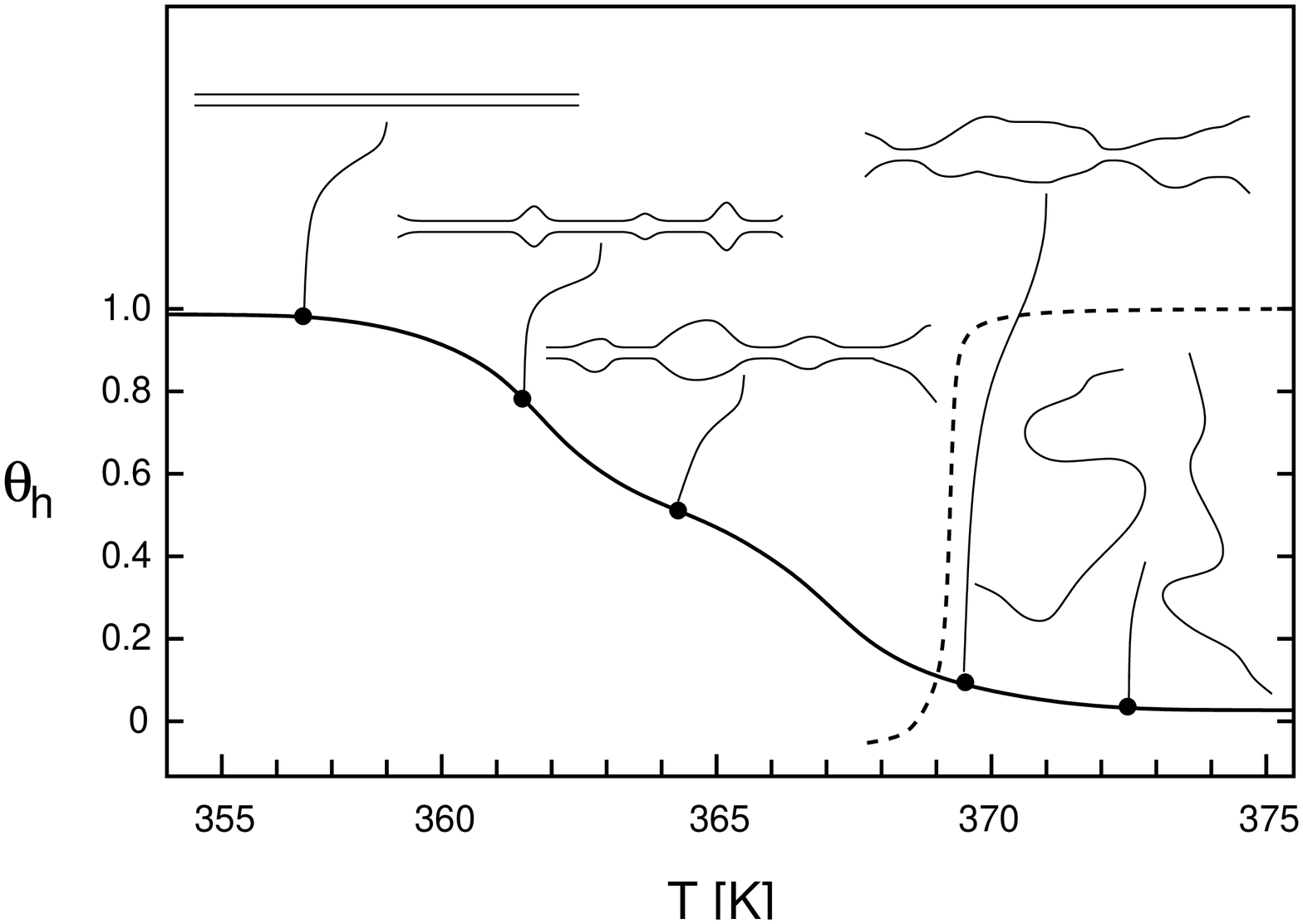}
\caption{Fraction $\theta_h$ of double-helical domains within the DNA as a
function of temperature. Schematic representation of $\theta_h(T)$,
showing the increased formation of bubbles and unzipping from the ends,
until full denaturation has been reached.
\label{melting_curve}}
\end{figure}

\subsection{Physiological background of DNA denaturation}
\label{breathintro}

The Watson-Crick double-helix is the
thermodynamically stable configuration of a DNA molecule under physiological
conditions (normal salt and room/body temperature). This stability is effected
(a) by Watson-Crick H-bonding, that is essential for the specificity of
base-pairing, i.e., for the key-lock principle according to which the
nucleotide Adenine exclusively binds to Thymine, and Guanine only to
Cytosine. Base-pairing therefore guarantees the high level of fidelity
during replication and transcription. (b) The second contribution to
DNA-helix stability comes from base-stacking between neighbouring bps:
through hydrophobic interactions between the planar aromatic bases, that
overlap geometrically and electronically, the bp stacking stabilises
the helical structure against the repulsive electrostatic force between the
negatively charged phosphate groups located at the outside of the DNA
double-strand. While hydrogen bonds contribute only little
to the helix stability, the major support comes from base-stacking
\cite{kornberg,delcourt}.

The quoted ease with which its component chains can come apart and rejoin,
without damaging the chemical structure of the two single-strands, is
crucial to many physiological
processes such as replication via the proteins DNA helicase and polymerase,
and transcription through RNA polymerase. During these processes, the proteins
unzip a certain region of the double-strand, to obtain access to the genetic
information stored in the bases in the core of the double-helix
\cite{kornberg,revzin,watsoncrick}. This unzipping corresponds to breaking
the hydrogen bonds between the bps. Classically, the so-called melting
and reannealing behaviour of DNA has been studied in solution in vitro by
increasing the temperature, or by titration with acid or alkali. During thermal
melting, the stability of the DNA duplex is related to the content of
triple-hydrogen-bonded G-C bps: the larger the fraction of G-C pairs,
the higher the required melting temperature or pH value. Thus, under thermal
melting, dsDNA starts to unwind in regions rich in A-T bps, and then
proceeds to regions of progressively higher G-C content
\cite{kornberg,delcourt}. Conversely, molten, complementary chains of
single-stranded DNA (ssDNA) begin to reassociate
and eventually reform the original double-helix under incubation at roughly
25$^\circ$ below the melting temperature $T_m$ \cite{kornberg}. The relative
amount of molten DNA in a solution can be measured by UV spectroscopy,
revealing large changes in absorption in the presence of perturbed
base-stacking \cite{wartell}. Careful melting studies allow one to obtain
accurate values for the stacking energies of the various combinations
of neighbouring bps, a basis for detailed thermodynamic modelling
of DNA-melting and DNA-structure per se \cite{blake,blossey}. In fact,
thermal melting data have been successfully
used to identify coding sequences of the genome due to the different
G-C content \cite{yeramian,yeramian1,carlon}.

\begin{figure}
\includegraphics[width=8cm]{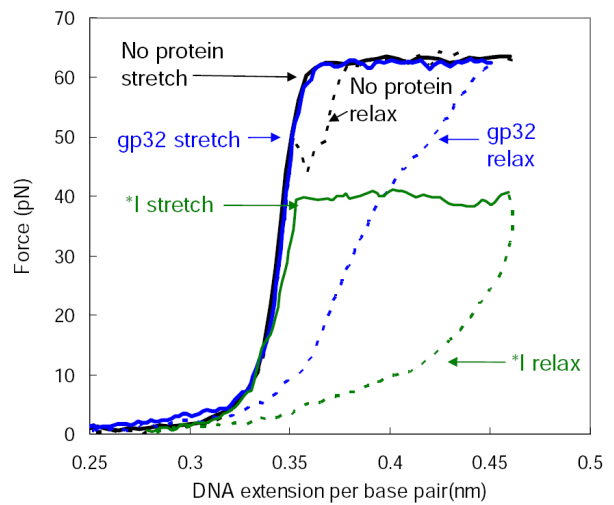}
\caption{Overstretching of double-stranded DNA. The black curve shows the
typical force-extension behaviour of DNA following the rapid worm-like chain
increase until at around 65 pN a plateau is reached. Crossing of the plateau
corresponds to progressive mechanical denaturation. See text for details.
Figure courtesy Mark C. Williams.}
\label{mark_stretch}
\end{figure}

Complementary to thermal or pH induced denaturation, dsDNA can be driven
toward denaturation mechanically, by applying a tensional stress along
the DNA in an optical tweezer trap \cite{williams}. As shown in figure
\ref{mark_stretch}, the force per extension increases in worm-like chain
fashion, until a plateau at approximately 65 pN is reached. This plateau
is sometimes interpreted as new DNA configuration, the S form \cite{busta}.
By a series of experiments, it appears more likely that the plateau corresponds
to the mechanical denaturation transition \cite{williams1}. To first order,
the effect of
the longitudinal pulling translates into an external torque $\mathfrak{T}$,
whose effect is a decrease in the free energy for melting a bp:
\begin{equation}
\Delta G_F=\Delta G_{F=0}-\mathfrak{T}\theta_0,
\end{equation}
where $\theta_0=2\pi/10.35$ is the twist angle per bp of the double helix
\cite{hwa}.

An important application of thermal DNA melting is the Polymerase Chain
Reaction (PCR). In PCR, dsDNA is melted at elevated temperatures into two
strands of ssDNA. By lowering the temperature in a
solution of invariable primers and single nucleotides, each ssDNA is completed
to dsDNA by the key-lock principle of base-pairing \cite{pcr,pcrrev}.
By many such cycles,
of the order of $10^9$ copies of the original DNA can be produced within the
range of hours.\footnote{Most proteins denature
at temperatures between 40 to 60$^\circ$C, including polymerases. In early
PCR protocols, after each heating step new polymerase had to be washed into
the reaction chamber. Modern protocols make use of heat-resistant polymerases
that survive the temperatures necessary in melting. Such heat-resistant proteins
occur, for instance, in bacteria dwelling near undersea thermal vents.}
Again, the error rate due to the underlying biochemistry can be considered
negligible for most purposes. In particular, from the viewpoint
of polymer physics/chemistry, the obtained sample is monodisperse and free
of parasitic reactions, creating (almost) ideal samples for physical studies,
in particular, as any designed sequence of bases can be custom-made in modern
molecular biology labs \cite{snustad}.

While the double-helix is the thermodynamically stable configuration
of the DNA molecule below $T_m$ (at non-denaturing pH), even at
physiological conditions there exist
local denaturation zones, so-called DNA-bubbles, predominantly in A-T-rich
regions of the genome \cite{wartell,poland}. Driven by ambient thermal
fluctuations, a DNA-bubble is a dynamical entity whose size varies by
thermally activated zipping and unzipping of successive bps at the two
forks where the ssDNA-bubble is bordered by the dsDNA-helix. This incessant
zipping and unzipping leads to a random walk in the bubble-size coordinate,
and to a finite lifetime of DNA-bubbles under non-melting conditions, as
eventually the bubble closes due to the energetic preference for the closed
state \cite{wartell,poland}. DNA-breathing typically
opens up a few bps \cite{gueron,krueger}. It has been demonstrated
recently that by fluorescence correlation methods the fluctuations of
DNA-bubbles can be explored on the single molecule level, revealing a
multistate kinetics that corresponds to the picture of successive
zipping and unzipping of single bps.\footnote{Essentially, the zipper
model advocated by Kittel \protect\cite{kittel}.} At room temperature, the
characteristic closing time of an unbounded bp was found to be in
the range 10 to 100 $\mu$sec corresponding to an overall bubble lifetime in
the range of a few msec \cite{altan}. The multistate nature of the
DNA-breathing was confirmed by a UV-light absorption study \cite{zocchi}.
The zipping dynamics of DNA is also investigated by NMR methods
\cite{nmr,nmr1,nmr2}, revealing considerably shorter time scales than the
fluorescence experiments. An interesting finding from NMR studies is the
dramatically different denaturation dynamics in B' DNA, where more than
three AT bps occur in a row \cite{russu}. It is conceivable that fluorescence
correlation and NMR probe
different levels of the denaturation dynamics. Our analysis of the single
DNA fluorescence data reported below demonstrates that, albeit the much
longer time scale, the dependence of the measured autocorrelation function
on the stacking along the sequence is very sensitive, and agrees well with
the quantitative behaviour predicted from the stability data.

The presence of fluctuating DNA-bubbles is essential to the
understanding of the binding of single-stranded DNA binding proteins
(SSBs) that selectively bind to ssDNA, and that play important roles
in replication, recombination and repair of DNA \cite{kornberg1}. One
of the key tasks of SSBs is to prevent the formation of secondary
structure in ssDNA \cite{alberts,snustad}. From the thermodynamical
point of view one would therefore expect SSBs to be of an effectively
helix-destabilising nature, and thus to lower $T_m$
\cite{vonhippel}. However, it was found that neither the gp32 protein
from the T4 phage nor E.coli SSBs do
\cite{vonhippel,vonhippel1,rich}. An explanation to this apparent
paradox was suggested to consist in a kinetic block, i.e., a kinetic regulation
such that the rate constant for the binding of SSBs is smaller than
the one for bubble closing \cite{vonhippel1,rich1}. This hypothesis
could recently be verified in extensive single molecule setups using
mechanical overstretching of dsDNA by optical tweezers in the presence
of T4 gene 32 protein \cite{pant,pant1,pant2}, as detailed below.

\subsection{The Poland-Scheraga model of DNA melting}

\begin{figure}
\includegraphics[width=8cm]{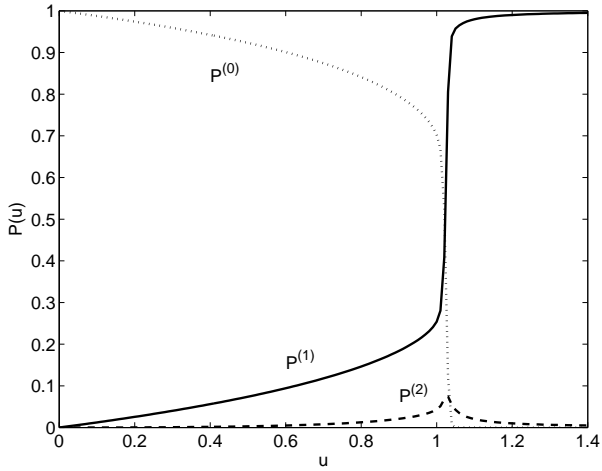}
\caption{Probability of having 0, 1, or 2 bubbles as a function of $u$
for a DNA region of chain length 400 bps. The cooperativity parameter
was $\sigma_0=10^{-3}$ and the loop correction exponent c=1.76 (see text).}
\label{400}
\end{figure}

The most widely used approach to DNA melting in bioinformatics is the
statistical, Ising model-like Poland-Scheraga model (sometimes also referred
to as Bragg-Zimm model) and its variations \cite{poland,cantor,wartell};
see also \cite{kafri,kafri1,richard,hanke}. It
defines
the partition function $\mathscr{Z}$ of a DNA molecule in a grand canonical
picture with arbitrary many bubbles. For simplicity, we will restrict the
following discussion to a single bubble. Below the melting temperature
$T_m$, the one bubble picture is a good approximation: due to the high
energy cost of bubble initiation, the distance between bubbles on a DNA
molecule is large, and bubbles behave statistically independently. In
typical experimental setups for measuring the bubble dynamics (see below),
the used DNA construct is actually designed to host an individual bubble.
For a homopolymer stretch of double-stranded DNA with 400 bps, figure \ref{400}
shows the probabilities to find zero, one or, or two bubbles as a function
of the Boltzmann factor $u=\exp(\Delta G/RT)$ for denaturation of a
single bp.\footnote{In biochemistry, energies are usually measured in calories
per mol. Instead of the Boltzmann factor $\beta=1/k_BT$ commonly used in physics
and engineering, it is therefore convenient to replace the Boltzmann constant
$k_B$ by the gas constant $R=k_BN_A$, where $N_A$ is the Avogadro-Loschmidt
number.} Even at the denaturation transition $\Delta G=0$, it is
quite unlikely to find two bubbles simultaneously.

The free energy $\Delta G$ to break an individual bp are constructed as
follows. We mention two different approaches. Common for both is the
Poland-Scheraga construction of the partition function. We start with
the case that a linear DNA molecule denatures from one of its ends. The
corresponding partition function is \cite{poland,wartell}
\begin{equation}
\label{endpart}
\mathscr{Z}_{\mathrm{end}}(m)=\prod_{x=1}^me^{\Delta G_{x,x+1}/RT}
\end{equation}
where $m$ is the number of broken bps, and $\Delta G_{x,x+1}$ is the stacking
free energy for disruption of the bp at position $x$ measured from the end
of the DNA. The notation explicitly refers to the stacking between the bp at
$x$ and $x+1$.
The first closed bp is located at $x=m+1$. For a homonucleotide, $\Delta
G_{x,x+1}=\Delta G$, while for a given sequence of bps, there come into play
the different stacking energies for the possible combinations of pairs of bps
in sequence\footnote{I.e., an AT bp followed by another AT as different from
an AT followed by a TA, etc.} The stacking energies $\Delta G_{x,x+1}$ have
the following contributions.

The more traditional way to determine the stacking interactions is by fit of
bulk melting curves of DNA constructs containing exclusively pairs of the
specific bp-bp combination such as $(\mathrm{AT/TA})_n$
(see, e.g., \cite{santalucia,delcourt} and references therein). The free energy
used in this automated fit procedure using the MELTSIM algorithm \cite{blake},
\begin{equation}
\label{eblake}
\Delta G^{\mathrm{Mix}}_{x,x+1}=\Delta H^{\mathrm{ST}}_{x,x+1}-T\Delta
S_{x,x+1},
\end{equation}
combines the stacking enthalpy difference $\Delta H^{\mathrm{ST}}_{x,x+1}$
for both hydrogen bond and actual stacking energies, and the entropy difference
$\Delta S_{x,x+1}$ chosen to explicitly depend on the nature of the broken bp.
A recent alternative to determine the stability parameters of DNA was
developed in the group of Frank-Kamenetskii, leading to the free energy
\cite{krueger} 
\begin{equation}
\label{ekrueger}
\Delta G^{\mathrm{Sep}}_{x,x+1}=\Delta G^{\mathrm{ST}}_{x,x+1}+\Delta G
^{\mathrm{HB}}_x
\end{equation}
where the Gibbs free energies $\Delta G^{\mathrm{ST}}_{x,x+1}$ and
$\Delta G^{\mathrm{HB}}_x$ measure the stacking of bps $x$ and $x+1$ and
the hydrogen bonding of bp $x$ \emph{including} the entropy release on
disruption. Note that $\Delta G^{\mathrm{HB}}_x$ is chosen such that it only
depends on the broken bp and has two values for AT and
GC bps, irrespective of the orientation (3' or 5'). The stacking free energies
$\Delta G^{\mathrm{ST}}$ were determined from denaturation at a DNA nick and
show a pronounced asymmetry between AT/TA and TA/AT bonds \cite{krueger}. For an
end-denaturing DNA both descriptions are equivalent (though somewhat different
when one puts numbers), as the breaking of each bp involves the disruption of
one hydrogen bonds of bp $x$ and one stacking with its neighbour.

The difference between the two approaches becomes apparent when we consider
the initiation of a bubble, i.e., a denatured coil enclosed by intact
double-helix. Now, the partition function for a bubble with left fork
position at $x_L$ and consisting of $m$ broken bps,
\begin{equation}
\label{midpart}
\mathscr{Z}_{\mathrm{mid}}(x_L,m)=\Lambda\Sigma(m)\prod_{x=x_L}^{x_L+m}
e^{\Delta G_{x,x+1}/RT},
\end{equation}
differs from (\ref{endpart}) in three respects: (i) While the bubble consists
of $m$ molten bps, $m+1$ stacking interactions need to be broken to create
two boundaries between intact double-strand and the single-strand in the
bubble; the extra stacking interaction is effectively incorporated into
$\Lambda$. (ii) The polymeric nature of the flexible single-stranded
bubbles involves the entropy loss factor $\Sigma(m)=1/(m+D)^c$ with
critical exponent $c$ and the parameter $D$ to take care of finite
size effects\footnote{Usually, $D=1$ is chosen.} \cite{richard,blake,fixman};
(iii) the factor $\Lambda$: In the standard notation, $\Lambda
\equiv\sigma_0=\exp(-F_s/RT)\simeq10^{-4}-10^{-5}$ \cite{lazurkin,blake,%
wartell,poland}, while according to \cite{krueger}, $\Lambda=\xi\simeq10^{-3}$
is called the ring factor.
Interestingly, the cooperativity parameter $\sigma_0$ is of the order of
what corresponds to the singular unbalanced stacking enthalpy for breaking
the first bp to initiate the bubble.
The new stability data lead to a more pronounced asymmetry in opening
probabilities between different bp-bp combinations. The analysis in
references \cite{tobias_prl,tobias_long} demonstrates that the parameters
from \cite{krueger} appear to support better the biological relevance
of the TATA motif\footnote{The four bp {\small $\begin{array}{l}\cdot\mathtt{
TATA}\cdot\\[-0.16cm]\cdot\mathtt{ATAT}\cdot\end{array}$} sequence is one of
the typical codes marking where RNA polymerase starts the transcription process
\cite{alberts,snustad}.} in natural sequences, that is, show a more pronounced
simultaneous opening probability for the TATA motif.

As demonstrated for the autocorrelation function measuring the breathing
dynamics in figure \ref{autocorr}, the description in terms of the
partition function $\mathscr{Z}$ based on the stability parameters from
\cite{krueger} reproduces the experimental data well. The analysis in
\cite{tobias_prl,tobias_long} also indicates that the accuracy of the
model predictions for the bubble dynamics is rather sensitive to the
parameters. It is therefore conceivable that improved fluorescence
measurement of the bubble dynamics may be employed to obtain accurate DNA,
stability parameters, complementing the more traditional melting, NMR, and
gel electrophoresis
bulk methods. It should be noted that the dynamics is strongly influenced
by local deviations from the B configuration of the DNA double helix. Thus,
in local stretches of more than three AT bps in sequence, the B' structure
is assumed, leading to pronouncedly different zipping dynamics \cite{russu}.

Two major questions remain in the thermodynamic formulation of DNA denaturation
and its dynamics. Namely, the exact origin of the bubble initiation factor
$\sigma_0$ (or, alternatively, the ring factor $\xi$ from \cite{krueger}), and
a method to properly calibrate the zipping rate $k$. The
factor $\sigma_0$ is related to the entropic imbalance on opening the
first bp of a bubble: While this requires the breaking of two stacking
interactions, only one bp has access to an increased amount of degrees
of freedom. Still, these degrees of freedom must be influenced by the
fact that the single open bp is coupled to two zipper forks. Currently,
$\sigma_0$ remains a fit parameter. The exact value of the zipping rate
$k$ remains open. While NMR experiments indicate much faster rates in
the nanosecond range ($\simeq 10^8\mbox{ sec}^{-1}$), the fluorescence
correlation measurements produce values in the microsecond range ($\simeq
10^4-10^5\mbox{ sec}^{-1}$). This large discrepancy may be based on the
fact that both methods have different sensitivity to the amplitude of
intra-bp separation. Currently, $k$ is taken as a fit parameter. In the
analysis in \cite{tobias_prl,tobias_long}, we use the stacking parameters
from \cite{krueger} including the value of the ring factor $\xi$, so that
$k$ (a shift along the logarithmic abscissa) is the only adjustable
parameter.

It has been under debate what exact value should be taken for the critical
exponent $c$ entering in the entropy loss factor for a denaturation bubble.
This is connected to the fact that $c>2$ would imply a first order denaturation
transition on melting, while $1<c<2$ would stand for a second order
transition \cite{fisher,richard}. Speculations about a possible first order
transition are connected to the rather distinct spikes in the differential
melting curves \cite{wartell}.\footnote{Due to the rather small DNA samples
used in melting experiments (5000 bp and less \cite{wartell}), claims about
the order of the underlying thermodynamical phase transition should be
considered with some care.} Theoretical polymer physics approaches to explain
a $c>2$ are either based on the inclusion of polymeric self-avoidance
interactions of the bubble with the remainder of the chain \cite{kafri}; or
built on a directed polymer model \cite{garel1}. Despite the elegance of both
approaches, it is an open question how truly they represent the detailed
denaturation behaviour of real DNA \cite{hame_comment,somendra_comment}.
Applying the MELTSIM algorithm to typical sequences, it was found that there
is a connection between the fit result for the cooperativity parameter $\sigma
_0$, whose value is reduced from $\approx 10^{-5}$ to $\approx 10^{-3}$
by assuming $c=2.12$ instead of $1.76$ \cite{blossey}. Below the melting
transition, the typical bubble size is only a few bps, and in that regime the
polymeric treatment of the loop entropy loss is of approximative nature.
Indeed, in the analysis of \cite{krueger} no entropy loss due to polymer
ring formation was included. For the breathing dynamics, we include $c$,
to cover higher temperatures with somewhat larger bubbles,
but find no significant change in the behaviour between $c<2$ and $c>2$,
as long as the exponent is sufficiently close to 2.
We therefore use the value $c=1.76$, that is consistent with the traditional
data fits employed in the determination of the stacking parameters.

\subsection{Fluctuation dynamics of DNA bubbles: DNA breathing}

\begin{figure}
\begin{center}
\includegraphics[width=6.8cm]{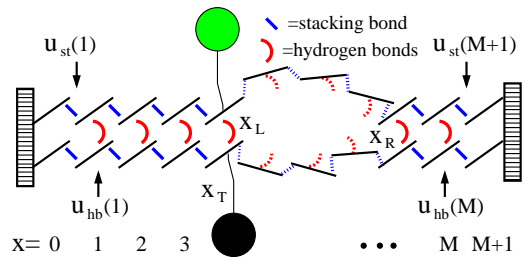}
\end{center}
\caption{Clamped DNA domain with internal bps $x=1$ to $M$,
statistical weights $u_{\rm hb}(x)$, $u_{\rm st}(x)$, and tag position
$x_T$. The DNA sequence enters through the statistical weights $u_{\rm
st}(x)$ and $u_{\rm hb}(x)$ for disrupting stacking and hydrogen bonds
respectively. The bubble breathing process consists of the initiation
of a bubble and the subsequent motion of the forks at positions $x_L$
and $x_R$. See \cite{tobias_long} for details.}
\label{fig:bubbles}
\end{figure}

Below the melting temperature $T_m$, DNA bubbles are intermittent, i.e.,
they form spontaneously due to thermal fluctuations, and after some time close
again. DNA-breathing can be thought of as a biased random walk in the phase
space spanned by the bubble size $m$ and its position denoted, e.g., by the
left zipper fork position $x_L$ \cite{tobias_prl,tobias_long}. The bubble
creation can be viewed
as a nucleation process \cite{landau}, whereas the bubble lifetime corresponds
to the survival time of the first passage problem of relaxing to the $m=0$
state after a random walk in the $m>0$ halfspace
\cite{hame_jpa,tobias_jpc,tobias_prl,tobias_long,tobias_pre}. Bubble
breathing on the
single DNA-bubble level was measured by fluorescence correlation spectroscopy
in \cite{altan}. This technique employs a designed stretch of DNA, in which
weaker AT bps form the bubble domain, that is clamped by stronger GC bonds.
In the bubble domain, a fluorophore-quencher pair is attached. Once the bubble
is created, fluorophore and quencher are separated, and fluorescence occurs.
A schematic of this setup is shown in figure \ref{fig:bubbles}.

The zipper forks move stepwise $x_{L/R}\rightarrow x_{L/R}\pm 1$ with rates
$\mathsf{t}^{\pm}_{L/R}(x_{L/R},m)$. We define for bubble size decrease
\begin{equation}
\label{eq:t_L_plus}
\mathsf{t}^+_L(x_L,m)=\mathsf{t}^-_R(x_L,m)=k/2 \qquad (m\ge 2)
\end{equation}
for the two forks.\footnote{Due to intrachain coupling (e.g., Rouse), larger
bubbles may involve an additional `hook factor' $m^{-\mu}$ \cite{tobias_jpc}.}
The rate $k$ characterises a single bp zipping.
Its independence of $x$ corresponds to the view that bp closure requires the
diffusional encounter of the two bases and bond formation; as sterically AT
and GC bps are very similar, $k$ should not significantly vary with bp stacking.
The rate
$k$ is the only adjustable parameter of our model, and has to be determined
from experiment or future MD simulations. The factor $1/2$ is introduced for
consistency \cite{tobias_jpc,tobias_pre,tobias_prl,tobias_long,tobias_leif}.
Bubble size increase is controlled by
\begin{eqnarray}
\nonumber
\mathsf{t}_{L}^{-}(x_L,m)&=&ku_{\rm st}(x_L) u_{\rm hb}(x_{L})
s(m)/2,
\label{eq:t_L_minus}\\
\mathsf{t}_{R}^{+}(x_L,m)&=&ku_{\rm st}(x_R+1) u_{\rm hb}(x_{R})
s(m)/2,
\label{eq:t_R_plus}
\end{eqnarray}
for $m\ge 1$, where $s(m)=\{(1+m)/(2+m)\}^c$. Finally, bubble initiation and
annihilation from and to the zero-bubble ground state, $m=0 \leftrightarrow
1$ occur with rates
\begin{eqnarray}
\nonumber
\mathsf{t}_G^+(x_L)&=&k\xi's(0)u_{\rm st}(x_L+1)u_{\rm hb}(x_L+1)u_{\rm
st}(x_L+2)\\
\mathsf{t}_G^-(x_L)&=&k.
\label{eq:t_G_plus}
\end{eqnarray}
The rates $\mathsf{t}$ fulfil detailed balance conditions. The annihilation
rate $\mathsf{t}_G^-(x_L)$ is twice the zipping rate of a single fork, since
the last open bp can close either from the left or right. Due to the clamping,
$x_L\ge 0$ and $x_R\le M+1$, ensured by reflecting conditions $\mathsf{t}
_L^-(0,m)=\mathsf{t}_R^+(x_L,M-x_L)=0$. The rates $\mathsf{t}$ together
with the boundary conditions fully determine the bubble dynamics.

\begin{figure}
\begin{center}
\includegraphics[width=9.1cm]{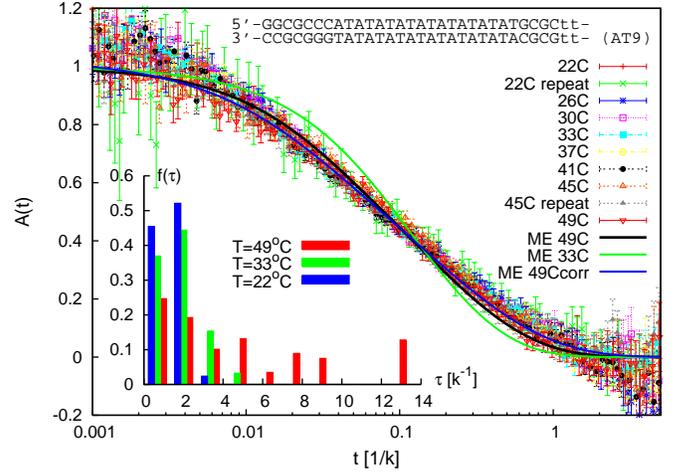}
\end{center}
\caption{Time-dependence of the autocorrelation function $A_t(x_T,t)$ for
the sequence AT9 measured in the FCS setup of reference \cite{altan} at 100mM
NaCl. The full
lines show the result from the master equation, based on the DNA stability
parameters from Krueger et al.~\cite{krueger}. The inset shows the broadening
of the relaxation time spectrum with increasing temperature.}
\label{autocorr}
\end{figure}

In the FCS experiment fluorescence occurs if the bps in a
$\Delta$-neighbourhood of the fluorophore position $x_T$ are open
\cite{altan}. Measured fluorescence time series thus correspond to
the stochastic variable $I(t)$, that takes the value 1 if at least all bps
in $[x_T-\Delta,x_T+\Delta]$ are open, else it is 0. The time averaged
($\overline{\,\,\cdot\,\,}$) fluorescence autocorrelation
\begin{equation}
\label{autocf}
A_t(x_T,t)=\overline{I(t)I(0)}-\overline{I(t)}^2
\end{equation}
for the sequence AT9 from \cite{altan} are rescaled in figure \ref{autocorr}.

We note that an alternative method to obtain precise DNA stability data
may be provided by a DNA construct with two AT-rich zones
between which a shorter GC-rich bridge is located. The first passage
problem corresponding to bubble merging at temperatures between the
melting temperatures of the AT and GC zones was recently calculated
\cite{hansen}, and provides the framework for modified fluorescence
correlation setups similar to the one from reference \cite{altan}.

\subsection{Probabilistic modelling---the master equation (ME)}

DNA breathing is described by the probability
distribution $P(x_L,m,t)$ to find a bubble of size $m$ located at $x_L$
whose time evolution follows the ME
\cite{tobias_jpc,tobias_pre,tobias_prl,tobias_long,tobias_leif}
\begin{equation}
\label{master}
\frac{\partial}{\partial t}P(x_L,m,t)=\mathbb{W} P(x_L,m,t).
\end{equation}
The transfer matrix $\mathbb{W}$ incorporates
the rates $\mathsf{t}$. Detailed balance guarantees equilibration toward
$\lim_{t\to\infty}P(x_L,m,t)=\mathscr{Z}(x_L,m)/\mathscr{Z}$, with
$\mathscr{Z}=\sum_{x_L,m}\mathscr{Z}(x_L,m)$
\cite{vankampen}. The
ME and the explicit construction of $\mathbb{W}$ are discussed at
length in references \cite{tobias_jpc,tobias_long}. Eigenmode analysis and matrix
diagonalisation produces all quantities of interest such as the ensemble
averaged autocorrelation function
\begin{equation}
\label{auto}
A(x_T,t)=\langle I(t)I(0)\rangle-(\langle I\rangle)^2.
\end{equation}
$\langle I(t)I(0)\rangle$ is proportional to the survival density that the
bp is open at $t$ and that it was open initially \cite{tobias_prl,tobias_long}.

In figure \ref{autocorr} the blue curve shows the predicted behaviour of $A(
x_T,t)$, calculated for $T=49^\circ$C with the parameters from
\cite{krueger}. As in the experiment we assumed that fluorophore and quencher
attach to bps $x_T$ and $x_T+1$, that both are required open to produce
a fluorescence signal. From the scaling plot, we calibrate the zipping
rate as $k=7.1\times 10^4/$s, in good agreement with the findings from
reference \cite{altan}. The calculated behaviour reproduces the data within
the error bars, while the model prediction at $T=35^\circ$C shows more
pronounced deviation. Potential causes are destabilising effects of the
fluorophore and quencher, and additional modes that broaden the decay
of the autocorrelation. The latter is underlined by the fact that for
lower temperatures the relaxation time distribution $f(\tau)$, defined
by $A(x_T,t)=\int\exp(-t/\tau)f(\tau)d\tau$, becomes narrower
(figure \ref{autocorr} inset).
Deviations may also be associated with the correction for diffusional motion
of the DNA construct, measured without quencher and neglecting
contributions from internal dynamics \cite{oleg1}. Indeed, the black
curve shown in figure \ref{autocorr} was obtained by a 3\% reduction of the
diffusion time;\footnote{For diffusion time $\tau_D=150\mu$s measured for an RNA
construct of comparable length in \cite{oleg1}.} see details in
\cite{tobias_long}.

A remark on a prominent alternative approach to DNA breathing appears in order. 
This is the Peyrard Bishop Dauxois (PBD) model \cite{peyrard,dauxois} based
on the set of Langevin equations \cite{peyrard_cm}
\begin{eqnarray}
\nonumber
m\frac{d^2y_n}{dt^2}&=&-\frac{dV(y_n)}{dy_n}-\frac{dW(y_{n+1},y_n)}{dy_n}-
\frac{dW(y_n,y_{n-1})}{dy_n}\\
&&-m\gamma\frac{dy_n}{dt}=\xi_n(t).
\label{peyrard}
\end{eqnarray}
Here, $V(y_n)=D_n\left[\exp(-a_ny_n)-1\right]^2$ is a Morse potential
for the hydrogen bonding, $D_n$ and $a_n$ assuming two different values
for AT and GC bps; $W(y,y')=\frac{k}{2}\left[1+\rho\exp\{-\beta(y+y')\}
\right](y-y')^2$ is a nonlinear potential to include bp-bp stacking
interactions between adjacent bps $y$ and $y'$. The parameters $k$, $\rho$,
$\beta$, $\gamma$, and $m$ are invariant of the sequence. Usually, the 
stochastic equations (\ref{peyrard}) is integrated numerically
\cite{peyrard_cm}.
Due to its formulation in terms of a set of Langevin equations, the DPB model
is very appealing, and it is a useful model to study some generic features of
DNA denaturation. The disadvantage of the current formulation of the DBP model
is the fact that it does not include enough parameters to account for the
known independent stability constants of double stranded DNA (in fact, only
two parameters are allowed to vary with the sequence, in contrast to the 12
independent parameters needed to fully describe the bp stacking and hydrogen
bonding \cite{krueger}).
Moreover, there appear to be certain ambiguities in the proper formulation
of boundary conditions in the stochastic integration \cite{the_prl_footnote},
and also with respect to the interpretation of the biological relevance and
computational limitations of the PBD model \cite{peyrard_comment}.
The master equation and Gillespie approach brought forth in references
\cite{tobias_jpc,tobias_pre,tobias_prl,tobias_long,tobias_leif} bridges the
gap between the thermodynamic data for
the bp stacking and hydrogen bonding obtained by various experimental methods,
and the dynamical nature of DNA breathing. It is hoped that both dynamic
models will synergetically be developed further and eventually lead to a
better understanding of DNA denaturation fluctuations.

\subsection{Stochastic modelling---the Gillespie algorithm}

\begin{figure*}
\includegraphics[width=14cm]{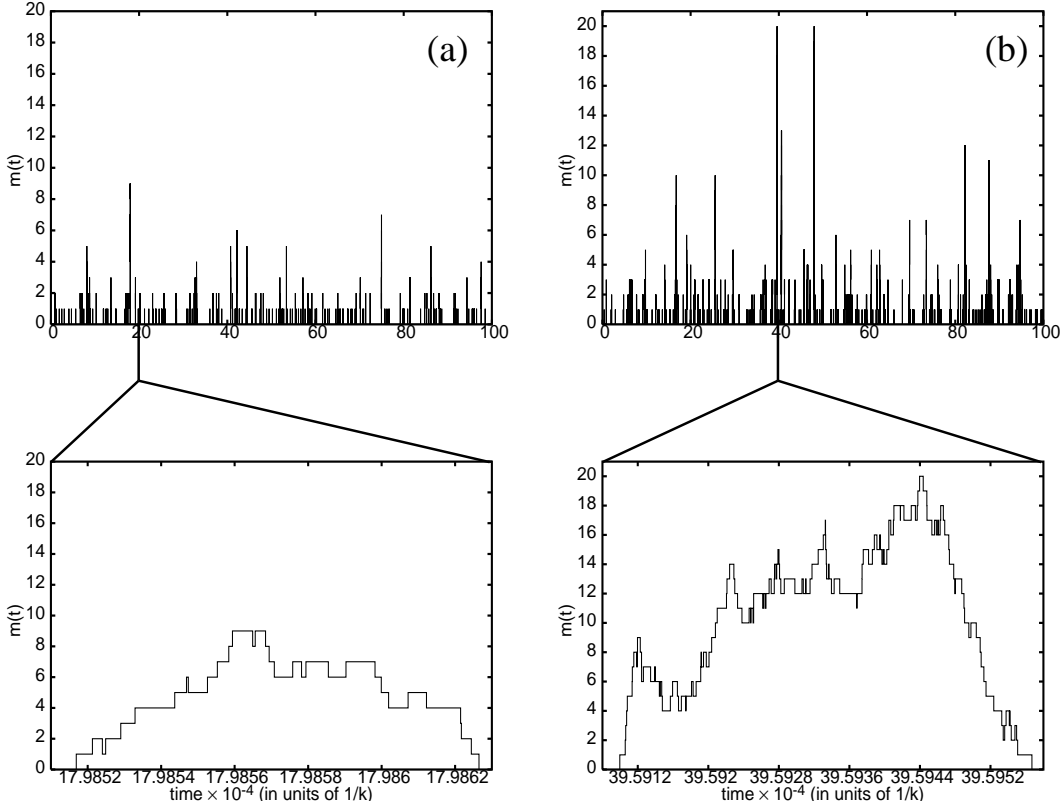}
\caption{Time series of single bubble-breathing dynamics for
$\sigma_0=10^{-3}$, $M=20$, and (a) $u=0.6$ and (b) $u=0.9$. The lower
panel shows a zoom-in of how single bubbles of size $m(t)$ open up and close.}
\label{suman}
\end{figure*}

Despite its mathematically simple form, the master equation (\ref{master})
needs to be solved numerically by inverting the transfer matrix
\cite{tobias_jpc,tobias_long}. Moreover, it produces
ensemble-averaged information. Given the access to single molecule data,
it is of relevance to obtain a model for the fully stochastic time evolution
of a single DNA-bubble, providing a description for pre-noise-averaged
quantities such as the step-wise (un)zipping. With this scope, we
introduced a stochastic simulation scheme for the (un)zipping dynamics,
using the Gillespie algorithm to update the state of the system by determining
(i) the random time between individual (un)zipping events, and (ii) which
reaction direction (zipping, $\leftarrow$, or unzipping, $\rightarrow$) will
occur \cite{suman}. This scheme is efficient computationally, easy to
implement, and amenable to generalisation.

To define the model, we denote a bubble state of $m$ broken bps by the
occupation numbers $b_m=1$ and $b_{m'}=0$ ($m'\neq m$). The stochastic
simulation then corresponds to the nearest-neighbour jump process
\begin{equation}
\label{eq1}
b_0\rightleftarrows b_1\rightleftarrows\ldots
\rightleftarrows b_m\rightleftarrows\ldots
\rightleftarrows b_{M-1}\rightleftarrows b_M,
\end{equation}
with reflecting boundary conditions at $b_0$ and $b_M$. Each jump away
from state $b_m$ occurs after a random time $\tau$, and in random direction
to either $b_{m-1}$ or $b_{m+1}$, governed by the reaction probability
density function\footnote{The original expression for the reaction probability
density function,
$P(\tau,\mu)=b_m\tr^{\mu}(m)$ $\exp\left(-\tau\sum_{m,\mu}b_m\tr^{\mu}(m)
\right)$, that is relevant for consideration of multi-bubble states,
simplifies here due to the particular choice of the
$b_m$.} \cite{gillespie,gillespie1}
\begin{equation}
\label{rpdf}
P(\tau,\mu)=\tr^{\mu}(m)e^{-\left(\tr^+(m)+\tr^-(m)\right)\tau},
\end{equation}
where $\mu\in\{+,-\}$ denotes the unzipping ($+$) or zipping ($-$) of a
bp, and the
jump rates $\tr^{\pm}(m)$ are defined below. From the joint probability
density function (\ref{rpdf}),
the waiting time probability
density function that a jump away from $b_m$ occurs is given by $\psi(\tau)
=\sum_{\mu}P(\tau,\mu)$, i.e., it is Poissonian. The probability that the
bubble size does not change in the time interval $[0,t]$ is given by
$\phi(t)=1-\int_0^t\psi(\tau)d\tau$. The fork position $x_L$ (and thereby
the sequence of bps) is straightforwardly incorporated
\cite{tobias_prl,tobias_long}.

We start the simulations from the completely zipped state, $b_0=1$ at
$t=0$, and measure the bubble size at time $t$ in terms of
$m(t)=\sum_{m=0}^Mmb_m(t)$. The time series of $m(t)$ for a single stochastic
realization is shown in figure \ref{suman}. It is distinct that the bubble
events are very sharp (note the time windows of the zoom-ins), and
most of the time the zero-bubble state $b_0$ prevails due to $\sigma_0\ll 1$.
Moreover, raising the temperature increases the bubble size and lifetime, as
it should. By construction of the simulation procedure, it is guaranteed that an
occupation number $b_m=1$ ($m\neq 0$) corresponds to exactly one bubble.

In a careful analysis, it was shown that the stochastic simulation method
provides accurate information of the statistical quantities of the bubble,
such as opening probability and autocorrelation function \cite{suman}. It
can therefore be used to obtain the same information as the master equation
with the advantage of also giving access to the noise in the system.
With the Gillespie technique, we also obtained the data points in the
graphs in this section.

\subsection{Bacteriophage T7 promoter sequence analysis.}

\begin{figure}
\begin{center}
\includegraphics[width=8.8cm]{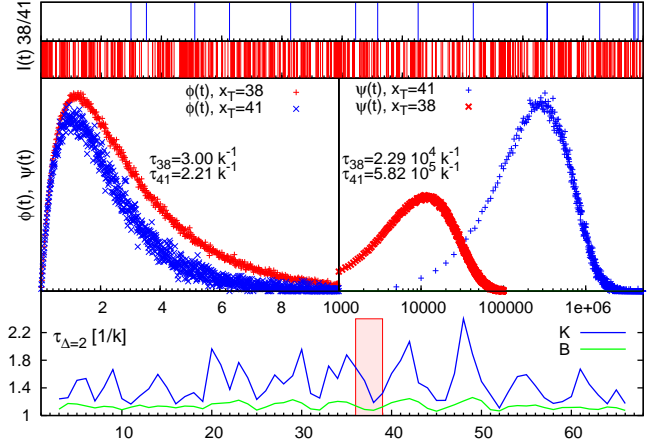}
\end{center}
\caption{Time series $I(t)$ for the T7 promoter, with
$x_T=38$, 41. Middle: Waiting time ($\psi(\tau)$) and bubble
survival time ($\varphi(\tau)$) densities. Bottom: Mean bubble
survival time, $\Delta=2$.}
\label{signal}
\end{figure}

An example from the analysis of the promoter sequence
\begin{equation}
\begin{array}{l}
\mbox{\small\texttt{\textcolor{white}{AAAA}1\textcolor{white}{%
AAAAAAAAAAAAAAAAAA}20}}\\[-0.1cm]
\mbox{\small\texttt{\textcolor{white}{AAAA}|\textcolor{white}{%
AAAAAAAAAAAAAAAAAA}|\textcolor{white}{AAAAAAAAA}\textcolor{white}{AAA}}}\\[-0.1cm]
\mbox{\small\texttt{5'-aTGACCAGTTGAAGGACTGGAAGTAATACGACTC}}\\
\mbox{\small\texttt{\textcolor{white}{AAA}AG}\textcolor{red}{
\texttt{TATA}}\texttt{GGGACAATGCTTAAGGTCGCTCTCTAGGAg-3'}}\\[-0.1cm]
\mbox{\small\texttt{\textcolor{white}{AAAAA}|\textcolor{white}{AAA}|
\textcolor{white}{AAAAAAAAAAAAAAAAAAAAAAAAAA}|\textcolor{white}{AAA}}}\\[-0.1cm]
\mbox{\small\texttt{\textcolor{white}{AAAAA}\textcolor{red}{38}%
\textcolor{white}{AA}\textcolor{blue}{41}\textcolor{white}{%
AAAAAAAAAAAAAAAAAAAAAAAAAA}68\textcolor{white}{AAA}}}
\end{array}
\label{promoter}
\end{equation}
of bacteriophage T7 is shown in figure \ref{signal} \cite{tobias_prl}. It
depicts the time series of $I(t)$ for the tag positions
$x_T=38$ at the beginning of TATA, and $x_T=41$ at the first GC bp after
TATA. It is distinct how frequent bubble events are in TATA in comparison
to the vicinal GC-rich domain (note that AT/TA bps are particularly weak
\cite{krueger}). This is quantified by the waiting time density $\psi(\tau)$,
whose characteristic time scale is more than an order of magnitude longer
for the $x_T=41$ position. In contrast, we observe almost identical
behaviour for the bubble survival density $\phi(\tau)$. Due to the proximity
of $x_T=41$ to TATA, the typical bubble sizes for both tag positions are
similar, and therefore the relaxation time. However, as shown in
figure \ref{signal} bottom, the variation of the mean lifetime obtained from
the master equation is quite small (within a factor 2) for the entire sequence.
The latter graph also shows the insignificant variation according to the
earlier stability parameters by Blake et al.~\cite{blake}.

The results summarised in figure \ref{signal} and further studies in
\cite{tobias_prl,tobias_long} may indicate that it is not solely the
increased opening probability at the TATA motif, as studied in \cite{choi}.
Given the rather short bubble opening times of order of a few $k^{-1}$,
it might be sufficient to induce binding of transcription enzymes (or
other single stranded DNA binding proteins) if only bubble events are
repeated often enough. In the present example, the waiting time between
individual bubble events is increased by a factor of 25 inside the TATA
motif. Guided by such results, detailed future studies combining optical
tweezers overstretching and monitoring transcription initiation may be a
step toward better understanding of this important biochemical process.

We note that the influence of noise (e.g., due to repetition of single
molecule experiments) on the bubble dynamics can also be studied in the
weak noise limit by a WKB method \cite{hans}. This model provides 
information, for instance, about the time it takes a DNA to denature under
temperatures above $T_m$ (mathematically corresponding to a finite time
singularity). Bubble breathing can be mapped on the Coulomb problem of
the Schr{\"o}dinger equation, and the corresponding phase transition
studied \cite{hans_short}.

\subsection{Interaction of DNA bubbles with selectively single-strand
binding proteins}

Let us now come back to the destabilising effect of single-stranded DNA
binding proteins (SSBs) mentioned in section \ref{breathintro}. In a
homopolymer approach, this was studied in a master equation approach
in references \cite{tobias_pre,tobias_jpc}. The quantity of interest is
the joint probability $P(m,n,t)$ to have a bubble consisting of $m$ broken
bps, and $n$ SSBs bound to the two arches of the bubble. In addition to the
rates $\mathsf{t}^{\pm}$ for bubble increase and decrease, the rates
$\mathsf{r}^{\pm}$ for SSB binding and unbinding are necessary to define
the breathing dynamics in the presence of SSBs. On the statistical level,
the effect of the SSBs becomes coupled to the motion of the zipper forks.
Thus, the rate for bubble size decrease is proportional to the probability
that no SSB is located right next to the corresponding zipper fork; and
the rate for SSB binding is proportional to the probability that there is
sufficient unoccupied space on the bubble. Binding is allowed to be
asymmetric, and is related to a parking lot problem in the following sense.
The number $\lambda$ of bps occupied by a bound SSB is usually (considerably)
larger than one. In order to be able to bind in between two already bound SSBs,
the distance between these two SSBs must be larger than $\lambda$. The larger
$\lambda$ the less efficient the SSB-binding becomes, similar to parking
large cars on a parking lot desgined for small vehicles. Apart from the
binding size $\lambda$ of the SSBs, two additional physical parameters
come into play: the unbinding rate $q$ of the SSBs, and their binding
strength $\kappa=c_0K^{\mathrm{eq}}$ consisting of the volume concentration
$c_0$ of SSBs and the equilibrium binding constant $K^{\mathrm{eq}}=v_0\exp
\left(\beta|E_{\mathrm{SSB}}|\right)$, with the typical SSB volume and binding
energy $E_{\mathrm{SSB}}$.

\begin{figure}
\includegraphics[width=8cm]{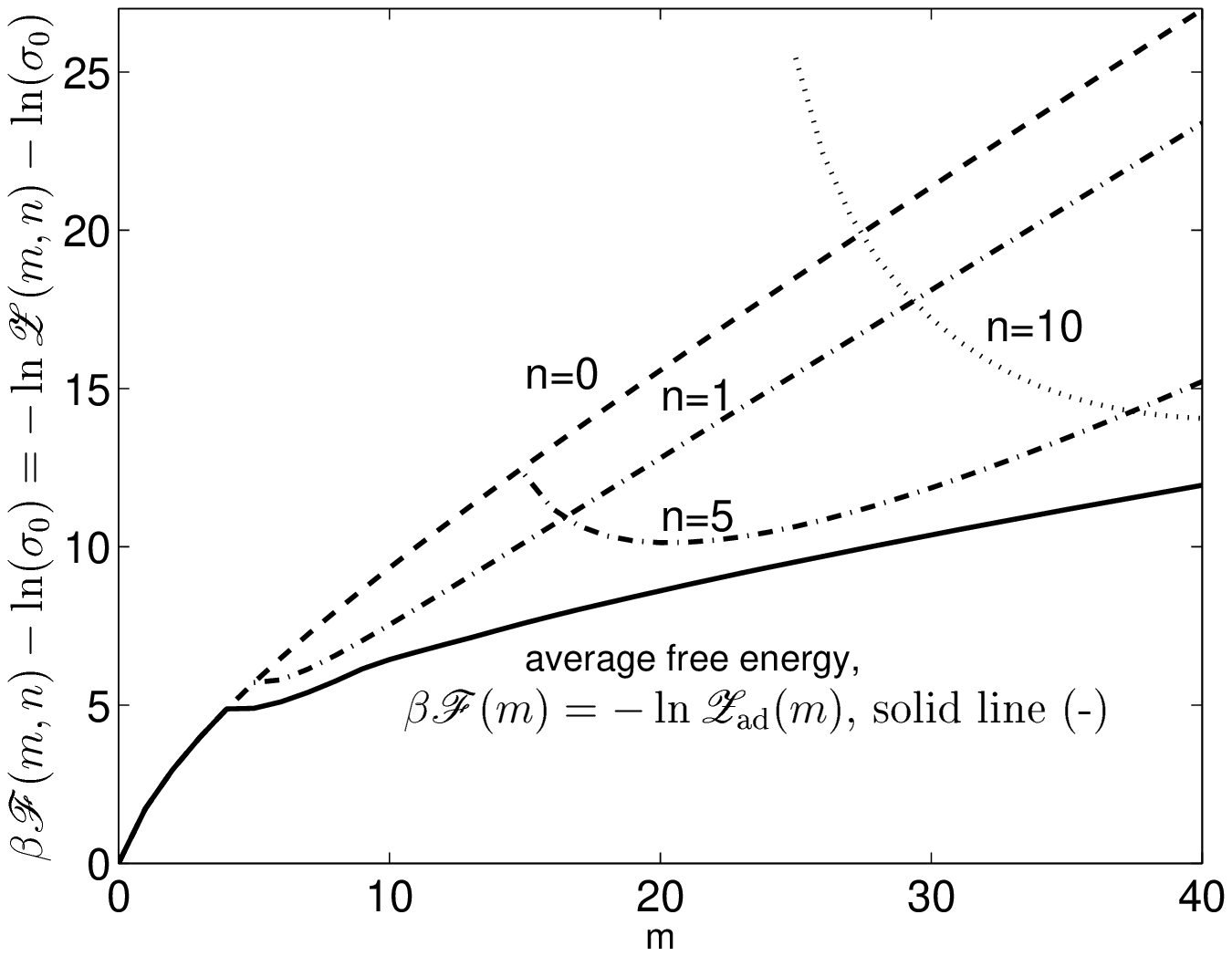}
\includegraphics[width=8cm]{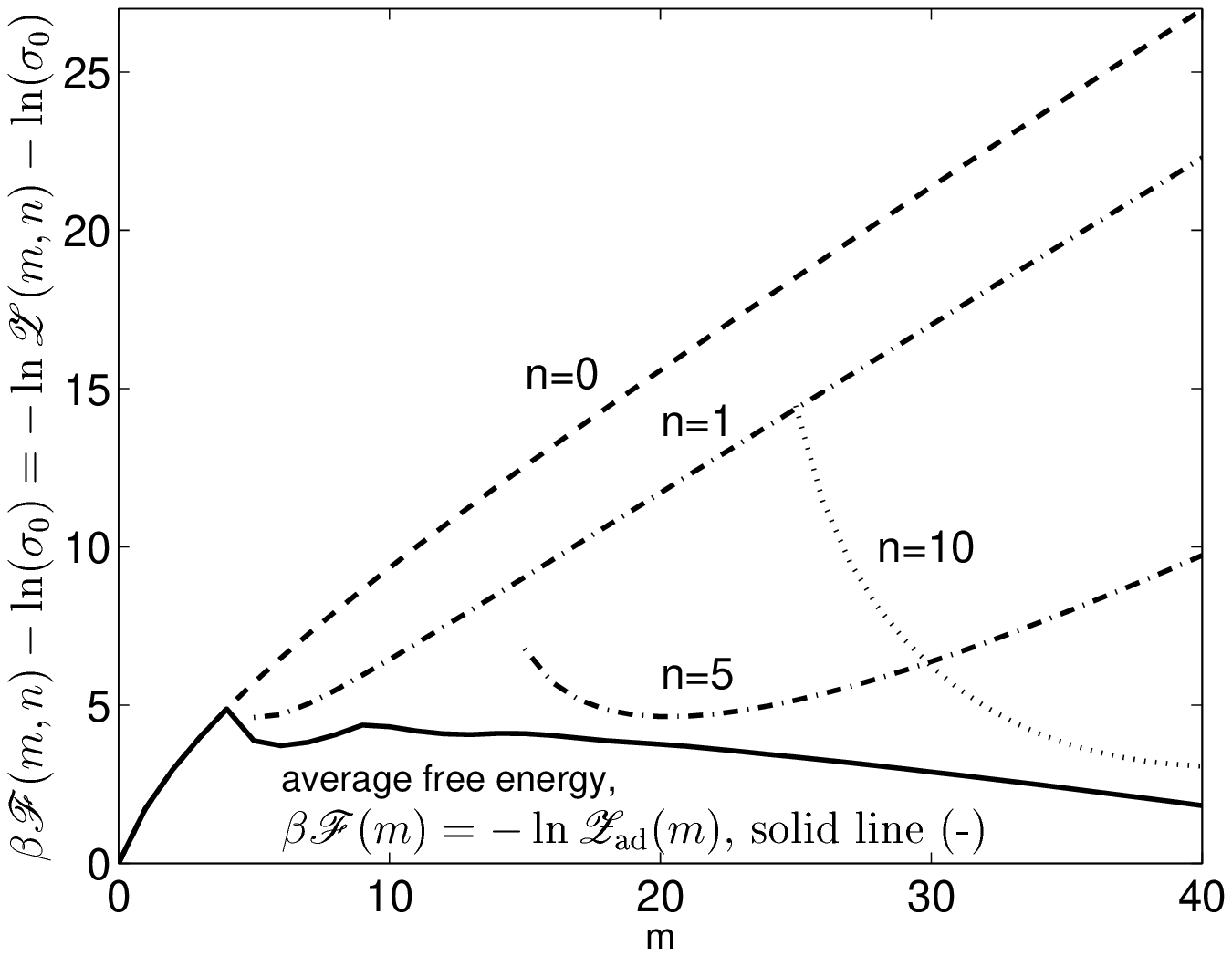}
\caption{Effective free energy in the limit $\gamma\gg 1$ (---),
and `free energy' for various
fixed $n$ ($u=0.6$, $M=40$, $c=1.76$, $\lambda=5$). Top: $\kappa=0.5$; bottom:
stronger binding, $\kappa=1.5$.
\label{ssb_breath}}
\end{figure}

The coupled dynamics of SSB-binding and bubble breathing is discussed in
references \cite{tobias_pre,tobias_jpc}; similar effects in end-denaturing
DNA was studied in \cite{tobias_jpca} in detail. Here, we report
the behaviour of the effective free energy landscape in the limit of fast
SSB-binding in the sense that the dimensionless parameter $\gamma\equiv q/
k$ of SSB-unbinding and bubble zipping rates is large, $\gamma\gg 1$. This
limit allows one to average out the SSB-dynamics and to calculate an effective
free energy, in which the bubble dynamics with the slow variable $m$ runs
off. The result for two different binding strengths $\kappa$ is shown in
figure \ref{ssb_breath}, along with the free energies corresponding to keeping
$n$ fixed. It is distinct that while for lower $\kappa$ the presence of SSBs
diminishes the slope of the effective free energy, for larger $\kappa$ the
slope actually becomes negative. In the first case, that is, the bubble
opening is more likely, but still globally unfavourable. In the latter
case, the presence of SSBs indeed leads to full denaturation. One observes
distinct finite size effects due to $\lambda>1$: only when the bubble reaches
a minimal size $m\ge\lambda$, SSB-binding may occur, a second SSB is allowed
to bind to the same arch only once $m\ge2\lambda$, etc. This effect also
produces the nucleation barrier for full denaturation in the lower plot of
figure \ref{ssb_breath}.
Similar finite size effects were investigated for biopolymer translocation
in references \cite{tobias_pb,amlomme}.
We note that the transition to denaturation could also be achieved by
reaching a smaller positive slope of the effective free energy in the
presence of SSBs, and additional titration or change of the effective
temperature through actual temperature change or mechanical stretching
as performed in the experiments reported in references \cite{pant,pant1,pant2}.

\section{Role of DNA conformations in gene regulation}
\label{generegulation}

Our current understanding of gene regulation to large extent is based on the
experiments by Andr{\'e} Lwoff at Institut Pasteur more than 50 years ago
\cite{lwoff}. Lwoff and his collaborators discovered that while a strain of
\emph{E.coli}, a common intestinal bacterium, divided regularly when
undisturbed,
an unexpected phenomenon occurred when the strain was exposed to UV light:
the bacteria stop growing and after some 90 minutes they burst (lyse),
releasing a load of viruses. These viruses then invade new
\emph{E.coli}.\footnote{Often these viruses are called phages or
bacteriophages---bacteria eaters.} Some
of the newly infected bacteria immediately lyse again, while the rest
divides normally---while carrying the virus in them. This dormant state
(lysogeny) of the bacterium can then be driven toward lysis by renewed UV
exposure.

\begin{figure}
\includegraphics[width=8.2cm]{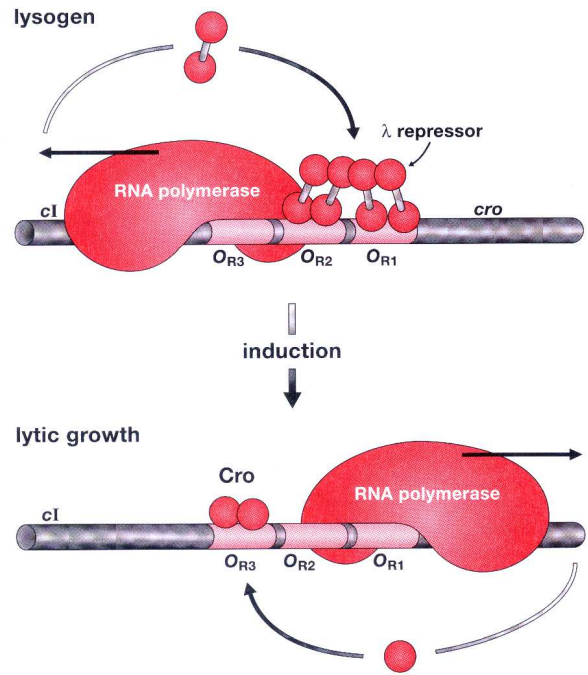}
\caption{Gene regulation, here the example of the (divergent) bacteriophage
$\lambda$ switch after infection of \ecoli. Figure from \cite{ptashne1},
with permission from M. Ptashne. This figure was modified by the author
from the corresponding figure in M. Ptashne, A Genetic Switch: Phage Lambda
and Higher Organisms, 2nd edition  \copyright Blackwell Science, Malden, MA
and Cell Press, Cambridge, MA, with permission.
\label{ptashne}}
\end{figure}

The UV exposure-induced transition from lysogeny to lysis occurs as sketched
in figure \ref{ptashne}. On infection, the bacteriophage $\lambda$ injects its
DNA into \ecoli. In the lysogenic pathway, the viral DNA is integrated into
the host DNA. During lysogeny, repressor dimers bind to certain operator
sites on the $\lambda$ part of the DNA, recruiting RNA polymerase to bind
to the overlapping promoter region(s) and blocking of the vicinal promoter
for the divergent \emph{cro\/} gene. RNA polymerase then transcribes the
\emph{cI} gene to the left of the operator, leading to the expression of
new repressor molecules. UV light, however, leads to cleavage of the
repressor dimers.\footnote{By activation of RecA proteins.} Now the basal
transcription of the gene \emph{cro}, opposite to the \emph{cI\/} gene with
respect to the operator region, leads to the expression of the Cro protein.
Cro bound to the operator then recruits RNA polymerase to the operator,
stabilising the Cro production and blocking \emph{cI}. Simultaneously, a whole
sequence of genes
is being expressed, and the virus reproduces itself inside \ecoli until
lysis occurs. UV light flips the switch from transcription of the
gene \emph{cI\/} maintaining the dormant lysogenic pathway, inducing lysis
that is fostered by transcription of the \emph{cro\/} gene
\cite{ptashne,ptashne1}.

The activity of a gene can be monitored even on the single genome level,
by combining the targeted gene \emph{gI} with the gene leading to synthesis
of GFP, the green fluorescent protein, i.e., when \emph{gI} is transcribed,
then so is the gene for GFP. Occurrence of fluorescence then reports
transcription of \emph{gI}. Connected to questions such as the stability of
a genetic pathway is the search process of a specific gene by regulatory
proteins, that is, how dynamically the binding protein actually locates
the operator on the genome. We address these points in what follows.

\subsection{Physiological background of gene regulation and expression}

The $\lambda$ switch from figure \ref{ptashne} is an example of a relatively
simple mechanism. Even simpler is the well-studied Lac repressor. There,
the \emph{lacZ} gene is expressed by recruitment through the CAP protein
when \ecoli is starved of glucose and exposed to lactose. This enables
\ecoli to digest lactose. In absence of lactose, \emph{lacZ} is blocked
by the rep protein. In general, the expression of a certain gene is just
one element in a cascade of simultaneous and/or hierarchical control units,
such as in the developmental regulatory network of the sea urchin embryo
\cite{davidson}. The basic physiological background is common to all of them:

Genes are the blueprints of proteins. They control physiological processes
but also developmental pathways: from a fertilised egg cell, eventually
all cell types (skin, hair, liver, brain, etc cells) of a human body emerge,
or a skin cell changes colour on sunlight exposure. A gene is but a stretch
of a DNA molecule, typically comprising some 200-1000 bps. Roughly
speaking, a gene is on when it is being transcribed by RNA polymerase,
otherwise it is off. RNA polymerase binds at the promoter region consisting
of some 60 bps close to the beginning of the gene. It then converts
the A, T, G, C code of the gene into a complementary messenger RNA (from
which, in turn, the protein is produced during translation). The stop of
transcription is triggered by a certain sequence at the end of the gene.
Depending on specific conditions of the recruitment by regulatory proteins,
RNA polymerase binding to the promoter of a certain gene is either blocked
(the gene is off), facilitated (high binding affinity of RNA polymerase due
to the (simultaneous) presence of certain protein(s)), or basal (in absence
of any bound regulatory protein, RNA polymerase can still have a minor
affinity to the promoter and then autonomously start transcription).
The transcription mechanism is part and parcel of the central dogma of
molecular biology summarised in figure \ref{crick}.

\begin{figure}
\includegraphics[width=8.6cm]{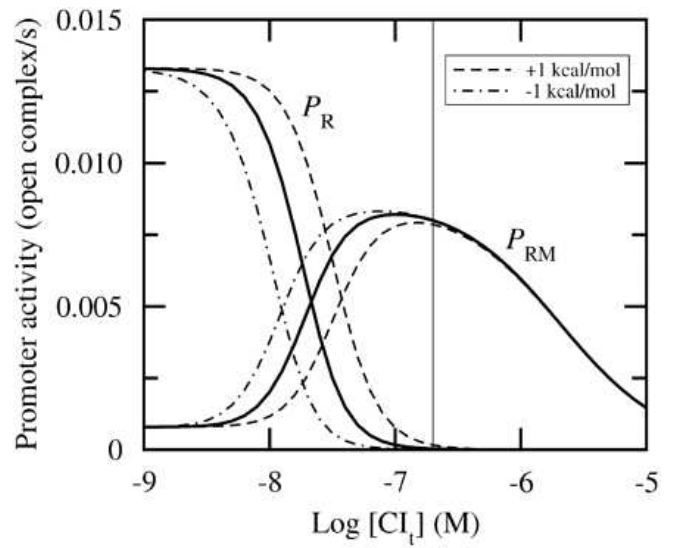}
\caption{Activity of the two $\lambda$ promoters as function of repressor
concentration for vanishing Cro concentration. The full line corresponds
to wild type data, whereas the dashed lines correspond to "mutations". The
thin vertical line corresponds to lysogenic CI concentration. See
reference \cite{bakk2} for details.
\label{bakk}}
\end{figure}

Molecular switches such as the $\lambda$ switch are surprisingly stable
against noise, despite the fact that there are only about 100 repressor
dimers in the entire bacteria cell \cite{reichardt}. Thus, apart from
external induction, lysis occurs by spontaneous induction due to absence of
CI from the operators \cite{metzi,mewo}.
Such noise-induced errors are estimated to occur once
in $10^7$ cell generations \cite{rozanov,little2}. The stability of the
$\lambda$ switch against noise was analysed in terms of a Wentzell-Freidlin
approach \cite{aurell} and by a simulation analysis \cite{aurell1}. The
latter confirmed that the currently known molecular mechanisms used in
modelling the $\lambda$ switch appear sufficient. While the classical
Shea-Ackers model based on a statistical mechanical approach \cite{shea}
is well established and studied numerically \cite{arkin},
it relies on the knowledge of 13 fundamental Gibbs
free binding energies composed to 40 different binding states of regulatory
proteins and RNA polymerase at the two promoters of $\lambda$. Simulation
of the complete $\lambda$ regulatory system proved the understanding of the
mechanisms of the switch \cite{arkin}. Two more recent studies show that the
$\lambda$ switch remains stable even when each of the fundamental Gibbs free
energies is varied within its (appreciable) experimental error. Moreover,
effects of potential mutations resulting in more significant changes of
the binding energies were studied, and it was shown that certain mutations
can even be compensated by parallel mutations influencing other binding
energies (suppressors) \cite{bakk2,bakk3}. A typical result is shown in
figure \ref{bakk}.

\subsection{Binding proteins: Specific and nonspecific binding modes}

Given their very specific function, DNA-binding proteins must recognise
a specific (cognate) sequence of nucleotides along the genome. In fact,
without opening the double helix, the outside of the DNA can be read by
proteins, as the edge of each bp is exposed at the surface. These
patterns are unique only in the major groove of the DNA, this being the
reason why gene regulatory proteins generally bind to the major groove.
Apart from single bp pattern recognition, the protein binding is
sensitive to the special surface features of a certain DNA region. This
local structure of DNA needs to be complementary to the protein structure.
Typical structure patterns (motifs) include helix-turn-helix, zinc 
fingers, leucine zipper, and helix-loop-helix motifs \cite{alberts}.
In bacteria, typical DNA-binding proteins cover some 20 bps or
less. For instance, the lac repressor has a cognate sequence of 21 base
pairs, the CAP protein 16, and the $\lambda$ repressor cI 17 bps
\cite{alberts}. Although the interaction with a single nucleotide within
such a DNA-protein bond is relatively weak, the sum of all matching
nucleotides reaches appreciable values for the overall binding enthalpy,
see below. Moreover, regulatory proteins bound simultaneously can 
significantly enhance the stability of their individual bonds.

A simple model for the binding interaction goes back to the work of Berg
and von Hippel \cite{bvh,berg}. Accordingly, the binding free energy is comprised
of two contributions: (i) the (average) non-specific binding free energy
due to electrostatic interaction with the DNA; and (ii) additional binding
free energy if the sequence of the binding site is sufficiently close to the
best (perfectly matching) sequence. The transition to the non-specific
binding is supposed to occur via a conformational change of the regulatory
protein from one that allows more hydrogen bond-formation to another that
permits closer contact between the positive charges of the binding protein
to the negatively charged DNA backbone \cite{wbvh}. This is supported by
more recent structural studies: While in the non-specific binding mode
the Lac repressor is bound to DNA in a rather loose and fuzzy way
\cite{kaladimos}, it appears much more ordered in the specific mode. In
fact, in the latter the protein induces a bend in the DNA \cite{lewis}.
In case (ii), the additional binding free energy can, to good approximation,
be considered independent and additive. reference \cite{gerland} provides a
review of these issues, and derives the following result.
Accordingly, to satisfy both the thermodynamic and kinetic constraints of
the DNA-binding protein interaction, each additional base mismatch in
comparison to the best sequence amounts to the loss of roughly 2 $k_BT$,
and the optimal value for the transition between best specific binding to
the cognate site and non-specific binding is shown to be some 16 $k_BT$
below the energy of the best binding. This value is quite close to the
$\approx 14$ $k_BT$ found for the difference between specific and nonspecific
binding in \cite{bakk,bakk1}.

The fact that regulatory proteins bind with varying affinity is an important
ingredient in gene regulation: Not all promoters should have the same activity,
because some proteins are required by the cell at much higher levels than
others. Thus, one given regulatory protein, that controls the recruitment to
the promoters of several genes, can act with different strength depending
on the degree of matching with the local sequence.

Non-specific binding can become quite appreciable. It was discovered in
reference \cite{kao} that in the case of the Lac repressor less than 10\% of
the proteins were unbound. In a more recent study using in vivo data of
the $\lambda$ switch, it was found that in a lysogen nearly 90\% of the
repressor protein cI is non-specifically bound. This implies that only
10-20 free cI dimers exist in the \ecoli cell at any time, pointing at
the important role of non-specific binding in the search process of the
cognate site addressed in the following subsection. Under different 
conditions, both cI and Cro are always non-specifically bound by more than
50\%. The corresponding non-specific binding energies were estimated as
7 $k_BT$ \cite{bakk,bakk1}.

We note that in contrast to regulatory proteins, restriction enzymes
have an approximate all-or-nothing matching
condition: If a defined sequence matches the restriction enzyme, it will cut,
otherwise not. Even a single mismatch reduces the action of the restriction
enzyme by orders of magnitude. This distinction from regulatory
protein makes sense as restriction enzymes are survival mechanisms and 
should not just cut the cell's own DNA \cite{stormo}. This does not mean
that restriction enzymes do not bind non-specifically---in fact, this is
an important ingredient of their search process in total analogy to 
regulatory proteins. However, their sole active role occurs on complete
matching.

\subsection{The search process for the specific target sequence}

\begin{figure}
\includegraphics[width=8cm]{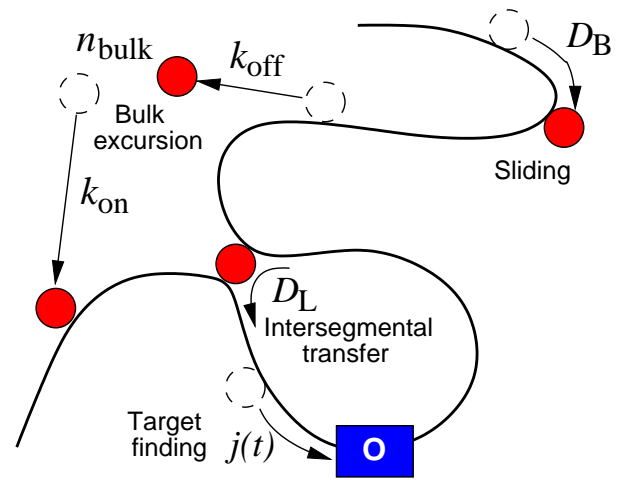}
\caption{Schematic of the search mechanisms in equation (\ref{eq:themodel}).}
\label{fig:ownmechanisms}
\end{figure}

To find their specific (cognate) binding site along the genome, DNA-binding
proteins such as restriction enzymes or transcription factors have to search
megabases along the DNA molecule. The high accuracy of gene expression control
by binding proteins such as in the $\lambda$-switch requires
a fast search and recognition of the target sequence by the proteins.
A simple 3-dimensional (3D) search of the target sequence by the proteins
is not sufficient to explain experimentally measured
target search rates. It has been suggested relatively early \cite{adam,eigen}
that additional search mechanisms such as 1D sliding along the genome are
needed to account for the actual efficiency of the search process. In their
pioneering work, Berg, von Hippel and coworkers established a statistical
model for target search comprising the four fundamental steps, as shown in
figure \ref{fig:ownmechanisms}: (i) 3D macrohops during which the 
protein fully detaches from the genome until after a volume excursion it
rebinds to the DNA (as a good approximation, the landing site on the DNA
after a macrohop can be assumed to be equidistributed and uncorrelated);
(ii) microhops
during which the protein detaches from the DNA but always stays very close
to it (i.e., the microhop takes place within a cylinder whose radius
corresponds to the escape distance of the protein from
the DNA, see \cite{berg}); (iii) 1D sliding along the genome (while
preserving a certain bonding to the DNA due to nonspecific binding);
and (iv) intersegmental jumps.
The latter are mediated by DNA-loops bringing two chemically remote segments
of the DNA close in Euclidean space, see, for instance, \cite{hame_looping} and
references therein. A protein like Lac repressor, which can establish bonds to
two different stretches of dsDNA simultaneously, can then jump from one to
the neighbouring segment.\footnote{Possibly, also other binding proteins are
able to perform intersegmental jumps.} This process might lead to a
paradoxical diffusion behaviour \cite{paradox,dirk}. However, if the
conformational changes in the DNA
are not too slow, both the bulk mediated macrohops and the intersegmental
transfer lead to fast mixing of the enzymes' positions along the chain (as
it was shown for the related problem in \cite{JLum}), and on the mean-field
level can be described by a desorption followed by the absorption at a
random place.

Recently, there has been renewed interest in the targeting problem, both
theoretically (see, for instance, \cite{gerland,slutsky,marko2,coppey}) and
experimentally (e.g., \cite{grillo,record}), including single molecule studies
\cite{shima,pant1,pant2}. Despite the extensive knowledge of specific binding
rates and both specific and non-specific binding free energies, the precise
relative
contributions of the different search mechanisms (and, to some extent, also
the stringent criteria to define these four elementary interactions) are not
fully resolved. Moreover, it has been suggested that under tight(er) binding
conditions, the sliding of the protein becomes subdiffusive due to the local
structure landscape of a heteropolymer DNA \cite{slutsky1}. This complication,
however, is expected to be relaxed in a more loosely bound search mode of the
searching protein \cite{slutsky}. We here adopt the latter view of normal
diffusion, which is corroborated by the experimental study in the next
subsection.

\subsection{A unique situation: Pure one-dimensional search of SSB mutants}

In previous studies, the 1D sliding problem had always been considered
as a problem of 3D diffusion which is enhanced by 1D diffusion. Thus,
workers such as Berg, Winter, and von Hippel \cite{berg} assumed that proteins
nonspecifically bound would on average unbind before finding their
specific binding sites. This results in an enhancement of specific binding
rates that is proportional to the 1D sliding rate, but the overall specific
binding rate depends linearly on protein concentration. These studies neglect
the possibility that the protein finds its specific site before unbinding.
Given the experimental conditions under which transcription factor binding has
been previously studied, this approximation is appropriate. However, as
demonstrated in \cite{igor_prot}, this mechanism, in which the unbinding rate is
much lower than the specific binding rate, occurs for the 1D search of DNA by
the single-stranded DNA binding protein T4 gene 32 protein (gp32). This fast 1D
search rate is essential for gp32 to be able to quickly find specific locations
on DNA molecules that are undergoing replication, and which have large sections
of single-stranded DNA exposed for gp32 binding. The resulting
nonlinear concentration dependence of gp32 binding will likely
have significant effects on gp32's ability to find its replication sites as
well as its ability to recruit other proteins during replication.
If these nonlinear effects also occur for TFs, this
characteristic will strongly affect regulatory processes governed by protein
binding.

\begin{figure}
\begin{center}
\includegraphics[width=6.2cm,angle=270]{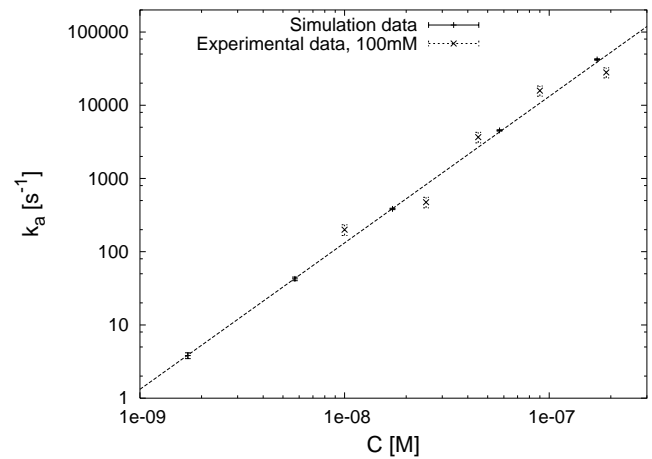}
\end{center}
\caption{Dimensional binding rate $k_a$ in 1/s as function of protein
concentration $C$ in M, for parameters corresponding to 100 mM salt. The
fitted 1D diffusion constant
for sliding along the dsDNA is $D_{\mathrm{1d}}=3.3\cdot 10^{-9}\mathrm{cm}^2/
\mathrm{sec}$, located nicely within the experimental value $10^{-8}\ldots
10^{-9}\mathrm{cm}^2/\mathrm{sec}$ \protect\cite{pant}.}
\label{rates}
\end{figure}

Results from the single DNA overstretching experiment are shown in
figure \ref{rates} along with the results from the theoretical and simulations
analysis from references \cite{igor_prot}. The scaling of search
rate as function of concentration is described by the relation
\begin{equation}
k_a=D_{\mathrm{1d}}n_0^2
\end{equation}
obtained for the pure 1D search of random walkers of line density $n_0^2$
searching along the DNA. For low concentrations, the McGhee and von Hippel
isotherm \cite{mcghee} predicts a linear relation between $n_0$ and the
volume concentration $C$; thus, $k_a\propto C^2$. The experimental evidence
for the purely linear search process, as shown in figure \ref{rates} for 100
mM salt, was found for a large range of salt concentrations, see
references \cite{igor_prot,pant2} for details. The case of high line density
of proteins was discussed in \cite{igor_finite}.

\subsection{L{\'e}vy flights and target search}

We now address the general search process with interchange of 1D and 3D
diffusion, and intersegmental jumps. To this end, we first quickly review
the definition of L{\'e}vy flights \cite{klafternature,physicstoday,report,%
report1,coffey}.

L{\'e}vy flights (LFs) are random walks whose
jump lengths $x$ are distributed like $\lambda(x)\simeq|x|^{-1-\alpha}$ with
exponent $0<\alpha<2$ \cite{hughes}. Their probability density
to be at position $x$ at time $t$ has the characteristic function
$P(q,t)\equiv\int_{-\infty}^{\infty}e^{iqx}P(x,t)dx=\exp\left(-D_{\mathrm{L}}
|q|^{\alpha}t\right)$, a consequence of the generalised central limit theorem
\cite{levy,gnedenko}; in that sense, LFs are a natural extension of normal
Gaussian diffusion ($\alpha=2$). LFs occur in a wide range of systems
\cite{report}; in particular, they represent an optimal search
mechanism in contrast to locally oversampling Gaussian search
\cite{stanley}. Dynamically, LFs can be described by a space-fractional
diffusion equation $\partial P/\partial t=D_{\mathrm{L}}\partial^{\alpha}
P(x,t)/\partial|x|^{ \alpha}$, a convenient basis to introduce additional
terms, as shown below. $D_{\mathrm{L}}$ is a diffusion constant of
dimension $\mathrm{cm}^{\alpha}/\mathrm{sec}$, and the fractional derivative is
defined via its Fourier transform, $\mathscr{F}\{\partial^\alpha P(x,t)
/\partial |x|^\alpha\}=-|q|^\alpha P(q,t)$ \cite{report,report1,coffey}. LFs
exhibit
superdiffusion in the sense that $\langle|x|^{\zeta}\rangle^{2/\zeta}\simeq
D_{\mathrm{L}}t^{2/\alpha}$ ($0<\zeta<\alpha$) \cite{report}, spreading
faster than the linearly growing mean squared displacement of standard
diffusion ($\alpha=2$).
A prime example of an LF is linear particle diffusion to next neighbour
sites on a fast folding (`annealed') polymer that permits intersegmental
jumps at chain contact points (see figure \ref{fig:ownmechanisms}) caused
by polymer looping \cite{paradox,dirk}. In fact, the contour length $|x|$
stored in a loop between such contact points is distributed in 3D like
$\lambda(x)\simeq|x|^{-1-\alpha}$, where $\alpha=1/2$ for Gaussian chains
($\theta$ solvent), and $\alpha\approx 1.2$ for self-avoiding walk chains
(good solvent) \cite{duplantier}.

In our description of the target
search process, we use the density per length $n(x,t)$ of proteins on
the DNA as the relevant dynamical quantity ($x$ is the distance along
the DNA contour). Apart from intersegmental transfer, we include 1D
sliding along
the DNA with diffusion constant $D_{\mathrm{B}}$, protein dissociation
with rate $k_{\mathrm{off}}$ and (re)adsorbtion with rate $k_{\mathrm{on}}$
from a bath of proteins of concentration $n_{\mathrm{bulk}}$.
The dynamics of $n(x,t)$ is thus governed by the equation
\cite{michael}
\begin{eqnarray}
\nonumber
\frac{\partial}{\partial t}n(x,t)=&&\left(D_{\rm B} \frac{\partial^2}{\partial
x^2}+D_{\rm L} \frac{\partial^\alpha}{\partial |x|^\alpha}-k_{\rm
off}\right)n(x,t)\nonumber\\
&&+k_{\rm on}n_{\rm bulk}-j(t)\delta(x).
\label{eq:themodel}
\end{eqnarray}
Here, $j(t)$ is the flux into the target located at $x=0$. We determine
the flux $j(t)$ by assuming that the target is perfectly absorbing: $n(
0,t)=0$ \cite{chechkin_bvp}. Be initially the system at equilibrium, except
that the target is
unoccupied; then, the initial protein density is $n_0=n(x,0)=k_{\rm on}
n_{\rm bulk}/k_{\rm off}$.\footnote{Note that the dimension of the on and
off rates differ; while
$[k_{\mathrm{off}}]=\mathrm{sec}^{-1}$, we chose $[k_{\mathrm{on}}]=
\mathrm{cm}^2/\mathrm{sec}$.} The total number
of particles that have arrived at the target up to time $t$ is
$J(t)=\int_0^t d t'\;j(t')$. We derive explicit analytic
expressions for $J(t)$ in different limiting regimes, and study
the general case numerically. We use $J(t)$ to obtain
the mean first arrival time $T$ to the target; in
particular, to find the value of $k_{\rm off}$ that minimises $T$.

\begin{figure}
\includegraphics[width=6.8cm]{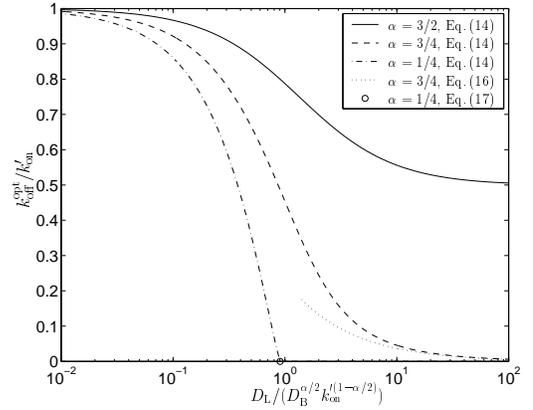}
\caption{Optimal choice of off rate $k_{\rm off}$ as function of the
LF diffusion constant, from numerical evaluation of the model in reference
\cite{michael}. The circle on the abscissa marks where $k_{\rm off}^{\rm opt}$
becomes 0 in the case $\alpha<1/2$.}
\label{searcheff}
\end{figure}

The various regimes of target search embodied in equation (\ref{eq:themodel})
are discussed in detail in references \cite{michael,michael_long}. The main
result for the efficiency of the related search process is summarised in
figure \ref{searcheff}, i.e., which protein unbinding rate $k_{\mathrm{off}}$
optimises the mean search time $T$. Three regimes can be distinguished:

(i) Without L{\'e}vy flights, we obtain $k_{\rm off}^{\rm opt}=k_{\rm on}'$:
the proteins should spend equal amounts of time in bulk and on the DNA.
This corresponds to the result obtained for single protein searching on
a long DNA \cite{slutsky,coppey}. 

(ii) For $\alpha>1$, i.e., when DNA is in the self-avoiding regime,
we find
\begin{equation}
k_{\rm off}^{\rm opt}\sim (\alpha-1)k_{\rm on}':
\end{equation}
The optimal off rate shrinks linearly with decreasing $\alpha$.

(iii) For $\alpha<1$, i.e., when DNA leaves the self-avoiding phase
(e.g., by lowering the temperature or introducing attractive interactions)
the value of $k_{\rm off}^{\rm opt}$ approaches zero as the frequency of
intersegmental jumps ($\propto D_L$) increases: The L{\'e}vy flight
mechanism becomes so efficient that bulk excursions become irrelevant.
At $\alpha=1/2$, the case of the ideal Gaussian chain, we observe a
qualitative change:
When $\alpha<1/2$, the rate $k_{\rm off}^{\rm opt}$ reaches
zero for \emph{finite} values of the rate for intersegmental jumps.
Note that when $\alpha<1$, the spread of the L{\'e}vy flight ($\simeq
t^{1/\alpha}$) grows
faster than the number of sites visited ($\simeq t$), rendering the mixing
effect of bulk excursions insignificant. A scaling argument to understand
the crossover at $\alpha=1/2$ relates the probability density of first
arrival with the width ($\simeq t^{1/\alpha}$) of the Green's function of a
L{\'e}vy flight
$p_{\mathrm{fa}}\simeq t^{-1/\alpha}$. We see that the associated mean
arrival time becomes finite for $0<\alpha<1/2$, even for the infinite chain
limit considered here.

We remark that this model is valid for an annealed DNA only. This means
that the chain can equilibrate (at least, locally) on the typical time
scale between intersegmental jumps. Even though real DNA in solution
might not be fully annealed, features of this analysis will reflect on
the target search. A more detailed study of different regimes of DNA
is under way.

\subsection{Viruses---extreme nanomechanics}

Viruses have played an important role in the discovery of the mechanisms
underlying gene regulation, see, for instance, reference \cite{lwoff}. From
a nanoscience perspective, viruses are of interest on their own part.
During the assembly of many viruses, the viral DNA of several $\mu$m length
is packaged into the capsid, the protein container making up most of the virus, 
by a motor protein. This motor packages the DNA by exerting forces of up to
60 pN or more, causing pressures building up in the capsid of the order
of 6 MPa \cite{smith,zandi}. The size of the capsid spans few tens
of $\mu$m, and is therefore comparable to the persistence length of DNA
\cite{feiss,morita,smith,hud1,hud}. Therefore, fluctuation-based undulations are
suppressed, and the chain can be approximately thought of as being wound
up helically like thread on a bobbin, or like a ball of
yarn. Ultimately, a relatively highly ordered 3D
configuration of the DNA inside the capsid is achieved, which under certain
conditions may even lead to local crystallisation of the DNA
\cite{feiss,morita,pack,hud,kindt,kondev1,ali}.
It is generally argued that this
ordered
arrangement helps to avoid the creation of entanglements or even knots of the
wound-up DNA, thus enabling easy ejection, i.e., release of the DNA once
the phage docks to a new host cell; this ejection is not assisted by the
packaging motor, but it can be facilitated by host cellular DNA polymerase,
which starts to transcribe the DNA and thereby pulls it out of the capsid
\cite{alberts,snustad,kindt,arsuaga}.
Details on the packaging energetics can be found in reference \cite{pack}, and
the works cited therein. Model calculations for the
entropy loss, binding and twist energy, and electrostatic forces that need
to be overcome on packaging reveal, that at higher packaging ratios the
packaging force almost exclusively comes from the electrostatic repulsion.

\section{Functional molecules and nanosensing}

Complex molecules can be endowed with the distinct feature
that they contain subunits which are linked to each other
mechanically rather than chemically \cite{schill}. The investigation of
the structure and properties of such interlocked {\em topological
molecules\/} is
subject of the growing field of chemical topology \cite{frisch};
while speculations about the possibility of catenanes
\footnote{{\em catena\/} (lat.), the chain.} (Olympic rings) date
back to the early 20th century lectures of Willst{\"a}tter, the
actual synthesis of catenanes and rotaxanes
\footnote{{\em rota\/} (lat.), the wheel; {\em axis\/} (lat.),
the axle.} succeeded in 1958 \cite{schill}. Modern organic chemistry
has seen the development of refined synthesis methods to generate topological
molecules.

\subsection{Functional molecules}

In parallel to the miniaturisation in electronics \cite{bishop} and the
possibility of manipulating single (bio)molecules \cite{strick1},
supramolecular chemistry which makes use of chemical topology properties is
coming of age \cite{lehn,dietrich}. Thus, rotaxane-type molecules are believed
to be the building blocks for certain nanoscale machines and motors
\cite{blanco}, so-called hermaphrodite molecules have been shown to perform
linear relative motion (``contraction and stretching'') \cite{jimenez},
and pirouetting molecules have been synthesised \cite{raehm}.
Moreover, topological molecules are thought to become
components for molecular electronics switching devices in memory and
computing applications \cite{pease,lehn1}.
These molecular machines are usually of lower molecular weight, and their
behaviour is essentially energy-dominated in the sense that their
conformations and dynamical properties are governed by external and thermal
activation in an energy landscape. The understanding of the physical
properties and the theoretical modelling of such designer molecules and their
natural biological counterparts has increasingly gained momentum, and the
stage is already set for the next generation of applications
\cite{bishop,strick1,lehn,dietrich,blanco,jimenez,raehm,pease,lehn1,%
montemagno,soong,yurke,metz,motors,frey}.

In reference \cite{hame_cpl} we introduced some basic
concepts for functional molecules whose driving force is entropic rather
than energetic, see also the more recent publications in chemistry journals
\cite{functional,functional1}.
Entropy-functional molecules will be of higher molecular weight
(hundred monomers or above)
in order to provide sufficient degrees of freedom such that entropic
effects can determine the behaviour of the molecule.
The potential for such {\em entropy-driven functional molecules\/}
can be anticipated from the classical Gibbs Free energy
\begin{equation} \label{free}
\mathscr{F} = U - T S \, ;
\end{equation}
in functional molecules, $\mathscr{F}$ is minimised mainly by variation of the
internal energy $U$ representing the shape of the energy landscape of
the functional unit. New types of molecules were proposed for which
$\mathscr{F}$ is minimised by variations of the entropy $S$, while the energies
and chemical bondings are left unchanged \cite{hame_cpl}. The entropy-functional
units of such molecules can be specifically controlled by external parameters
like temperature, light flashes, or other electromagnetic fields
\cite{lehn,dietrich}. We note that DNA is already being studied as
a macromolecular prototype
building block for molecular machines \cite{yurke}.

\begin{figure}
\includegraphics[width=7.8cm]{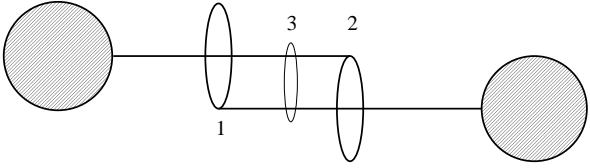}
\caption{Molecular muscle consisting of two interlocked rings
1 and 2 with attached rod-like molecules. Within this structure, sliding
rings, 3, can be placed, which, if activated, tend to contract the muscle
by entropic forces.}
\label{muscle}
\end{figure}

A typical example is the molecule shown in figure \ref{muscle}.
According to the arrangement of the sliding rings 1 and 2, this compound
exhibits the unique feature of a molecule that it can slide {\em laterally\/}.
Suggested as precursors of molecular muscles \cite{jimenez}, this compound
could be propelled with internal entropy-motors, which entropically
adjust the elongation of the muscle. In the configuration shown in
figure \ref{muscle}, the sliding ring 3 creates, if activated, an
entropic force which tends to contract the ``muscle'';
at $T = 300 \, {\mbox K}$ and on a typical scale $x = 10 \, \mbox{nm}$,
the entropic force $k_B T / x$
is of the order of pN, and thus comparable to the force created
in biological muscle cells \cite{gittes}. Molecular muscles of such a make
can be viewed as the nano-counterpart of macroscopic muscle models proposed
by de Gennes \cite{degennesm}, in which the contraction is based on
the entropy difference between the isotropic and nematic phases in liquid
crystalline elastomer films \cite{thomsen}.

Similarly, one might speculate
whether the DNA helix-coil transition \cite{poland}
in multiplication setups could be
facilitated in the presence of pre-ring molecules which in vitro attach
to an opened loop of the double strand and close, creating an entropy
pressure which tends to open up the vicinal parts of the DNA which are
still in the helix state. Finally, considering molecular motors,
it would be interesting to design an externally controllable,
purely entropy-driven rotating nanomotor.

Numerous additional nanoapplications of biopolymers appear in current
literature. An interesting example is the nanomotor created by a DNA
ring in a periodically driven external field, for instance, a focused
light beam inducing localised temperature variations \cite{kulic_schiessel,%
kulic_schiessel_jpc}. The speeds possibly attained by such a device are
of the order of those reached by biological organisms. Such a nanorotor
could be used to stir smallest volumes in higher viscous environments.

\subsection{Nanosensing}

The advances in minituarization of reactors and devices also brings along
the need of probes, by which smallest volumes can be tested. For instance,
microarrays used in genomics require sensors to detect the presence of
certain proteins (often at small concentrations) in a microdish, without
disturbing the environment in the small volume too much. Similarly, single
molecule experiments require specific local detection possibilities.

\begin{figure}
\includegraphics[width=6.8cm]{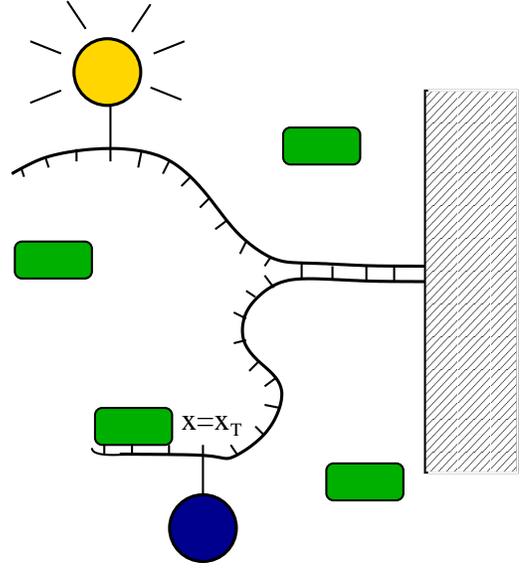}
\caption{Molecular beacon based on local DNA denaturation. The green blobs
may represent single-stranded DNA binding proteins, or more specifically
binding proteins binding or other molecule to a custom designed DNA sequence
along the denaturation fork. Bound proteins stabilize the denatured fork
and change the spectrum of the beacon.}
\label{beacon}
\end{figure}

A fine example for a potential nanosensore is
the blinking behaviour of a fluorophore-quencher pair mounted on
the denaturation wedge as shown in figure \ref{beacon}. This setup, similar
to the ones described in references \cite{altan,oleg1} works as follows.
As long as the dsDNA is intact, fluorophore and quencher are in close
proximity. Once they come apart from one another when the denaturation
wedge opens up, the incident laser light causes fluorescence of the dye.
The on/off blinking of this "molecular beacon" can be monitored in the
focus of a confocal microscope, or, depending on the intensity of the
emitted light, by a digital camera.
The blinking renders immediate information about the state of the bp,
that is tagged by the dye-quencher pair. Fluorescence, that is, indicates that
the bp is currently broken. It is therefore advantageous to define
the random variable $I(t)$ with the property
\begin{equation}
I(t)=\left\{\begin{array}{ll}
0 & \mbox{if base-pair at $x=x_T$ is closed}\\
1 & \mbox{if base-pair at $x=x_T$ is open}\end{array}\right.,
\end{equation}
and in experiments one typically measures the corresponding blinking
autocorrelation function
\begin{equation}
A(t)=\langle I(t)I(0)\rangle-\langle I\rangle_{\mathrm{eq}^2},
\label{A_t}
\end{equation}
where $\langle I\rangle_{\mathrm{eq}}$ is the (ensemble) equilibrium value,
or its spectral decomposition
\begin{equation}
A(t)=\int_0^{\infty}f(\tau)\exp\left(-\frac{t}{\tau}\right)d\tau,
\end{equation}
where
\begin{equation}
f(\tau)=\sum_{p\neq 0}T_p^2\delta\left(\tau-\tau_p\right).
\end{equation}
is called the relaxation time spectrum.

\begin{figure}
\includegraphics[width=7.8cm]{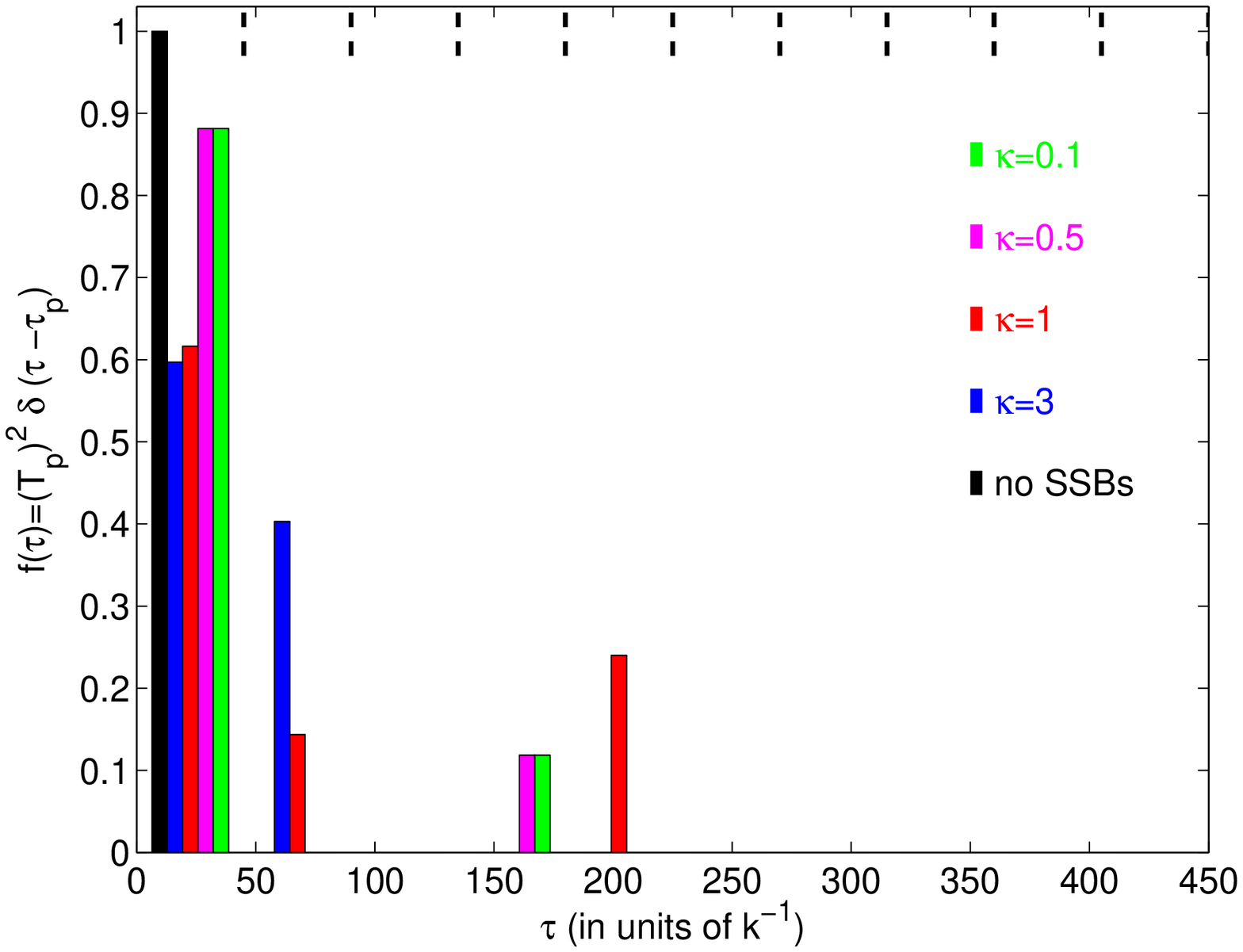}
\includegraphics[width=7.8cm]{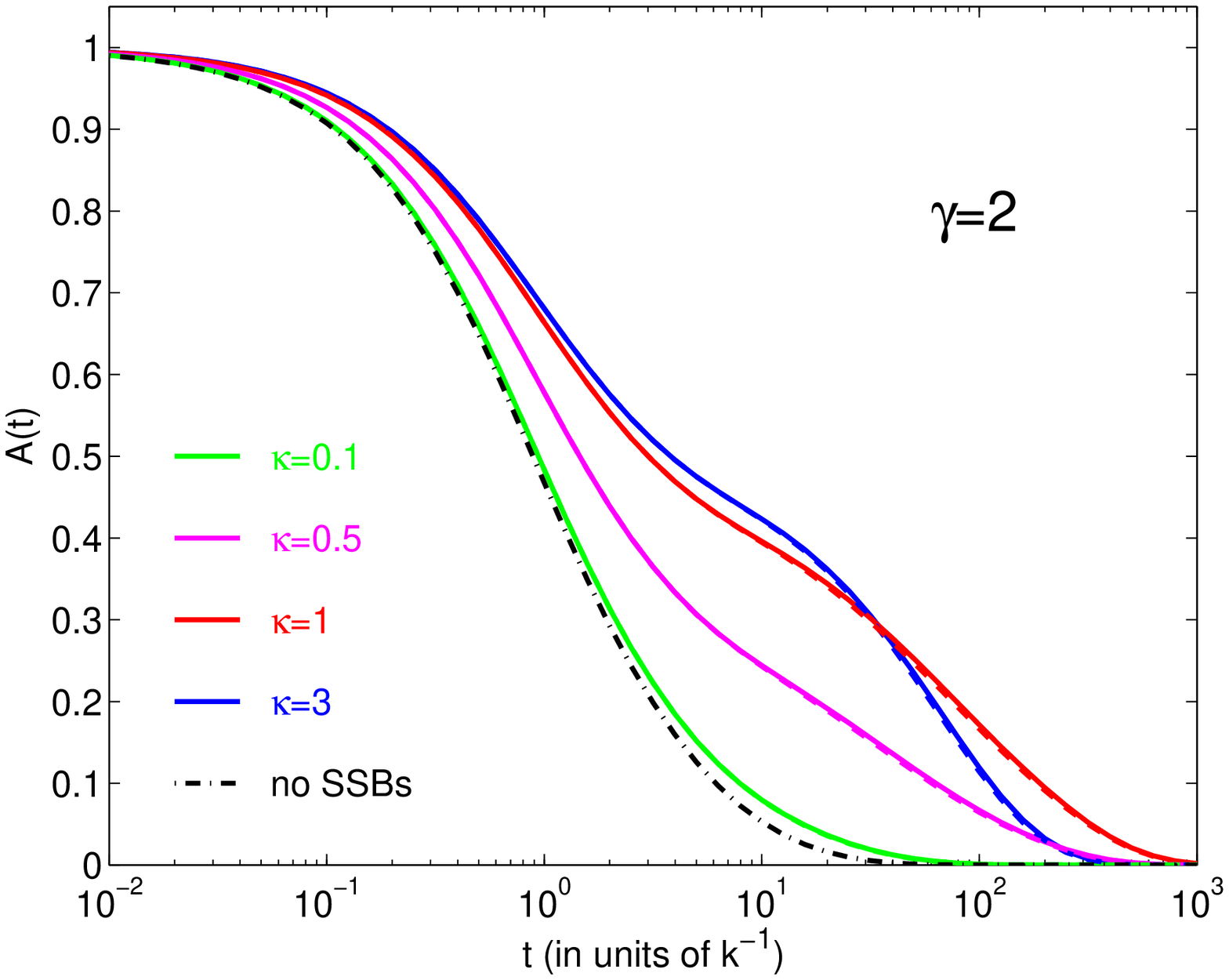}
\caption{Spectral response of the denaturation beacon in the presence of
single-stranded DNA binding proteins. Top: Relaxation time spectrum, bottom:
blinking autocorrelation function.}
\label{blinkbeac}
\end{figure}

Figure \ref{blinkbeac} shows an example for the achievable sensitivity of
such nanobeacons, in an example where the denaturation wedge is in
solution together with a certain concentration (proportional to $\kappa$,
compare Sec. IVG) of selectively single-stranded
DNA binding proteins, as discussed previously. It is distinct how both
measurable signals, $A(t)$ and $f(\tau)$ change with varying SSB-concentration.

\section{Summary}

Biopolymers such as DNA, RNA, and proteins are indispensable for their
specificity and robustness in all forms of life. Given their detailed
physical properties such as DNA's persistence length of some 50nm or
its local denaturation in nano-bubbles already at room temperature,
and biochemically relevant interfaces such as 10-20 bps, they 
deeply stretch into the nanoscience domain. This statement is twofold
in the following sense. Firstly, nanotechniques such as atomic force
microscopes become important tools to manipulate and probe biomolecules
and their interaction even on the single molecule level. Secondly,
biomolecules are entering the stage as nanotools such as nanosensors,
functional molecules, or highly sensitive force transducers.

The possibility to perform controlled experiments on biomolecules, for
instance, to measure the force-extension curves of single biopolymers,
also opens up novel possibilities to test new physical theories. The
foremost examples may be the exploration of persistence lengths and
other polymer physics properties, and the statistical mechanical
concepts relevant for small system sizes. The latter are known under
the keyword of the Jarzynski relation connecting the non-equilibrium
work performed on a physical system with the difference in the thermodynamic
(i.e., equilibrium) potential between initial and final states
\cite{jarzinsky,udo}.
However, there exist by now several similar theories addressing different
physical quantities, such as the concept of entropy production along a single
particle trajectory \cite{udo1}.

This review summaries fundamental physical properties of DNA, and their
relevance for both biological processes and technological applications.
The extensive list of references will be useful for further studies on
specific topics covered herein. We are confident that the role of 
biomolecules in technology, not at least for biomedical applications,
will experience a dramatic increase during the coming years and will
enable us to extend current physical understanding of fundamental
processes.

\acknowledgments

RM and TA acknowledge many helpful and enjoyable discussions with
Jozef Adamcik, Audun Bakk,
Suman Banik, Erika Ercolini, Giovanni Dietler, Hans Fogedby, Yacov Kantor,
Mehran Kardar, Joseph Klafter, Oleg Krichevsky, Michael Lomholt, Maxim
Frank-Kamenetskii, Igor Sokolov, Andrzej Stasiak, Francesco Valle, and
Mark Williams.
RM acknowledges partial funding from the Natural
Sciences and Engineering Research Council (NSERC) of Canada and the Canada
Research Chairs programme. TA acknowledges partial funding from the Wallenberg
foundation.
AH acknowledges funding by the AFOSR (FA9550-05-1-0472) and by the NIH
(SCORE program GM068855-03S1).
SDL acknowledges funding by the Joint DMS/NIGMS
Mathematical Biology Initiative (NIH GM 67242)
and by the National Foundation for Cancer Research
through the Yale-NFCR Center for Protein and Nucleic Acid Chemistry.

\clearpage

\section*{\large Biographical notes}

\begin{minipage}[t]{3.6cm}
\unitlength=1cm
\begin{picture}(4,0)
\put(-5.48,-11.72){\includegraphics{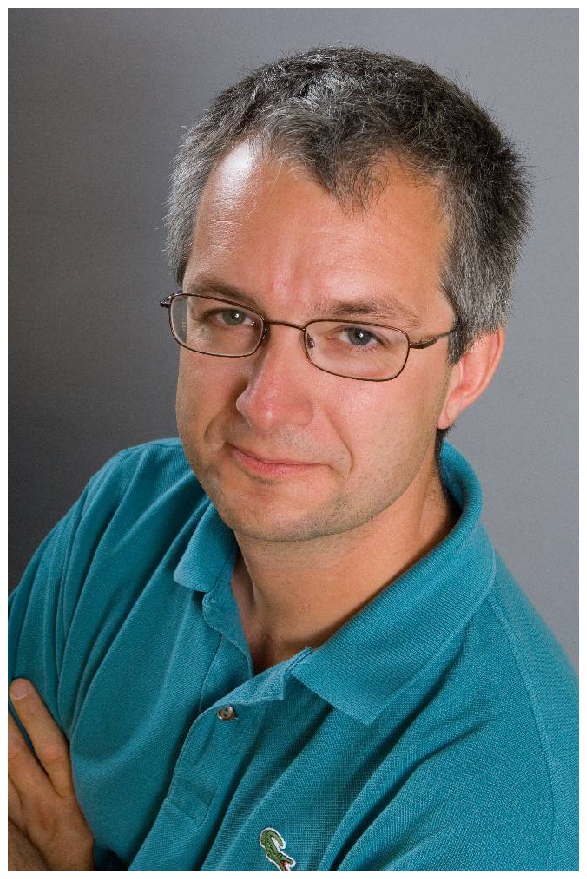}}
\end{picture}
\end{minipage}
\begin{minipage}[t]{4.6cm} \noindent
\textbf{Ralf Metzler} received his doctoral degree in physics from the
University of Ulm, FRG. He then went to Tel Aviv University as Humboldt
Feodor Lynen fellow and later as Minerva Amos de Shalit fellow, to work
with Joseph Klafter. As DFG Emmy Noether fellow, after a period as visiting
scientist at the University of Illinois at Urbana-Champaign with Peter Wolynes,
Ralf moved to the Massachusetts Institute of\\\vspace*{-0.26cm}
\end{minipage}
\noindent Technology (MIT) in Cambridge,
MA, where he worked with Mehran Kardar. In 2002 Ralf was appointed Assistant
Professor at
the Nordic Institute for Theoretical Physics (NORDITA) in Copenhagen,
Denmark. In summer 2006, Ralf assumed his post as
Associate Professor and Canada Research Chair in Biological Physics at
the University of Ottawa, Canada.
Ralf works extensively on biological physics problems and anomalous
stochastic processes. These include DNA physics such as DNA denaturation,
DNA topology and the role of DNA conformations in gene regulation, as well
as anomalous diffusion in biological systems. The latter questions are connected
with L{\'e}vy statistics leading to, respectively, subdiffusion or L{\'e}vy
flights. In his spare time, Ralf enjoys the company of his daughter and wife,
he listens to classical music and is a keen reader of murder mysteries. In
Canada, Ralf is looking forward to nature walks, cross-country skiing and
ice-skating.

\hspace{0.2cm}

\begin{minipage}[t]{3.6cm}
\unitlength=1cm
\begin{picture}(4,0)
\put(-10.82,-19.78){\includegraphics{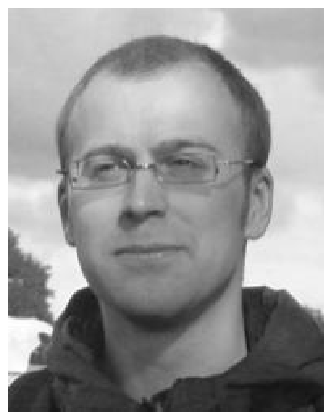}}
\end{picture}
\end{minipage}
\begin{minipage}[t]{4.6cm} \noindent
\textbf{Tobias Ambj{\"o}rnsson} received his PhD from Chalmers and
Gothenburg University, where he worked on the electromagnetic response of
matter. In 2003 he became a NORDITA postdoctoral fellow, working with
Ralf Metzler on the modelling of single biomolecule problems such as
DNA breathing and biopolymer translocation through nanopores. Tobias
re-\\\vspace*{-0.26cm}
\end{minipage}
\noindent
recently received a prestigious fellowship from the Wallenberg foundation,
allowing him to join the group of Robert Silbey at MIT in autumn 2006,
to pursue studies on nanosensors. Tobias enjoys listening to pop music,
travelling, and spending time with friends.

\hspace{0.2cm}

\begin{minipage}[t]{3.6cm}
\unitlength=1cm
\begin{picture}(4,0)
\put(-5.32,-11.12){\includegraphics{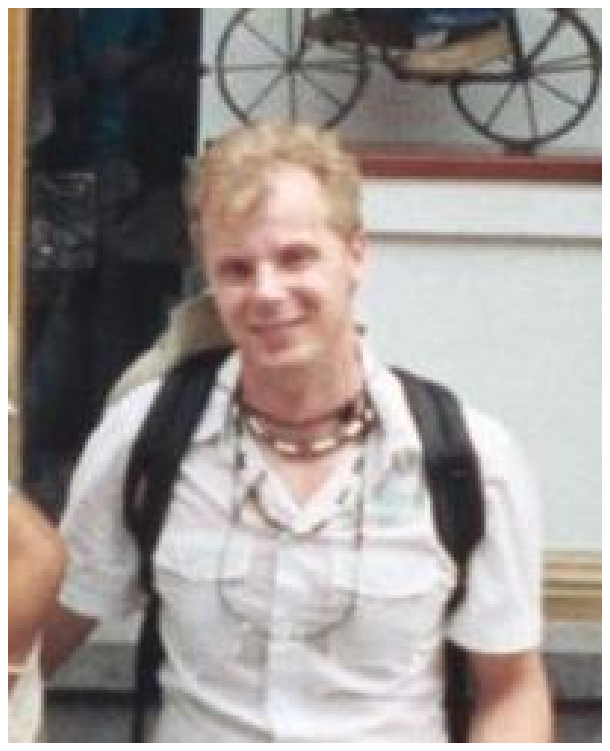}}
\end{picture}
\end{minipage}
\begin{minipage}[t]{4.52cm} \noindent
\textbf{Andreas Hanke}
received his doctoral degree in physics from the University of
Wuppertal in Germany, FRG. He then went to the Massachusetts Institute of
Technology (MIT) for his postdoctoral studies with Mehran Kardar.
After a period as visiting scientist with Michael Schick at the University of
Washington, Se\-attle, he moved to a postdoc-\\\vspace*{-0.26cm}
\end{minipage}
\noindent
toral position in John Cardy's
group at the University of Oxford, UK, followed by postdocs with Udo Seifert
in Stuttgart, FRG, and again with John Cardy in Oxford. In 2004 Andreas
became Assistant Professor of Physics at the University of Texas at
Brownsville. He is also Adjunct Assistant Professor at the Department of
Physics at The University of Texas at Dallas and a member of the Institute
of Biomedical Sciences and Technology at UT Dallas. In addition, his
research includes summer appointments at the UT Dallas NanoTech Institute.
Andreas has worked in the fields of mesoscopic quantum systems, soft
condensed matter physics, and biological physics. Currently he is building
up a theory division in Molecular Biophysics and Nanoscience at the
University of Texas at Brownsville. In his spare time, Andreas enjoys
latin and salsa music and dancing, all
kinds of outdoor activities, and excursions to Mexico.

\hspace{0.2cm}

\begin{minipage}[t]{4.26cm}
\unitlength=1cm
\begin{picture}(4,0)
\put(-7.72,-15.28){\includegraphics{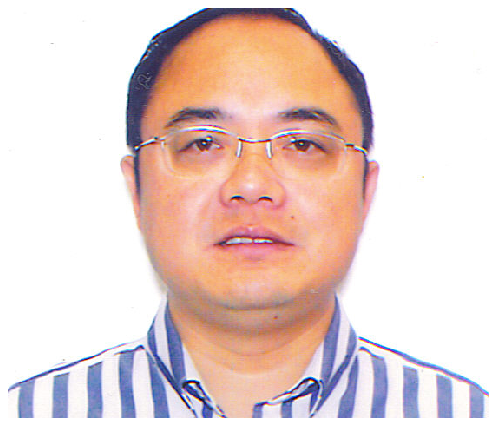}}
\end{picture}
\end{minipage}
\begin{minipage}[t]{3.82cm} \noindent
\textbf{Yongli Zhang} is a postdoctoral fellow of the Jane Coffin Childs
Memorial Fund for Medical Research in Carlos Bustamante’s group at University
of California at Berkeley. He received his Ph.D. in Molecular Biophysics
and Biochem-\\[-0.26cm]
\end{minipage}
\noindent
Biochemistry from Yale University in 2003, under supervision of Donald
Crothers. In December 2006, he will move to the Department of Physiology
and Biophysics at Albert Einstein College of Medicine as an assistant
professor. Yongli has been doing both experimental and theoretical
research in DNA mechanics, chromatin dynamics, and mechanism of chromatin
remodeling. His lab will mainly use single-molecule techniques, such as
optical tweezers and atomic force microscopy, to study mechanism of
molecular motors and dynamics of macromolecular assemblies.

\hspace{0.2cm}

\begin{minipage}[t]{4.26cm}
\unitlength=1cm
\begin{picture}(4,0)
\put(-6.72,-14.48){\includegraphics{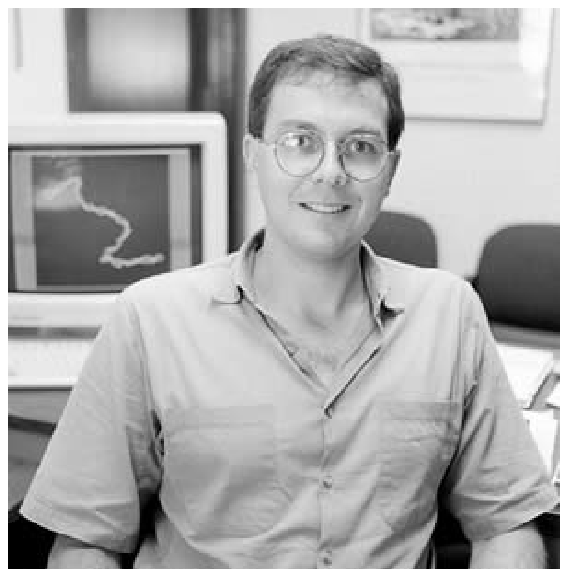}}
\end{picture}
\end{minipage}
\begin{minipage}[t]{3.74cm} \noindent
\textbf{Stephen Levene}
is Associate Professor of Molecular and Cell Biology at
The University of Texas at Dallas. Dr. Levene received his Ph.D. in
Chemistry from Yale University in 1985 and was an American Cancer Society
postdoctoral fellow with Bruno Zimm at University of
California,\\\vspace*{-0.26cm}
\end{minipage}
\noindent
San Diego
until 1989. He then spent one year as a staff scientist at the Human Genome
Center at Lawrence Berkeley Laboratory and was a Program in Mathematics and
Molecular Biology Fellow at University of California, Berkeley in Nicholas
Cozzarelli’s laboratory. Dr. Levene's research interests are in the area of
nucleic-acid structure and flexibility, mechanisms of DNA recombination, the
structural organization of human telomeres, and applications of these areas
to biotechnology. He leads the focus group in Molecular Diagnostics and
Bioimaging in the UT-Dallas Institute of Biomedical Sciences and Technology,
is recipient of an Obermann Interdisciplinary Research Fellowship, a member
of the Biophysical Society, and has served on several NIH study sections.
In addition to scientific pursuits, Dr. Levene is an avid snow skier and
cyclist, having previously competed in both disciplines.

\clearpage

\begin{appendix}

\section*{A polymer primer.}
\label{appA}

In this section, we introduce some basic concepts from polymer physics.
Starting from the random walk model, we define the fundamental measures
of a polymer chain, before introducing excluded volume. For more details,
we refer to the monographs \cite{degennes,doi,grosberg,flory}.

The simplest polymer model is due to Orr \cite{orr}. It models the polymer
chain a a random walk on a periodic lattice with lattice spacing $a$. Then,
each monomer of index $i$ is characterised by a position vector $\mathbf{R}_i$
with $i=0,1,\ldots,N$. The distance between monomers $i$ and $i+1$ is called
$\mathbf{a}_{i+1}=\mathbf{R}_{i+1}-\mathbf{R}_i$. Consequently, the end-to-end
vector of the polymer is
\begin{equation}
\mathbf{r}=\sum_i\mathbf{a}_i.
\end{equation}
Different $\mathbf{a}_i$ have completely independent orientations, such that
we immediately obtain the average ($\langle\cdot\rangle$ over different
configurations) squared end-to-end distance
\begin{equation}
\mathbf{R}_0^2=\langle\mathbf{r}^2\rangle=\sum_{i,j}\langle\mathbf{a}_i\cdot
\mathbf{a}_j\rangle=\sum_i\langle\mathbf{a}_i^2\rangle=Na^2.
\end{equation}
$R_0\simeq N^{1/2}a$ is a measure for the size of the random walk.
An alternative measure of the size of a polymer chain is provided by its radius
of gyration $R_g$, which may be measured by light scattering experiments. It is
defined by
\begin{equation}
\label{appgyr}
R_g^2=\frac{1}{1+N}\sum_{i=0}^N\langle\left(\mathbf{R}_i-\mathbf{R}_G\right)^2
\rangle,
\end{equation}
and measures the average squared distance to the centre of gravity,
\begin{equation}
\mathbf{R}_G=\frac{1}{1+N}\sum_{i=0}^N\mathbf{R}_i.
\end{equation}
Expression (\ref{appgyr}) can be rewritten as
\begin{equation}
R_g^2=(1+N)^{-2}\sum_{i=0}^{N-1}\sum_{j=i+1}
^N\langle\left(\mathbf{R}_i-\mathbf{R}_j\right)^2\rangle.
\end{equation}
With $\mathbf{R}_j-\mathbf{R}_i=\sum_{n=i+1}^j\mathbf{a}_n$, one can easily
show that $R_g^2=a^2N(N+2)/[6(N+1)]$. For large $N$, that is, $R_g\simeq
\frac{a^2}{6}N$, and therefore:
\begin{equation}
R_g\sim R_0\sim aN^{1/2}.
\end{equation}

On a cubic lattice in $d$ dimensions, each step can go in $2d$ directions,
and for a general lattice, each vector $\mathbf{a}_i$ will have $\mu$ possible
directions. The number of distinct walks with $N$ steps is therefore $\mu^N$.
Denote $\mathfrak{N}_N(\mathbf{r})$ the number of distinct walks with
end-to-end vector $\mathbf{r}$, the probability density function for a
given $\mathbf{r}$ is
\begin{equation}
p(\mathbf{r})=\frac{\mathfrak{N}_N(\mathbf{r})}{\sum_{\mathbf{r}}\mathfrak{N}_N(
\mathbf{r})}.
\end{equation}
For large $N$, due to the independence of individual $\mathbf{a}_i$, this
probability density function will acquire a Gaussian shape,
\begin{equation}
p(\mathbf{r})=\left(\frac{d}{2\pi Na^2}\right)^{d/2}\exp\left(-\frac{dr^2}{
2Na^2}\right),
\end{equation}
where the normalisation is such that $\langle\mathbf{r}^2\rangle=Na^2$.
From this expression, we can deduce that the number of degrees of freedom
of a closed random walk chain is proportional to $N^{-d/2}$, the entropy
loss suffered by a chain subject to the constraint $\mathbf{r}=0$. On a
general lattice,
\begin{equation}
\omega\simeq \mu^NN^{-d/2}
\end{equation}
with the connectivity constant $\mu$, a measure for in how many different
directions the next bond vector can point ($\mu=2d$ in a cubic lattice).
At fixed end-to-end distance, the entropy of the random walk becomes
$S(\mathbf{r})=S_0-dr^2/(2Na^2)$ where $S_0$ absorbs all constants.
For the free energy $\mathscr{F}(\mathbf{r})=E-k_BTS(\mathbf{r})$ we therefore
obtain
\begin{equation}
\mathscr{F}(\mathbf{r})=\mathscr{F}_0+\frac{dk_BTr^2}{2R_0^2},
\end{equation}
i.e., the random walk likes to coil, the restoring force $-\nabla
\mathscr{F}(\mathbf{r})$ being linear in $\mathbf{r}$. This is often called
the entropic spring character of a Gaussian polymer. Note that the `spring
constant' increases with temperature (`entropy elasticity').

\begin{figure}
\includegraphics[height=6cm]{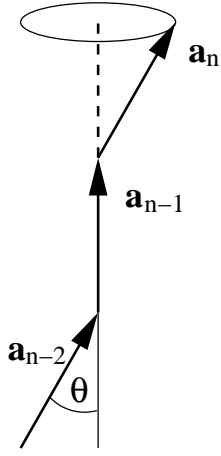}
\caption{Freely jointed chain, in which successive bond vectors include
an angle $\theta$.
\label{fjc}}
\end{figure}

In this random walk model of a polymer chain, it is straightforward to define
the persistence length of the chain. By this we mean that successive vectors
$\mathbf{a}_i$ are not independent, but tend to be parallel. Over long
distance, this correlation is lost, and the chain behaves like a random
walk. Due to the quantum chemistry of the monomers, an adjacent pair of
vectors $\mathbf{a}_i,\mathbf{a}_{i+1}$ includes preferred angles, for
carbon chains leading to the trans/gauche configurations. This feature is
captured schematically in the freely rotating chain as depicted in figure
\ref{fjc}.
Following \cite{doi}, we can obtain the correlation $\langle\mathbf{a}_n
\cdot\mathbf{a}_m\rangle$ as follows. If we fix all vectors $\mathbf{a}_m,
\ldots,\mathbf{a}_{n-1}$, then the average $\langle\mathbf{a}_n\rangle_{
\mathbf{a}_m,\mathbf{a}_{m+1},\ldots,\mathbf{a}_{n-1}\mbox{ fixed}}=
\mathbf{a}_{n-1}\cos\theta$. Multiplication by $\mathbf{a}_m$ produces
\begin{equation}
\langle\mathbf{a}_m\cdot\mathbf{a}_n\rangle_{\mathbf{a}_m,\ldots,\mathbf{a}
_{n-1}\mbox{ fixed}}=\mathbf{a}_m\cdot\mathbf{a}_{n-1}\cos\theta.
\end{equation}
Averaging
over the $\mathbf{a}_m,\ldots,\mathbf{a}_{n-1}$ leads to the recursion
relation $\langle\mathbf{a}_m\cdot\mathbf{a}_n\rangle=\langle\mathbf{a}_m
\cdot\mathbf{a}_{n-1}\rangle\cos\theta$. With the initial condition
$\langle\mathbf{a}^2\rangle=a^2$, we find
\begin{equation}
\langle\mathbf{a}_m\cdot\mathbf{a}_n\rangle=a^2\cos^{|n-m|}\theta.
\end{equation}
Thus, if $\theta=0$, we obtain a rigid rod behaviour, while for $\theta\neq 0$,
there occurs an exponential decay of the correlation between any two bond
vectors $\mathbf{a}_n$ and $\mathbf{a}_m$. This defines a length scale
\begin{equation}
\ell_p\equiv\frac{a}{\log\cos\theta},
\end{equation}
the `persistence length' of the chain. It diverges for $\theta\to 0$, while
for $\theta=90^\circ$, it vanishes, corresponding to the random walk model
discusses above (`freely jointed chain'). As
\begin{equation}
\sum_{k=-\infty}^{\infty}\langle\mathbf{a}_{n+k}\cdot\mathbf{a}_n\rangle=
a^2\left(1+2\sum_{k=1}^{\infty}\cos^k\theta\right)=a^2\frac{1+\cos\theta}{
1-\cos\theta},
\end{equation}
we find $R_0^2=a^2N(1+\cos\theta)/(1-\cos\theta)$, i.e., statistically, the
freely jointed chain behaves the same as the random walk chain, but with a
rescaled monomer length. The statistical unit in a polymer chain is often
taken to be the Kuhn length $\ell_K=2\ell_p$.

Above chain models are often referred to as being phantom, i.e., the chain
can freely cross itself. A physical polymer possesses an excluded volume
and behaves like a so-called self-avoiding chain. Mathematically, this can
be modelled by self-avoiding walks. To include the major effects, it is
sufficient to follow a simple argument due to Flory. Consider a chain with
unknown radius $R$ and internal monomer concentration $c_{\rm int}\simeq N/
R^d$. Assuming that the self-avoiding character is due to monomer-monomer
interactions, the repulsive energy is proportional to the squared concentration,
i.e.,
\begin{equation}
\mathscr{F}_{\rm rep}=\frac{1}{2}Tv(T)c^2,
\end{equation}
with the excluded volume parameter $v(T)$ ($v(T)\equiv(1-2\chi)a^d$ in
Flory's notation, where the $\theta$ condition $\chi=1/2$ corresponds to
ideal chain behaviour). To obtain the total averaged repulsive energy
$\mathscr{F}_{\rm rep|tot}$, we need to average over $c^2$. In a mean field
approach, we take $\langle c^2\rangle\longrightarrow\langle c\rangle^2\sim
c_{\rm int}^2$. We therefore obtain
\begin{equation}
\mathscr{F}_{\rm rep|tot}\simeq Tv(T)c_{\rm int}^2R^d=Tv(T)\frac{N^2}{R^d},
\end{equation}
favouring large values of $R$. This `swelling' competes with the entropic
elasticity contribution $\mathscr{F}_{\rm el}\simeq TR^2/(Na^2)$. The total
free energy becomes
\begin{equation}
\frac{\mathscr{F}}{T}\simeq v(T)\frac{N^2}{R^d}+\frac{R^2}{Na^2},
\end{equation}
with a minimum at $R_F^{d+2}=v(T)a^2N^3$, so that the Flory radius scales
like
\begin{equation}
R_F\sim AN^{\nu} \,\,, \, \mbox{therefore} \,\, \nu=\frac{3}{2+d} \,.
\end{equation}
The values of the exponent $\nu(d=2)=3/4$ and $\nu(d=3)=3/5$ are extremely
close to the best known values $0.75$ and $0.588$.\footnote{An interesting
discussion about the flaws underlying this reasoning can be
found in reference \protect\cite{degennes}.}

\subsection*{Polymer networks.}
\label{duplantier}

A linear excluded volume polymer chain has the size
\begin{equation}
R_g^2\simeq AN^{2\nu}
\end{equation}
with $\nu=0.75$ in $d=2$, and $\nu=0.588$ in $d=3$. Its number of degrees of freedom is
given in terms of the configuration exponent $\gamma$ such that
\begin{equation}
\omega\simeq\mu^NN^{\gamma-1},
\end{equation}
where $\gamma=1.33$ in $d=2$ and $\gamma=1.16$ in $d=3$.

Remarkably, similar critical exponents can be obtained for a general polymer
network of the type shown in figure \ref{network}, as originally
by Duplantier \cite{duplantier,duplantier1},
compare also and in references \cite{ohno,schaefer}: In a network ${\cal G}$
consisting of ${\cal N}$ chain segments of lengths $s_1,\ldots,s_{\cal N}$ and
total length $L=\sum_{i=1}^{\cal N}s_i$, the number of configurations
$\omega_{\cal G}$ scales as
\begin{equation}
\label{network}
\omega_{\cal G}(s_1,\ldots,s_{\cal N})=\mu^{L}s_{\cal N}^{\gamma_{\cal G}-1}
{\cal Y}_{\cal G}\left( \frac{s_1}{s_{\cal N}},\ldots,\frac{s_{{\cal N}-1}}{s_
{\cal N}}\right),
\end{equation}
where ${\cal Y}_{\cal G}$ is a scaling function, and $\mu$ is the effective
connectivity constant for self-avoiding walks. The exponent
$\gamma_{\cal G}$ is given by $\gamma_{\cal G}=1-d\nu{\cal L}+\sum_{N
\ge 1}n_N\sigma_N$, where $\nu$ is the swelling exponent, ${\cal L}$
is the number
of independent loops, $n_N$ is the number of vertices with $N$ outgoing legs,
and $\sigma_N$ is an exponent associated with such a vertex.
In $d=2$, this exponent is given by \cite{duplantier,duplantier1}
\begin{equation}
\label{top_coeff}
\sigma_N=\frac{(2-N)(9N+2)}{64}.
\end{equation}

In the dense phase in 2D
\cite{duplantier2,duplantier3,owczarek,owczarek1,duplantier4,duplantier5},
and at the $\Theta$ transition \cite{duplantier6}, analogous results can be
obtained.

First, consider the dense phase in 2D. If all segments have equal
length $s$ and $L = {\cal N} s$, the configuration number
$\omega_{\cal G}$ of such a network scales as
\cite{duplantier2,duplantier3}
\footnote{Note that due to the factor $\omega_0(L)$ the exponent of
$s$ is $\gamma_{\cal G}$, and not $\gamma_{\cal G}-1$ like in the
expressions used in the dilute phase \cite{duplantier} or at the
$\Theta$ point, for which  $\omega_0(L) \sim L^{- d \nu}$. However,
for 2D dense polymers one has $d \nu = 1$, so that both definitions of
$\gamma_{\cal G}$ are equivalent, cf.~section 3 in reference
\cite{duplantier3}.}
\begin{equation} \label{noc_md}
\omega_{\cal G}(s) \sim \omega_0(L) \, s^{\gamma_{\cal G}} \, \, ,
\end{equation}
where $\omega_0(L)$ is the configuration number of a simple ring of
length $L$. For dense polymers, and in contrast to the dilute phase
or at the $\Theta$ point, $\omega_0(L)$ (and thus $\omega_{\cal G}$)
depends on the boundary conditions and even on the shape of the system
\cite{duplantier3,owczarek,owczarek1,duplantier4,duplantier5}. For example,
for periodic boundary conditions (which
we focus on in this study) corresponding to a 2D torus, one finds
$\omega_0(L) \sim \mu^L \, L^{\Psi - 1}$ with a connectivity
constant $\mu$ and $\Psi = 1$ \cite{duplantier3}. However, the network
exponent
\begin{equation} \label{nexp}
\gamma_{\cal G} = 1 - {\cal L} + \sum_{N\ge 1}n_N\sigma_N
\end{equation}
is {\em universal\/} and depends only on the topology of the
network by the number ${\cal L}$ of independent loops, and by
the number $n_N$ of vertices of order $N$ with vertex exponents
$\sigma_N = (4 - N^2)/32$ \cite{duplantier2,duplantier3}. For a linear chain,
the corresponding exponent $\gamma_{\rm lin} = 19/16$ has been
verified by numerical simulations \cite{duplantier3,grassberger}. For a network
made up of different segment lengths $\left\{s_i\right\}$ of total
length $L = \sum_{i=1}^{\cal N} s_i$, equation (\ref{noc_md})
generalises to (cf.~section 4 in reference \cite{duplantier3})
\begin{equation} \label{noc}
\omega_{\cal G}(s_1, \ldots, s_{\cal N})
\, \sim \, \omega_0(L) \, s_{\cal N}^{\gamma_{\cal G}} \,
{\cal Y}_{\cal G}\left(\frac{s_1}{
s_{\cal N}},\ldots,\frac{s_{{\cal N}-1}}{s_{\cal N}}
\right) \,  ,
\end{equation}
which involves the scaling function ${\cal Y}_{\cal G}$.

\begin{figure}
\includegraphics[width=6.8cm]{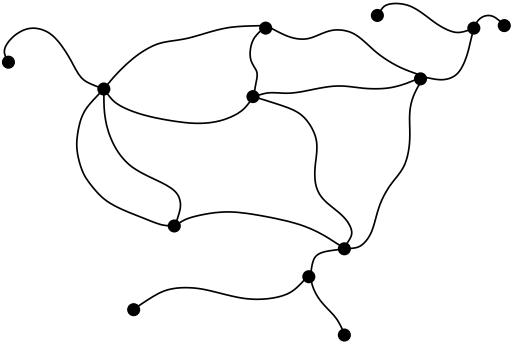}
\caption{Polymer network ${\cal G}$ with vertices ($\bullet$) of different
order $N$, where $N$ self-avoiding walks are joined ($n_1=5$, $n_3=4$, $n_4=3$,
$n_5=1$).
\label{netw}}
\end{figure}
For polymers in an infinite volume and endowed with an attractive
interaction between neighbouring monomers, a different scaling behaviour
emerges if the system is not below but right at the $\Theta$ point
\cite{duplantier3}. In this case the number of configurations of a general
network ${\cal G}$ is given by
\begin{equation} \label{noc_theta}
\overline{\omega}_{\cal G}(s_1, \ldots, s_{\cal N})
\sim \mu^L \, s_{\cal N}^{\overline{\gamma}_{\cal G} - 1} \,
\overline{{\cal Y}}_{\cal G}\left(\frac{s_1}{s_{\cal N}},
\ldots,\frac{s_{{\cal N}-1}}{s_{\cal N}}
\right),
\end{equation}
with the network exponent
\begin{equation} \label{nexp_theta}
\overline{\gamma}_{\cal G} =
1 - d \nu {\cal L} + \sum_{N\ge 1}n_N\overline{\sigma}_N \, \, .
\end{equation}
Overlined symbols refer to polymers at the $\Theta$ point.
In $d = 2$, $\nu = 4/7$ and $\overline{\sigma}_N = (2-N)(2N+1)/42$
\cite{duplantier3}.

\end{appendix}

\end{document}